\documentclass[10pt, oneside]{article}
\usepackage{hyperref}
\usepackage{enumitem} 
\usepackage{easylist}  	
\usepackage{graphicx}			
\usepackage{amssymb}
\usepackage[usenames]{color}
\usepackage{epsf}
\usepackage[usenames,dvipsnames]{pstricks}
\usepackage{epsfig}
\usepackage{pst-grad} 
\usepackage{url}
\usepackage{wrapfig}
\usepackage{float} 
\newenvironment{s-enumerate}{
\begin{itemize}
  \setlength{\itemsep}{1pt}
  \setlength{\parskip}{0pt}
  \setlength{\parsep}{0pt}
}{\end{itemize}}

\let\OLDthebibliography\thebibliography
\renewcommand\thebibliography[1]{
  \OLDthebibliography{#1}
  \setlength{\parskip}{.5ex}
  \setlength{\itemsep}{0pt plus 0.3ex}
}
\title{X-Rule's Precursor is also Logically Universal}
\author{Jos\'e Manuel G\'omez Soto%
\thanks{jmgomezuam@gmail.com, http://matematicas.reduaz.mx/$\sim$jmgomez}%
\hspace{2ex}{\it \small Universidad Aut\'onoma de Zacatecas.}\\
\hspace{2ex}{\it \small  Unidad Acad\'emica de Matem\'aticas. Zacatecas, Zac. M\'exico.}\\
Andrew Wuensche%
\thanks{andy@ddlab.org,  http://www.ddlab.org}%
\hspace{2ex}{\it \small Discrete Dynamics Lab.}\\
\\
{\normalsize Dedidated to the memory of Harold V. McIntosh our friend and teacher.}\\ 
{\normalsize 1929-2015}
}

\date{\small 26 November 2016 (to appear in Journal of Cellular Automata)}

\begin{document}

\maketitle

\vspace{-3ex}
\begin{abstract}

\noindent We re-examine the isotropic Precursor-Rule (of the
anisotropic \mbox{X-Rule}\cite{Gomez2015}) and show that it is also
logically universal.  The Precursor-Rule was selected from a sample of
biased cellular automata rules classified by
input-entropy\cite{Wuensche99}. These biases followed most
``Life-Like'' constraints --- in particular isotropy, but not simple
birth/survival logic.  The Precursor-Rule was chosen for its
spontaneously emergent mobile and stable patterns, gliders and
eaters/reflectors, but glider-guns, originally absent, have recently
been discovered, as well as other complex structures from the
Game-of-Life lexicon.  We demonstrate these newly discovered
structures, and build the logical gates required for universality in
the logical sense.

\end{abstract}

\begin{center}
{\it keywords: universality, cellular automata, glider-gun, logical gates.}
\end{center}

\section{Introduction}
\label{Introduction}

Since the publication of Conway's
\mbox{Game-of-Life}\cite{Gardner1970}, many rules have been found
with, to a degree, similarly interesting behavior\cite{Eppstein2010}. 
Most of these rules are
Game-of-Life variants and  ``Life-Like'' in that they follow a simple
birth/survival logic based on the total of 1s in the outer
neighborhood, which for Life is defined as birth=3, survival=2 or 3 (B3S23). 
The variants are useful to study the nature and
context of the Game-of-Life, to underline why the Game-of-Life itself is so
special, and why the birth/survival scheme  is able in some cases
to produce gliders, glider-guns, logic gates, and universal computation.

\begin{figure}[htb]
\begin{center}
\begin{minipage}[c]{.7\linewidth}
\begin{minipage}[c]{.19\linewidth}
\fbox{\includegraphics[width=1\linewidth, bb=121 10 216 329, clip=]{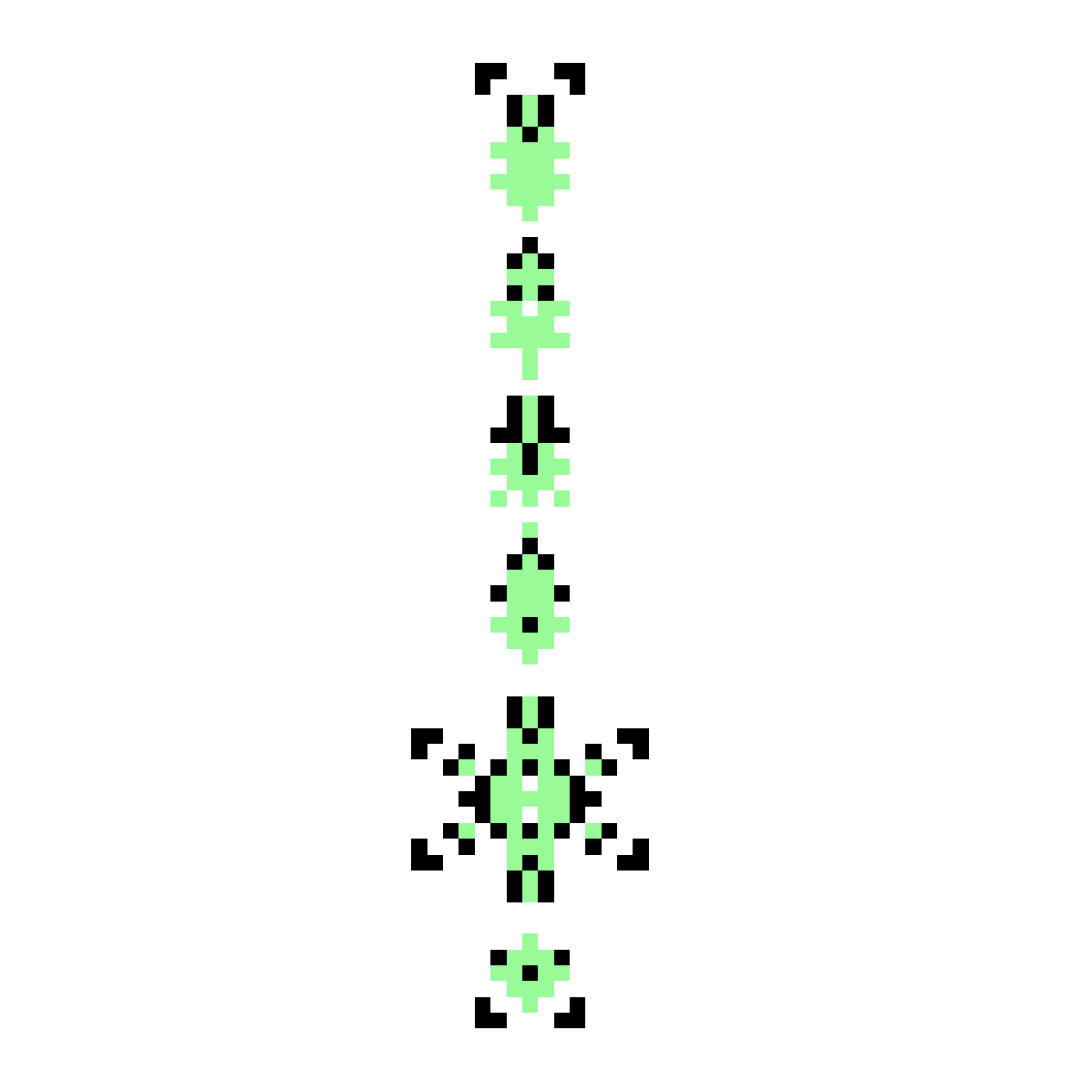}}(a)
\end{minipage}
\hfill
\begin{minipage}[c]{.62\linewidth}
\fbox{\includegraphics[width=1\linewidth, bb=13 18 323 323, clip=]{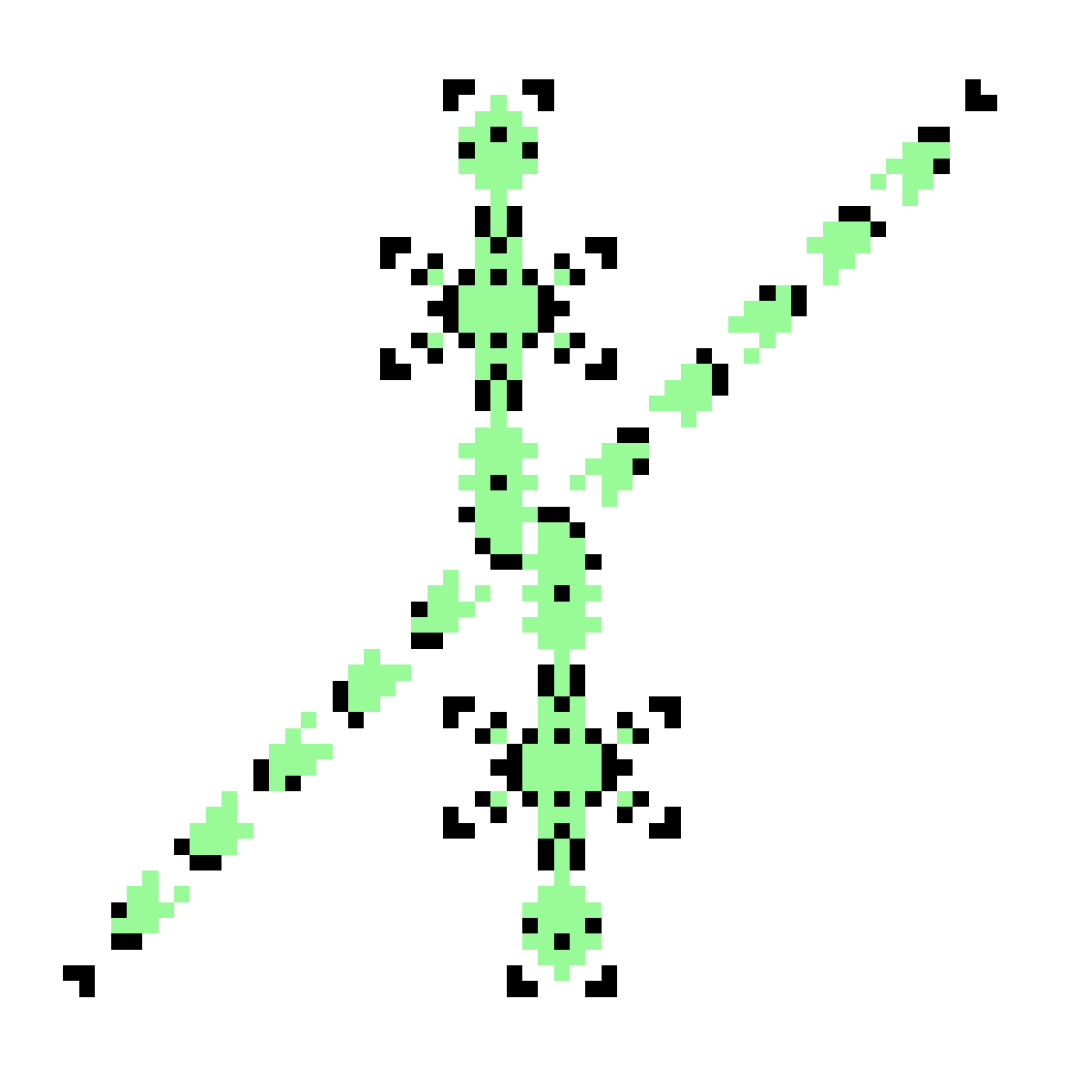}}(b)
\end{minipage}
\end{minipage} 
\end{center} 
\vspace{-3ex}
\caption[Two basic glider-guns in the Precursor-Rule, GGc and GGa]
{\textsf{Snapshots of the two basic glider-guns in the Precursor-Rule, 
GGc\cite{Wildmyron} and GGa\cite{jmgomez}, both have period 19.
(a) GGc shoots 4-phase orthogonal
gliders Gc (shown moving North) with speed $c$/2, where $c$ 
is the speed of light.
(b) GGa (made from two interacting GGc's) 
shoot 4-phase diagonal gliders Ga,
(shown moving NE and SW) with speed $c$/4.
Stable eaters/reflectors stop the glider streams.
Green ``dynamic trails'' are utilised throughout this paper
to highlight the history of travel of moving objects.
The time-step length of dynamic trails, set in DDLab\cite{Wuensche2016,Wuensche-DDLab},
is usually indicated. For (a) dynamic trails=7, for (b) dynamic trails=10.
\label{GGc-GGa-basic}
}}
\end{figure}

To generalise these questions, another approach is to consider rules
without the birth/survival scheme, to study their characteristics, and
thus to enrich the landscape that makes universal computation possible
in binary 2D cellular automata with a Moore neighborhood.  Rules have
been found that do not follow simple birth/survival but are
nevertheless candidates for universality.
To mention two examples, the isotropic R-Rule discovered by
Sapin\cite{Sapin2004} and the anisotropic
\mbox{X-Rule}\cite{Gomez2015} discovered by the authors of this
paper. The Precursor-Rule, defined in 
figures~\ref{precursor-table512} to \ref{precursor28}, belongs to this latter class
of cellular automata, not following birth/survival, but still isotropic, where 
all rotations/flips of a given Moore neighborhood map to the same output.

Gliders and stable ``eaters'' emerge spontaneously in the Game-of-Life, 
but a glider-gun was originally absent and only
subsequently discovered by \mbox{Gosper\cite{Gardner1970,Berlekamp1982}}. 
In a curious imitation of this order of events, glider-guns in the Precursor-Rule have only
recently been discovered. Thanks to these glider-guns, its possible to
build the logical gates for negation, conjunction and disjunction and
satisfy the third of Conway's three conditions for
universality\cite{Berlekamp1982} to demonstrate universal computation in the logical sense.

\begin{figure}[htb]
\begin{center}
\includegraphics[width=.9\linewidth]{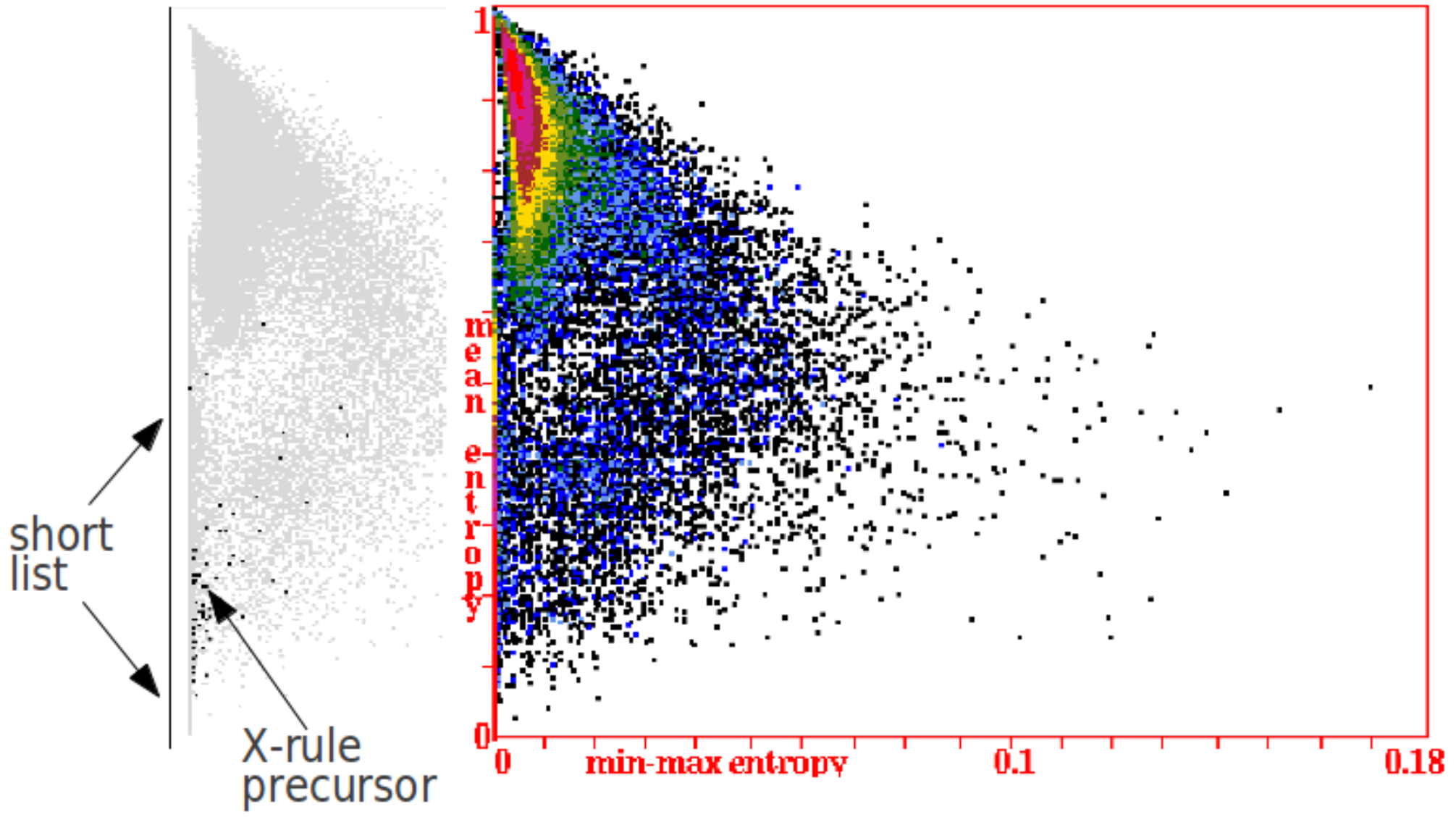}   
\end{center}
\vspace{-5ex}
\caption[Scatter-plot landscape]
{\textsf{The scatter-plot of a sample of 93000+ rules,
plotting min-max entropy variability against mean entropy, coarse-grained
on a 256$\times$256 grid where
the dot colors represent rule frequency intervals, 
+256,128,$\dots$,4,2,1 
\raisebox{.2ex}{\includegraphics[height=1.5ex]{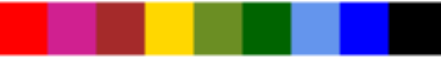}}\\
The left panel shows the location of the shortlist of
71 rules in the ordered region with low mean and  max-min entropy.
The Precursor-Rule is indicated.
\label{scatter-plot}
}}
\end{figure}  

The Precursor-Rule was selected from a sample of biased rules
classified by the input-entropy method\cite{Wuensche99,Wuensche2005},
giving the scatter-plot in figure~\ref{scatter-plot}.  These biases
followed ``Life-Like'' constraints though not simple birth/survival
logic, to the extent that the rules are binary, with a Moore
neighborhood, and in particular that they are isotropic, and where the
$\lambda$~parameter, the density of 1s in the look-up table, is similar
to the Game-of-Life where $\lambda=0.273$.  Excluding the chaotic
sector of the sample (the most heavily populated) a short list of 71 rules
with spontaneously emergent gliders and eaters (also called eaters/reflectors)
were selected from the ordered sector which has low entropy variability.

The Precursor-Rule itself was selected from this short list, firstly
because it featured two spontaneously emergent glider types, moving
orthogonally (Gc, figure~\ref{glider-Gc}) and diagonally (Ga,
figure~\ref{glider-Ga}), and secondly because it was possible to
construct oscillating behavior where glider Gc was made to bounce
between stable reflectors
(figures~\ref{SROs}, \ref{RBOs}).  This became the basis
for the design of the glider-guns in the anisotropic
X-rule\cite{Gomez2015}, a close mutant of its isotropic precursor.
Isotropic behaviour, where gliders and any other dynamical mechanisms
operate equivalently in any direction, has arguably an advantage over
anisotropy in that it simplifies and makes the design of the
mechanisms more flexible.

Glider-guns are the key components for logical gates and thus
universality. However, at the time we were unable to discover or
construct glider-guns in the Precursor-Rule.  Lately, with the
collaboration of members of the ConwayLife forum~\cite{ConwayLife-forum},
glider-guns have now been created for gliders Gc and Ga
\mbox{(figure~\ref{GGc-GGa-basic})}.  In addition, the forum
contributed a plethora of complex structures from the Game-of-Life
lexicon, including other glider-guns, oscillators, ships, puffer-trains, rakes, and
breeders, which enrich the Precursor-Rule's behavior and complexity. 
There is another orthogonal glider (Gb, figure~\ref{glider-Gb}) less likely to
emerge spontaneously because of its more complicated phases, but
as yet a glider-gun for Gb has not been discovered.

The Precursor-Rule has a number of glider-guns, any of which can be 
used to build logical gates, however we have chosen to use
the basic glider-gun in figure~\ref{GGc-GGa-basic}(b) to demonstrate logical universality, 
using analogous methods to Conway\cite{Berlekamp1982} and the X-Rule\cite{Gomez2015}.

\begin{figure}[htb]
\begin{center}
\fbox{\includegraphics[width=.9\linewidth,bb=34 74 459 427, clip=]{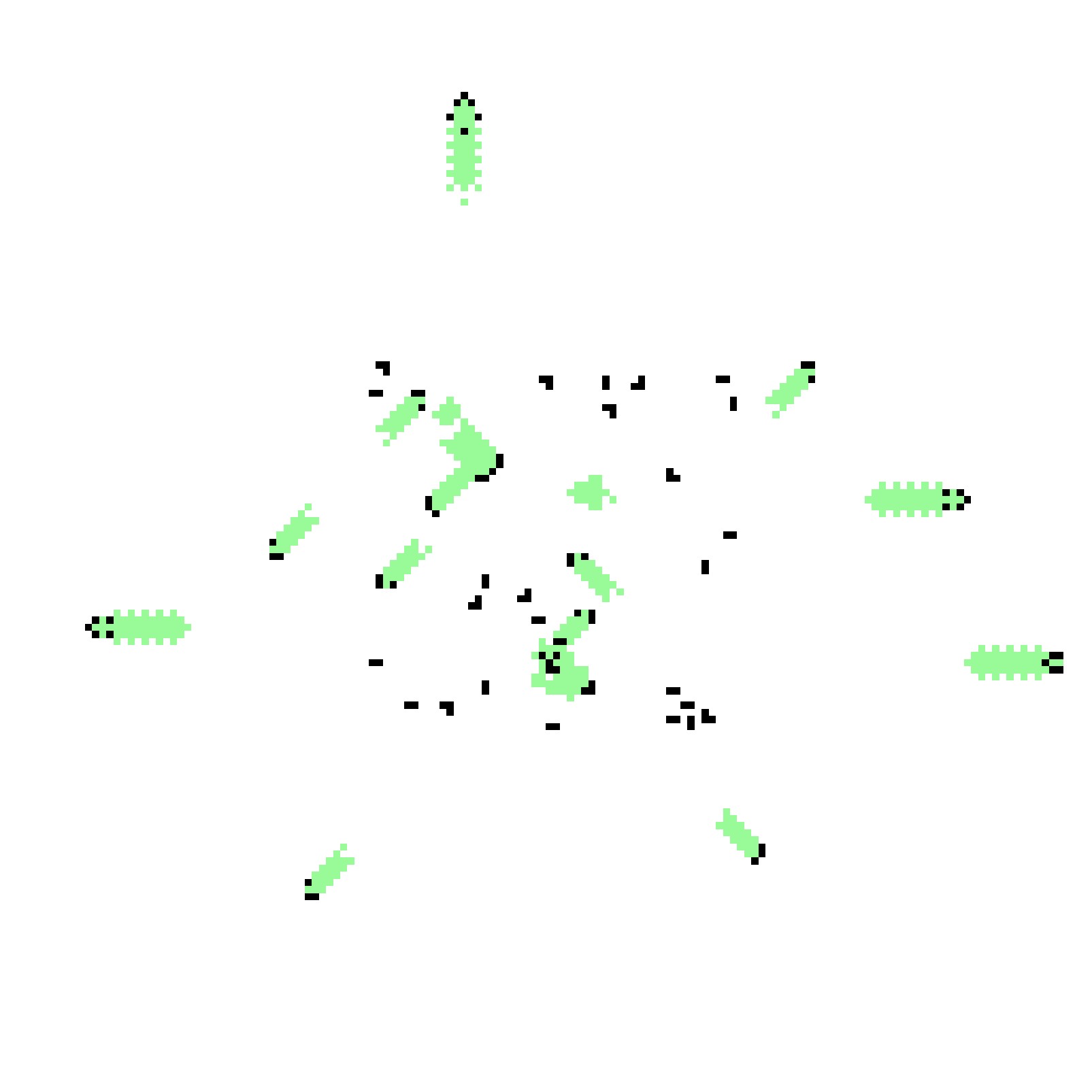}}  
\end{center} 
\vspace{-3ex}
\caption[Typical evolution from a random zone]
{\textsf{
A typical evolution emerging after 99 time-steps from a 50x50 central
random zone within a 150x150 space. 
Gliders Ga and Gc emerge, leaving stable eaters at the
center. Dynamic trails=20.
\label{Typical evolution}
\vspace{-2ex}
}}
\end{figure}

The paper is organised into the following further sections,
(\ref{The Precursor-Rule definition}) the Precursor-Rule definition,  
(\ref{Gliders, Eaters, and Collisions}) a description of gliders, eaters, and collisions,
(\ref{Basic glider-guns}) the basic glider-guns for gliders Gc and Ga, 
(\ref{Logical Universality}) logical universality by logical gates
using glider-gun GGa, 
(\ref{Precursor-Rule Universe}) a review of alternative glider-guns and other 
dynamical structures discovered to date, and
(\ref{Concluding remarks})  the concluding remarks.

\section{The Precursor-Rule definition}
\label{The Precursor-Rule definition}

\begin{figure}[htb]
\begin{center}
\includegraphics[width=.8\linewidth,bb= 2 -2 349 43]{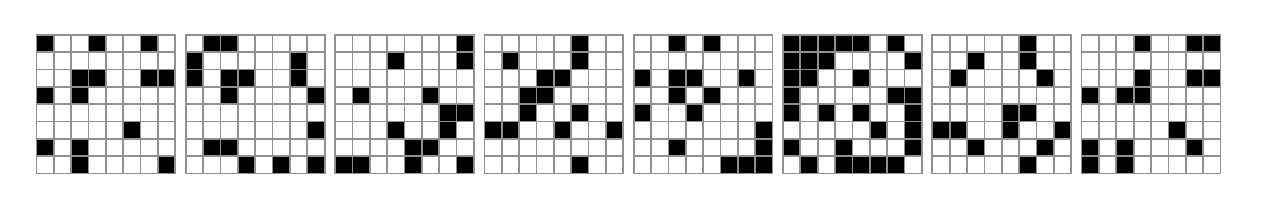}
\end{center} 
\vspace{-6ex}
\caption[precursor's rule-table]
{\textsf{The precursor's rule-table 
in descending order of their values\cite{Wolfram83}. 
\label{precursor-table512}
}}
\end{figure}

\begin{figure}[htb]
\begin{center}
\includegraphics[width=1\linewidth]{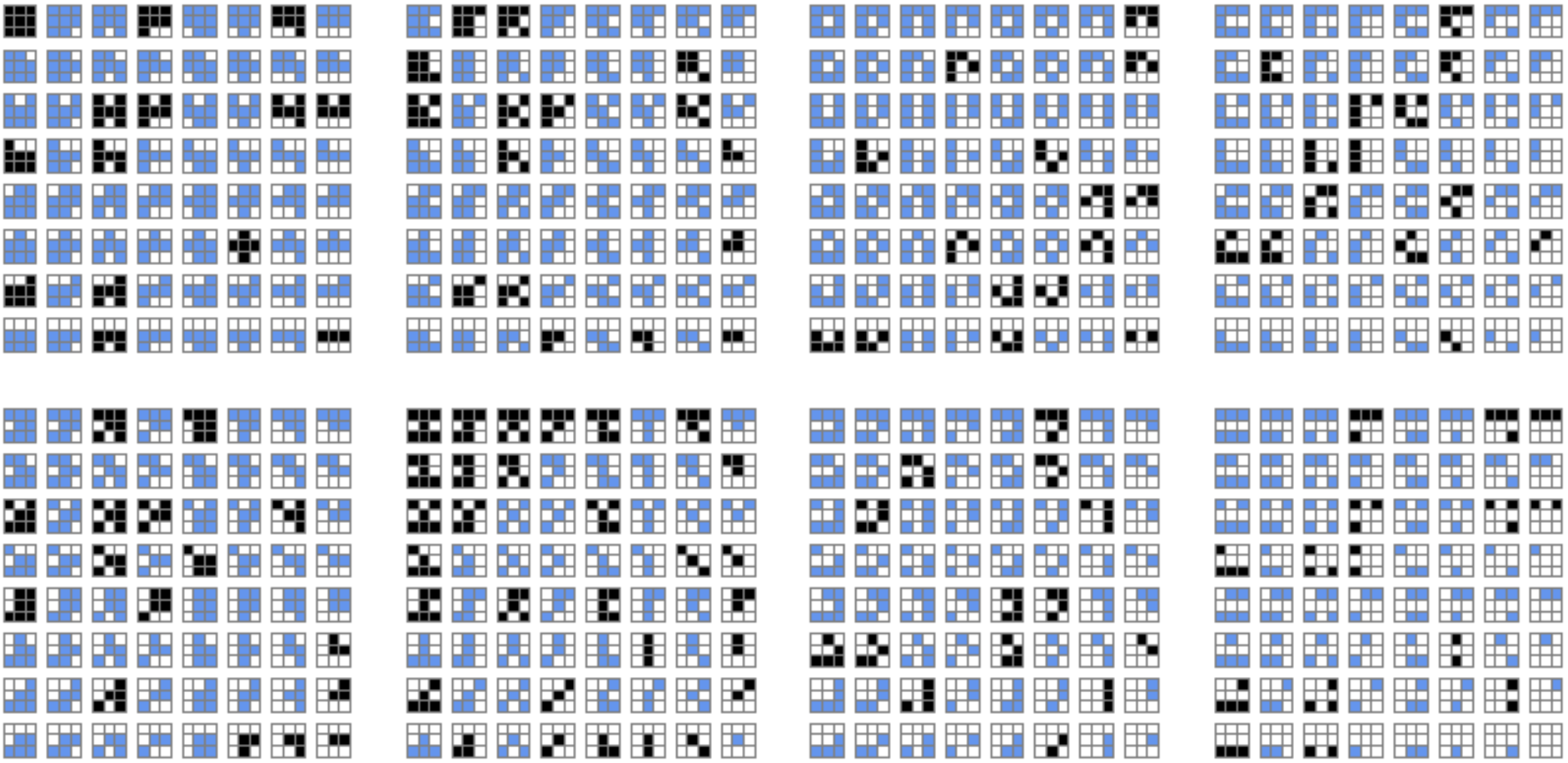} 
\end{center} 
\vspace{-4ex}
\caption[precursor's 512 neighborhoods]
{\textsf{The layout in figure~\ref{precursor-table512} expanded to show
all 512 neighborhoods.
134 black neighborhoods map to~1, 378 blue neighborhoods map to 0.
\label{precursor-neighborhoods-512}
}}
\end{figure}

\begin{figure}[htb]
\begin{center}
\includegraphics[width=.75\linewidth]{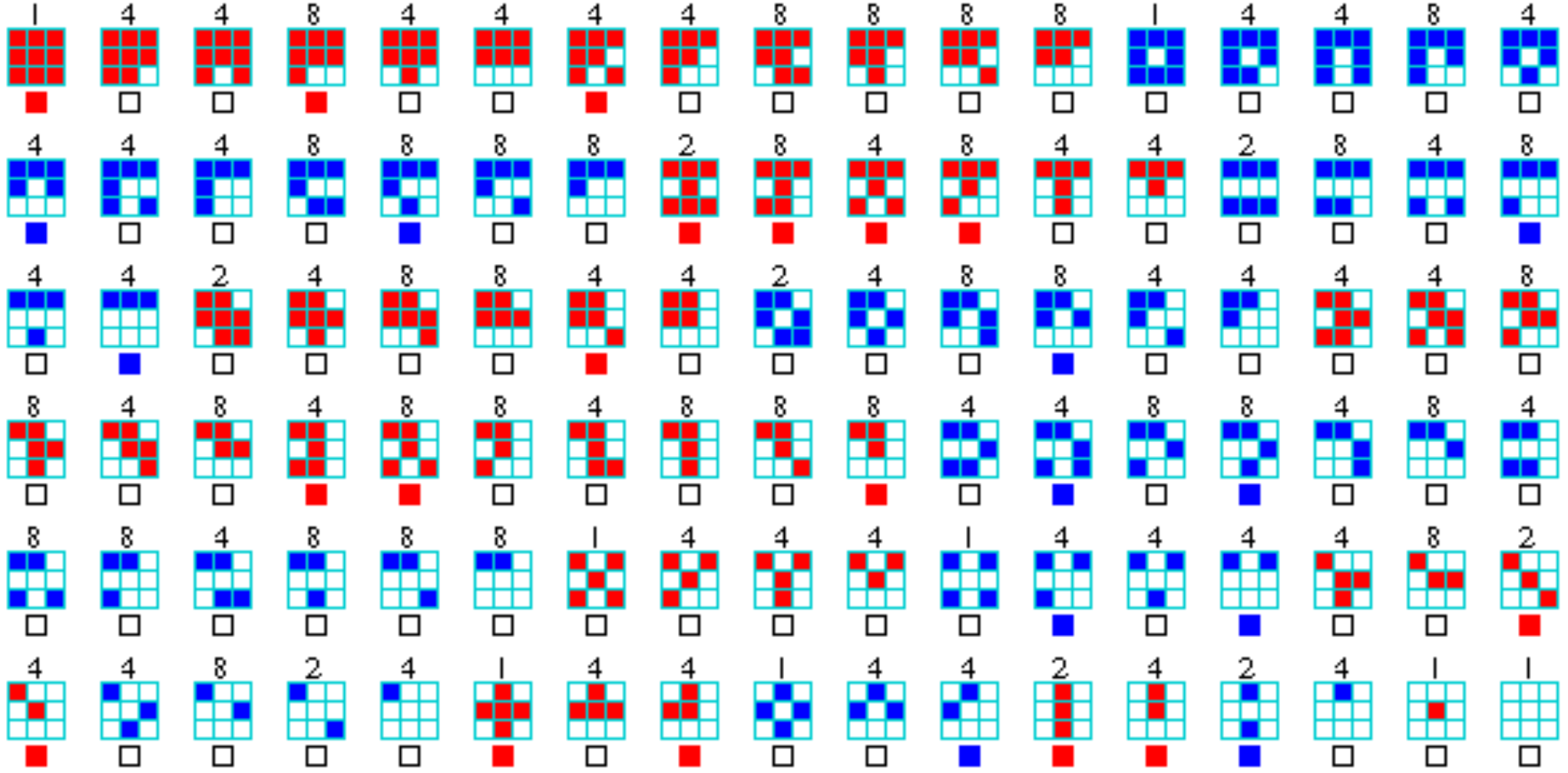}   
\end{center} 
\vspace{-4ex}
\caption[precursor's isotropic rule-table]
{\textsf{The rule-table based on 102 isotropic
    neighborhoods --- one (maximum value) prototype of each spin/flip group
    shown in descending order of their values\cite{Wolfram83}
    with outputs (squares) below --- empty=0, color=1.
    Blue or red depends on 0 or 1 as the neighborhood's
    central cell value, so a blue output for a blue neighborhood
    signals ``birth'', a red output for a red neighborhoods signals
    ``survival''.  The size of each group [1, 2, 4, 8] is shown above each prototype.
\label{precursor102}
}}
\end{figure}

\begin{figure}[htb]
\includegraphics[height=.052\linewidth]{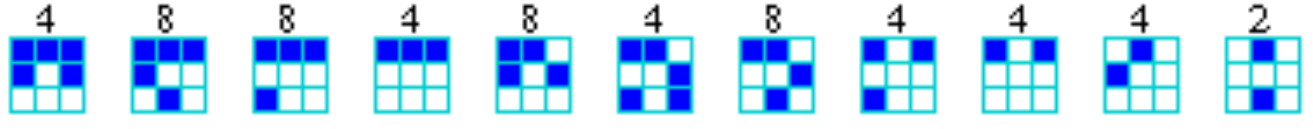} \raisebox{1.3ex}{$\rightarrow$ birth}\\
\includegraphics[height=.052\linewidth]{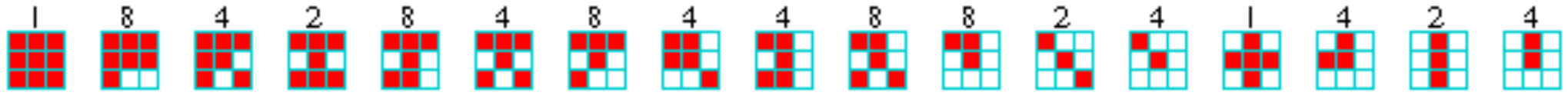} \raisebox{1.3ex}{$\rightarrow$ survival}
\vspace{-2ex}
\caption[Precursor birth]
{\textsf{The 28 isotropic prototypes with output 1 are shown here
separated into 11 cases of ``birth'' and 17 of ``survival'' (the rest output 0).
A birth/survival logic is not discernible.
The size of each group [1, 2, 4, 8] is shown above each prototype.
\label{precursor28}
}}
\end{figure}

\enlargethispage{3ex}
Figures \ref{precursor-table512} to \ref{precursor28} define the
Precursor-Rule in four ways; the rule-table, the rule-table expanded
to show all 512 neighborhoods, as a 102-bit isotropic
rule-table\footnote{Ongoing investigation shows that a small but
significant proportion of rule-table outputs are quasi-neutral
(wildcards) --- their mutations have little or no effect on most
glider-guns featured in this paper, making the Precursor-Rule
part of a cluster of very similar rules.}, and in terms of
birth/survival where a simple logic is not evident.\\

\clearpage

\clearpage

\section{Gliders, Eaters, and Collisions}
\label{Gliders, Eaters, and Collisions}
A glider is a special kind of oscillator, a mobile pattern that
recovers its form but in a displaced position, thus moving at a given
velocity. A rule with the ability to support a glider, together with a stable 
eater/reflector, and a diversity of interactions between gliders and eaters, 
provides the first hint of potential universality.

\begin{figure}[h]
\textsf{\small
\begin{center}
\begin{tabular}[t]{ @{}c@{} @{}c@{}  @{}c@{}  @{}c@{}  @{}c@{}   @{}c@{}   }
& \multicolumn{4}{c}{$\leftarrow$---------------- Ga ----------------$\rightarrow$} &\\[1ex]
  \includegraphics[height=.08\linewidth,bb= 0 0 30 30, clip=]{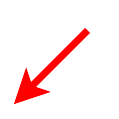}
&  \includegraphics[height=.08\linewidth,bb=-1 -2  42 35, clip=]{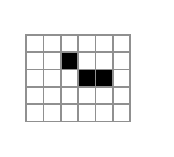}\color{white}-\color{black}%
& \includegraphics[height=.08\linewidth,bb=-3 -2  42 35,  clip=]{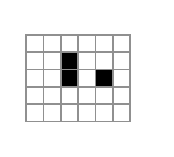}\color{white}-\color{black}%
& \includegraphics[height=.08\linewidth,bb=-3 -2  42 35,  clip=]{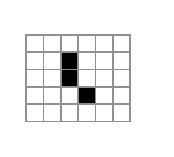}%
& \includegraphics[height=.08\linewidth,bb=-3 -2  42 35,  clip=]{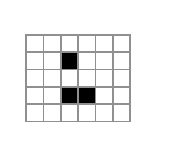}%
& \includegraphics[height=.08\linewidth,bb=-3 -2  42 35,  clip=]{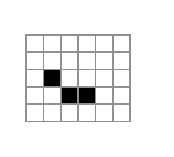}\\[-2ex]  
& 1 & 2 & 3 & 4 & 5 
\end{tabular}
\end{center}
}
\vspace{-4ex}
\caption[glider Ga]%
{\textsf{The 4 phases of glider Ga, moving SouthWest with speed $c/4$. 
}}
\label{glider-Ga}
\vspace{-4ex}
\end{figure}

\begin{figure}[h]
\textsf{\small
\begin{center}
\begin{tabular}[t]{ @{}c@{} @{}c@{}  @{}c@{}  @{}c@{}  @{}c@{}   @{}c@{}   }
 \multicolumn{4}{c}{$\leftarrow$-------------------------- Gc --------------------------$\rightarrow$} & &\\[1ex]
  \includegraphics[height=.1\linewidth,bb=-3 -2  62 45, clip=]{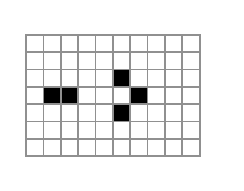}\color{white}-\color{black}%
& \includegraphics[height=.1\linewidth,bb=-3 -2  62 45,  clip=]{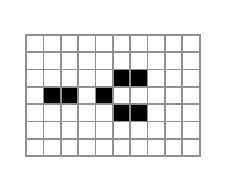}\color{white}-\color{black}%
& \includegraphics[height=.1\linewidth,bb=-3 -2  62 45,  clip=]{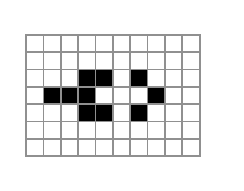}%
& \includegraphics[height=.1\linewidth,bb=-3 -2  62 45,  clip=]{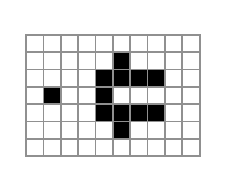}%
& \includegraphics[height=.1\linewidth,bb=-3 -2  62 45,  clip=]{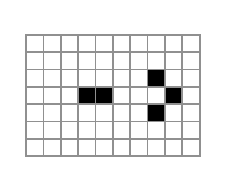}%
& \includegraphics[width=.09\linewidth,bb= 5 -3 32 22, clip=]{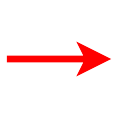}\\[-2ex]  
1 & 2 & 3 & 4 & 5 & 
\end{tabular}
\end{center}
}
\vspace{-4ex}
\caption[glider Gb]%
{\textsf{The 4 phases of glider glider Gb, moving East with speed $c/2$.
\label{glider-Gb} 
}}
\end{figure}

\begin{figure}[h]
\textsf{\small
\begin{center}
\begin{tabular}[t]{ @{}c@{} @{}c@{}  @{}c@{}  @{}c@{}  @{}c@{}   @{}c@{}   }
 \multicolumn{4}{c}{$\leftarrow$-------------------------- Gc --------------------------$\rightarrow$} & &\\[1ex]
  \includegraphics[height=.1\linewidth,bb=-3 -2  62 45, clip=]{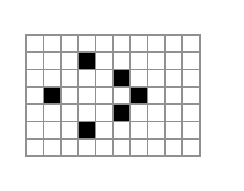}\color{white}-\color{black}%
& \includegraphics[height=.1\linewidth,bb=-3 -2  62 45,  clip=]{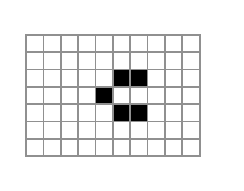}\color{white}-\color{black}%
& \includegraphics[height=.1\linewidth,bb=-3 -2  62 45,  clip=]{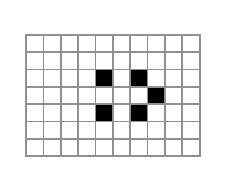}%
& \includegraphics[height=.1\linewidth,bb=-3 -2  62 45,  clip=]{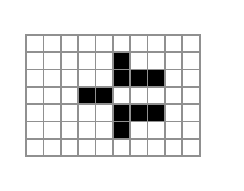}%
& \includegraphics[height=.1\linewidth,bb=-3 -2  62 45,  clip=]{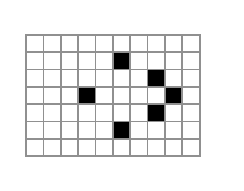}%
& \includegraphics[width=.09\linewidth,bb= 5 -3 32 22, clip=]{ArrowE.pdf}\\[-2ex]  
1 & 2 & 3 & 4 & 5 & 
\end{tabular}
\end{center}
}
\vspace{-4ex}
\caption[glider Gc]%
{\textsf{The 4 phases of glider Gc, moving East with speed $c/2$.
\label{glider-Gc}
}}
\vspace{-1ex}
\end{figure}

\begin{figure}[htb]
\begin{center}
\begin{minipage}[b]{.9\linewidth}
\textsf{\color{white}---\color{black}Ga\color{white}-----------\color{black}G2a\color{white}------------\color{black}G3a\color{white}------------\color{black}G4a\color{white}-----------\color{black}G5a\color{white}-----------\color{black}G6a\color{white}-----------\color{black}}\\[-2ex]
\includegraphics[width=1\linewidth,bb=-2 0 373 71, clip=]{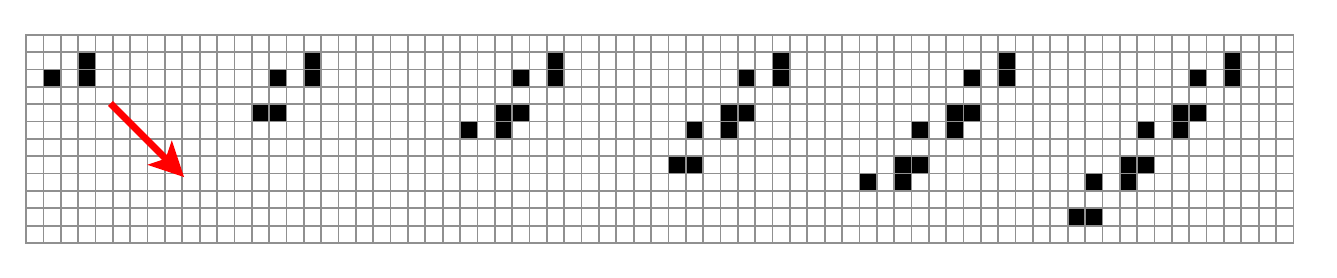}
\end{minipage} 
\end{center}
\vspace{-6ex}
\caption[Ga combined gliders SE]
{\textsf{Glider Ga (left) and combined Ga gliders of increasing size 2, 3, 4, 5 and 6,
orientated SE. They move in 4 phases at a speed of $c/4$. There is a one cell 
overlap between adjacent Ga gliders.
\label{Ga combined gliders SE}
}}
\end{figure}

\enlargethispage{5ex}
From a typical chaotic initial condition as in figure~\ref{Typical evolution},
and evolution subject to the Precursor-Rule, its easy to detect the 
spontaneous emergence of two eater types,
\raisebox{-1.7ex}{\includegraphics[height=3.7ex, bb=2 -1 20 20, clip=]{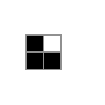}} and
\raisebox{-1.7ex}{\includegraphics[height=3.7ex,bb=2 -1  15 20, clip=]{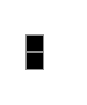}}
(and their spins/flips), and two glider types, glider Ga (Figure~\ref{glider-Ga}) and
glider Gc (Figure~\ref{glider-Gc}). A combined glider G2a, two Ga gliders joined together 
with a one cell overlap, can also emerge (figure~\ref{Ga combined gliders SE}).
Glider Gb (Figure~\ref{glider-Gb}) is not detected immediately but with a more patient
search it can be found. 
Figures~\ref{c-Ga1-0} to \ref{Gc+e0-7} describe some of the
collision results between Ga and Gc gliders, and between these gliders and
eaters.  Similar experiments could include G2a gliders and also the
oscillators in sections~\ref{Variable length/period oscillators} and \ref{Other oscillators}, 
to provide a more thorough collision catalog.

\vspace{-3ex}

\vspace{-2ex}

\vspace{-5ex}

\clearpage

\subsection{Gliders colliding with gliders}
\label{Gliders colliding with gliders}

The outcomes of collisions between gliders are very diverse,
depending on the phase, angle, and point of impact, and
include the destruction of either or both gliders,
a bounce, or a transformation to different or combined glider types. 
A residual pattern of eaters/reflectors may also be created.

The speed of a glider (or other periodic mobile structure) relative to
the speed of light $c$, is measured by the number of squares advanced
within its period. In general, orthogonal gliders based on Gc advance
2 squares in a period of 4 giving a speed of $c/2$, whereas diagonal
gliders based on Ga advance 1 square (on both axes) in a period of 4
giving an speed of $c/4$.

\begin{figure}[htb]
\begin{center}
\fbox{\includegraphics[width=1\linewidth,bb= 13 22 412 82, clip=]{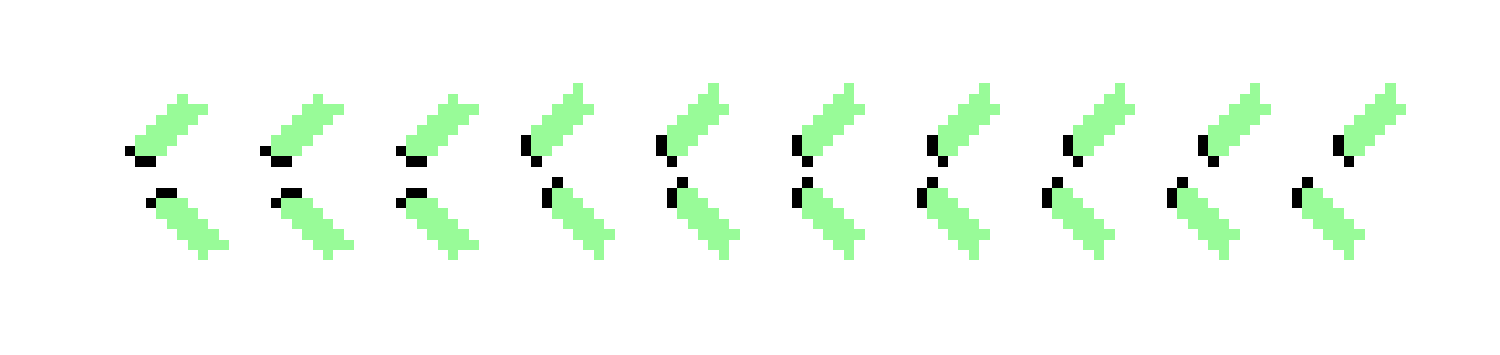}}\\[1ex] 
\fbox{\includegraphics[width=1\linewidth,bb= 13 7 412 97, clip=]{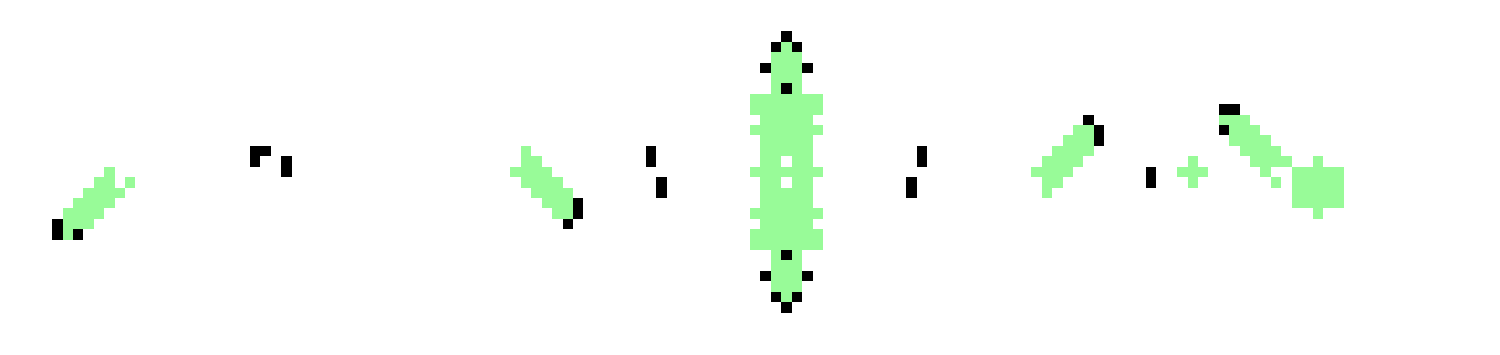}}
\end{center}
\vspace{-3ex}
\caption[two Ga gliders colliding at 90$^{\circ}$]  
{\textsf{Two Ga gliders colliding at 90$^{\circ}$, +29 time-steps.
\label{c-Ga1-0}
}}
\end{figure}

\begin{figure}[htb]
\begin{center}
\fbox{\includegraphics[width=1\linewidth,bb= 4 26 412 79, clip=]{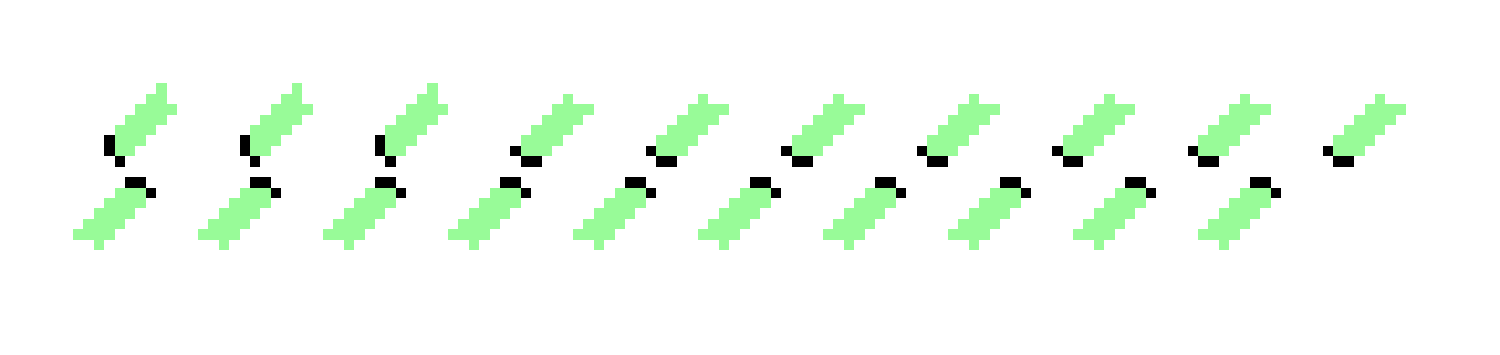}}\\[1ex] 
\fbox{\includegraphics[width=1\linewidth,bb= 4 8 412 97, clip=]{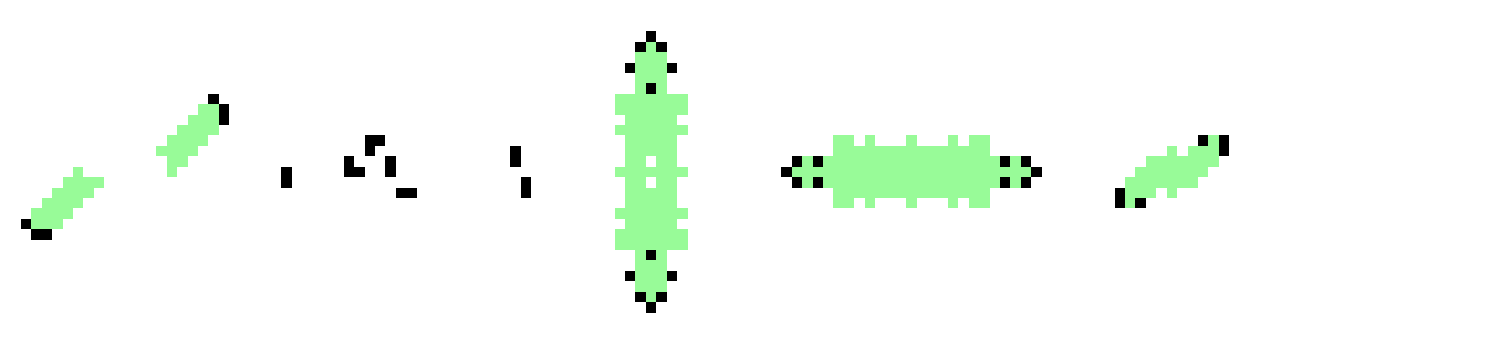}}
\end{center}
\vspace{-3ex}
\caption[Two Ga gliders colliding head-on]  
{\textsf{Two Ga gliders colliding head-on, +30 time-steps.
\label{ch-Ga1-0}
}}
\end{figure}

\begin{figure}[htb]
\begin{center}
\fbox{\includegraphics[width=.8\linewidth,bb= 3 7 407 126, clip=]{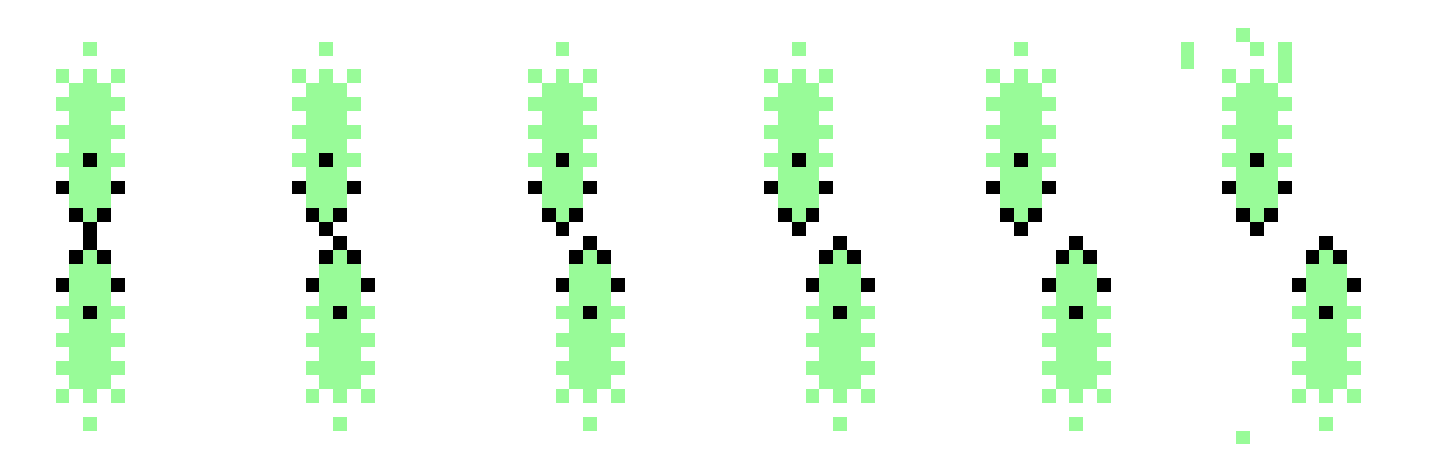}}\\[1ex] 
\fbox{\includegraphics[width=.8\linewidth,bb= 3 13 407 116, clip=]{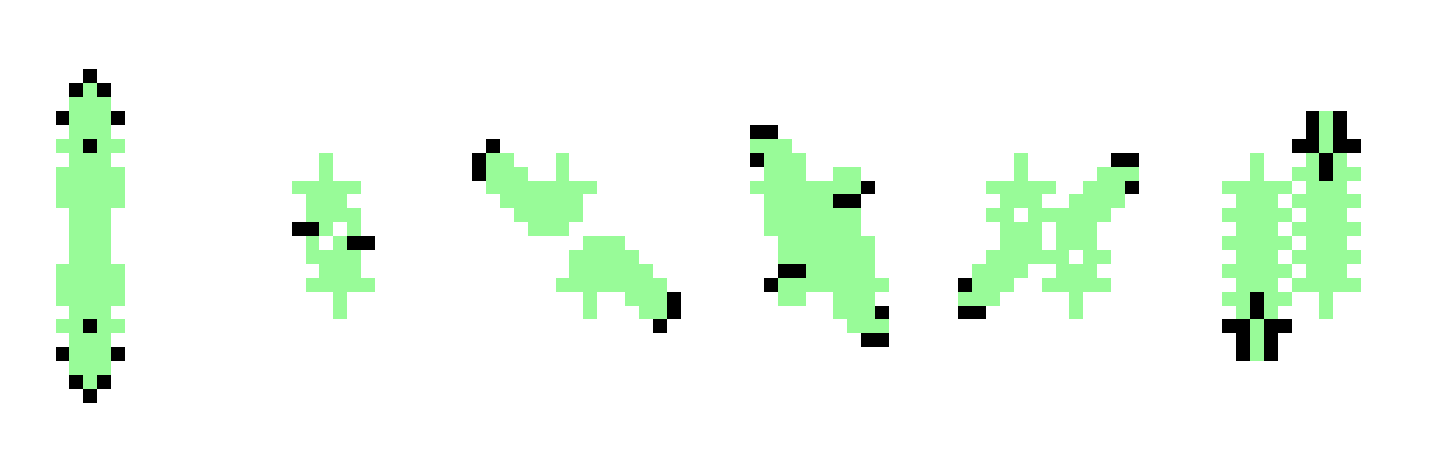}}
\end{center}
\vspace{-3ex}
\caption[two Gc gliders colliding head-on]
{\textsf{Two Gc gliders colliding head-on, +19 time-steps.
\label{c-Gc2-0}
}}
\end{figure}

\begin{figure}[htb]
\begin{center}
\fbox{\includegraphics[width=1\linewidth,bb= 47 28 323 74, clip=]{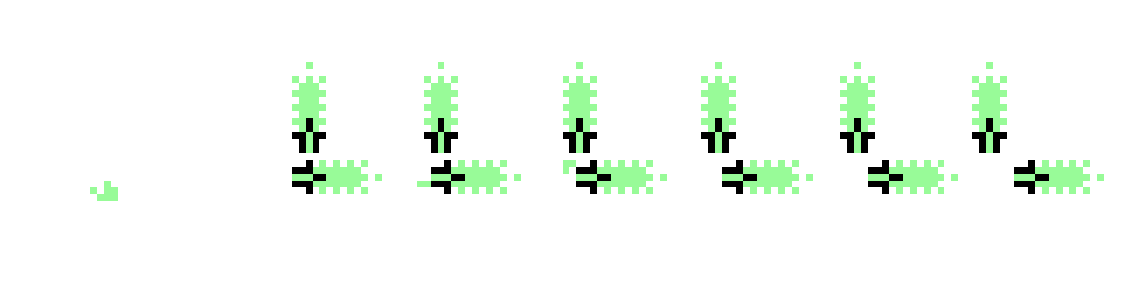}}\\[1ex] 
\fbox{\includegraphics[width=1\linewidth,bb= 47 16 323 74, clip=]{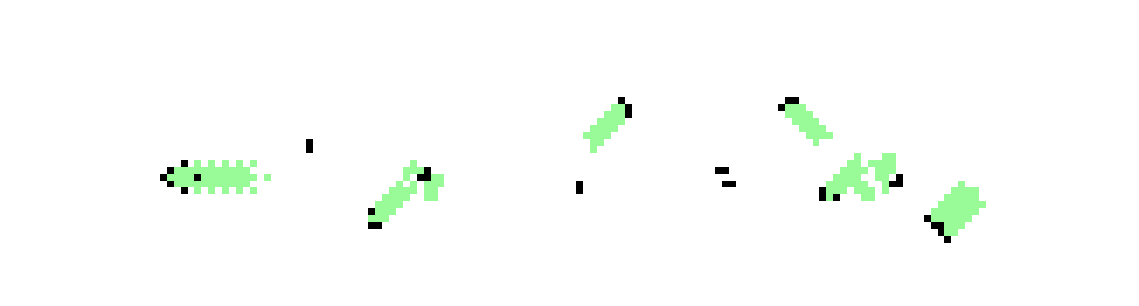}}
\end{center}
\vspace{-3ex}
\caption[two Gc gliders colliding at 90$^{\circ}$]
{\textsf{Two Gc gliders colliding at 90$^{\circ}$, +37 time-steps.
\label{cR+Gc-0}
}}
\end{figure}

\begin{figure}[htb]
\begin{center}
\fbox{\includegraphics[width=.8\linewidth,bb= 3 16 407 116, clip=]{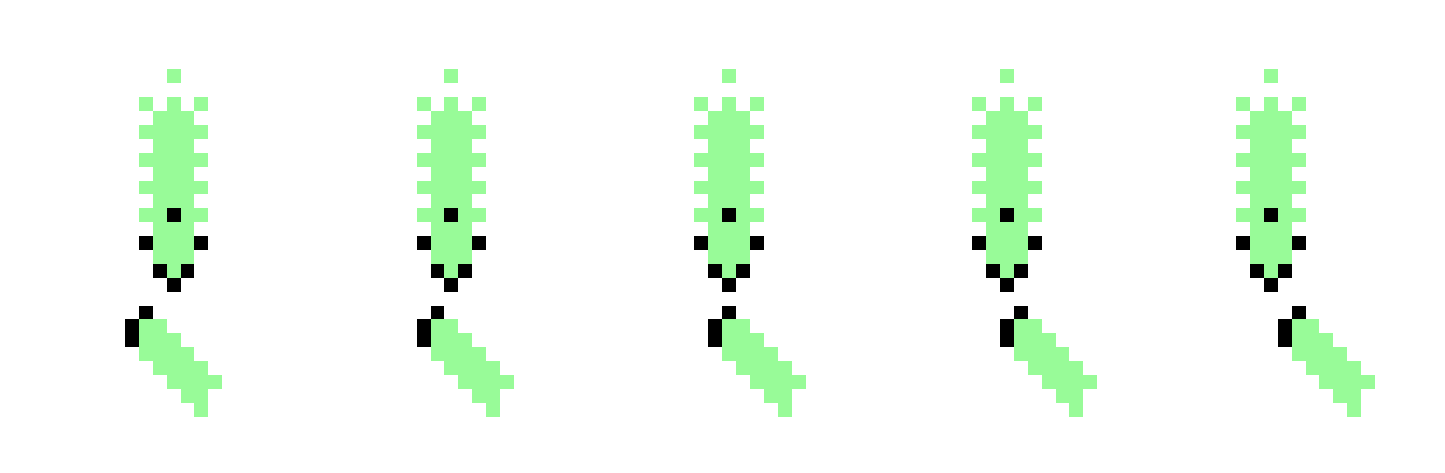}}\\[1ex] 
\fbox{\includegraphics[width=.8\linewidth,bb= -3 13 407 75, clip=]{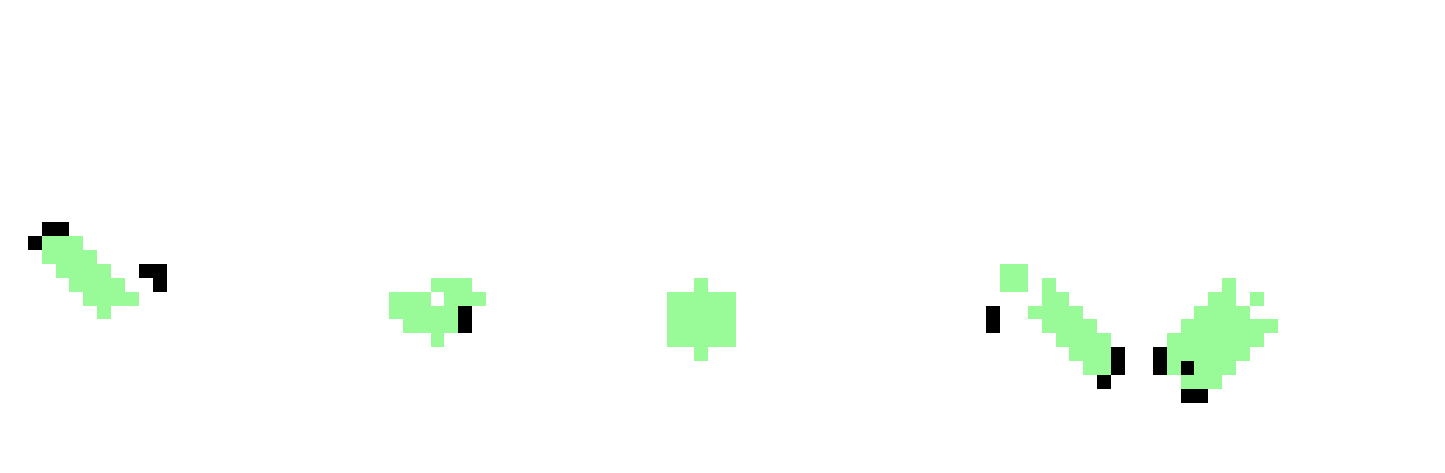}}
\end{center}
\vspace{-3ex}
\caption[glider Gc colliding with Ga]  
{\textsf{Glider Gc colliding with Ga, +26 time-steps.
\label{c-Gac-0}
}}
\end{figure}

Figures~\ref{c-Ga1-0} to \ref{c-Gac-0} show gliders about to collide
with each other at various points of impact (top panel), and the outcomes
after a given number
of time-steps (lower panel), giving just a flavour
of the diversity of behaviour, with dynamic trails=20.\\

\vspace{-30ex}

\clearpage

\subsection{Gliders colliding with eaters/reflectors}
\label{Gliders colliding with eaters/reflectors}

Stable structures emerge spontaneously in the
Precursor-Rule which may destroy and/or reflect colliding gliders.
The two basic eaters/reflectors (also known as ``still life'') are isolated patterns
consisting of 3 cells in an ``L'' shape,
and two adjoining cells,
giving the following with all rotations/flips:

\begin{center}
\includegraphics[height=4ex,bb=-1 -1 20 20, clip=]{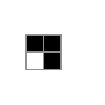}
\includegraphics[height=4ex,bb=-1 -1 20 20, clip=]{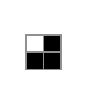}
\includegraphics[height=4ex,bb=-1 -1 20 20, clip=]{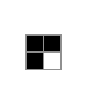}
\includegraphics[height=4ex,bb=-1 -1 20 20, clip=]{Ea4.pdf}
\includegraphics[height=4ex,bb=-1 -1 20 20, clip=]{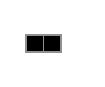}
\includegraphics[height=4ex,bb=-1 -1 20 20, clip=]{Ea6.pdf}
\end{center}

\vspace{-2ex}
The eaters/reflectors may themselves be destroyed or transformed in the collision,
and the glider may be destroyed, bounce, and transform to a different or combined glider.
As with collisions between gliders, the outcomes of collisions between a
glider and an eater/reflector are very diverse, depending on the
phase, angle, and point of impact.  

\begin{figure}[htb]
\begin{center}
\fbox{\includegraphics[width=1\linewidth,bb= 15 28 412 76, clip=]{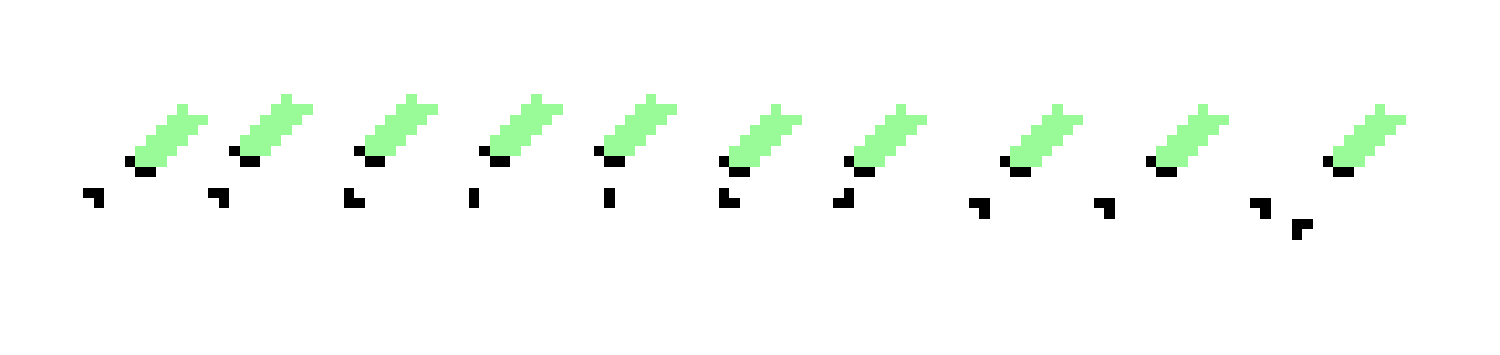}}\\[1ex] 
\fbox{\includegraphics[width=1\linewidth,bb= 15 8 412 91, clip=]{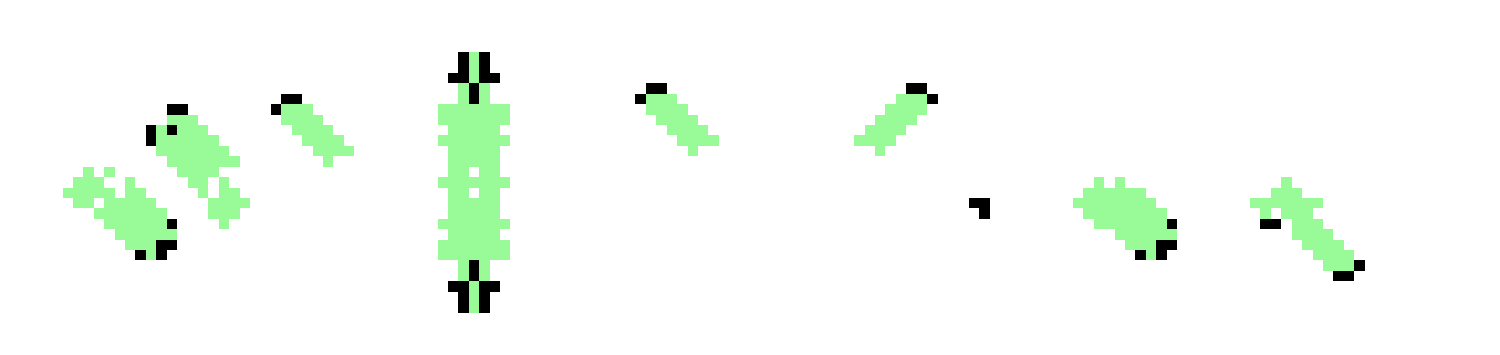}}
\end{center}
\vspace{-3ex}
\caption[glider Ga colliding with eaters]    
{\textsf{Glider Ga colliding with eaters/reflectors, +33 time-steps.
\label{Ga+e0.ps Ga}
}}
\end{figure}

\begin{figure}[htb]
\begin{center}
\fbox{\includegraphics[width=1\linewidth,bb= 6 16 391 79, clip=]{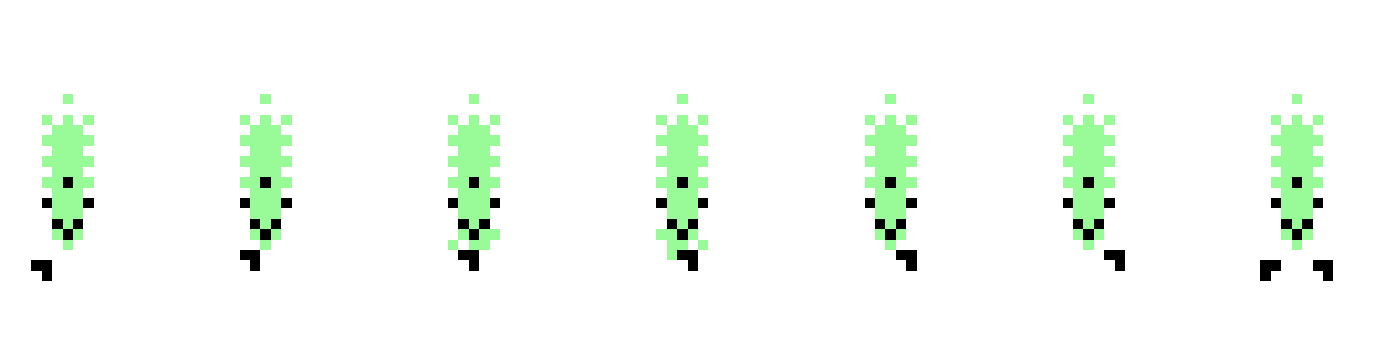}}\\[1ex] 
\fbox{\includegraphics[width=1\linewidth,bb= 6 13 391 88, clip=]{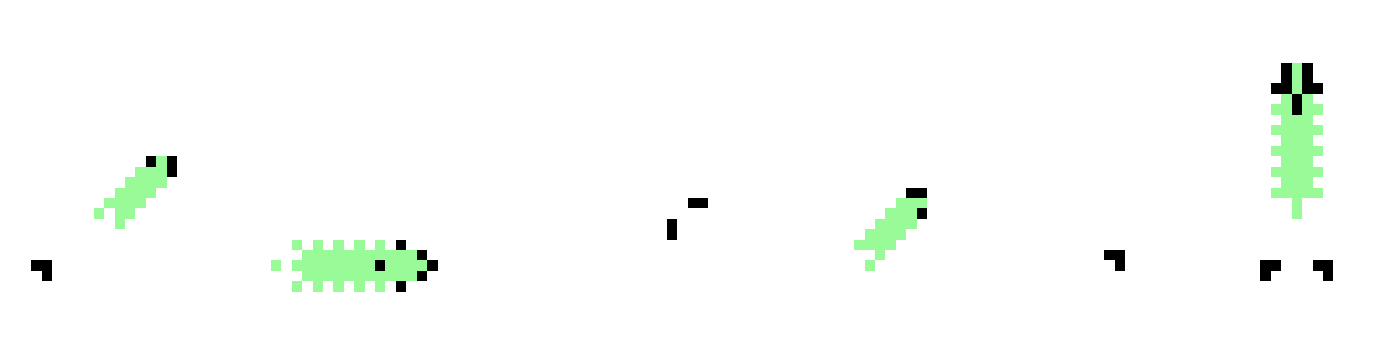}}
\end{center}
\vspace{-3ex} 
\caption[glider Gc colliding with eaters]    
{\textsf{Glider Gc colliding with eaters, +38 time-steps.
\label{Gc+e0-7}
}}
\end{figure}

Figures~\ref{Ga+e0.ps Ga} and \ref{Gc+e0-7}
show gliders about to collide with eaters
at various points of impact (top panel), and the outcomes after a given number
of time-steps (lower panel), showing just a flavour
of the diversity of behaviour, with dynamic trails=20.
\enlargethispage{3ex}

\clearpage

\section{Basic glider-guns}
\label{Basic glider-guns}
Although a diversity of interactions between gliders
and eaters provides the first hint of potential universality,
the essential ingredient is a glider-gun, a dynamic structure that
ejects gliders periodically into space. A glider-gun can also be seen as an
oscillator that adds to its form periodically to shed gliders. 
In some rules a glider-gun may emerge spontaneously\cite{Sapin2004,Wuensche2006},
but not in the Game-of-Life, the X-Rule, or the Precursor-Rule ---  in these cases 
the glider-gun is a complex structure  with a negligible probability of
emerging from a random pattern --- it has to be found, discovered or somehow constructed.

\begin{figure}[htb]
\begin{center}
\begin{minipage}[c]{1\linewidth}
\begin{minipage}[c]{.5\linewidth} 
\fbox{\includegraphics[width=1\linewidth,bb= 124 47 264 145, clip=]{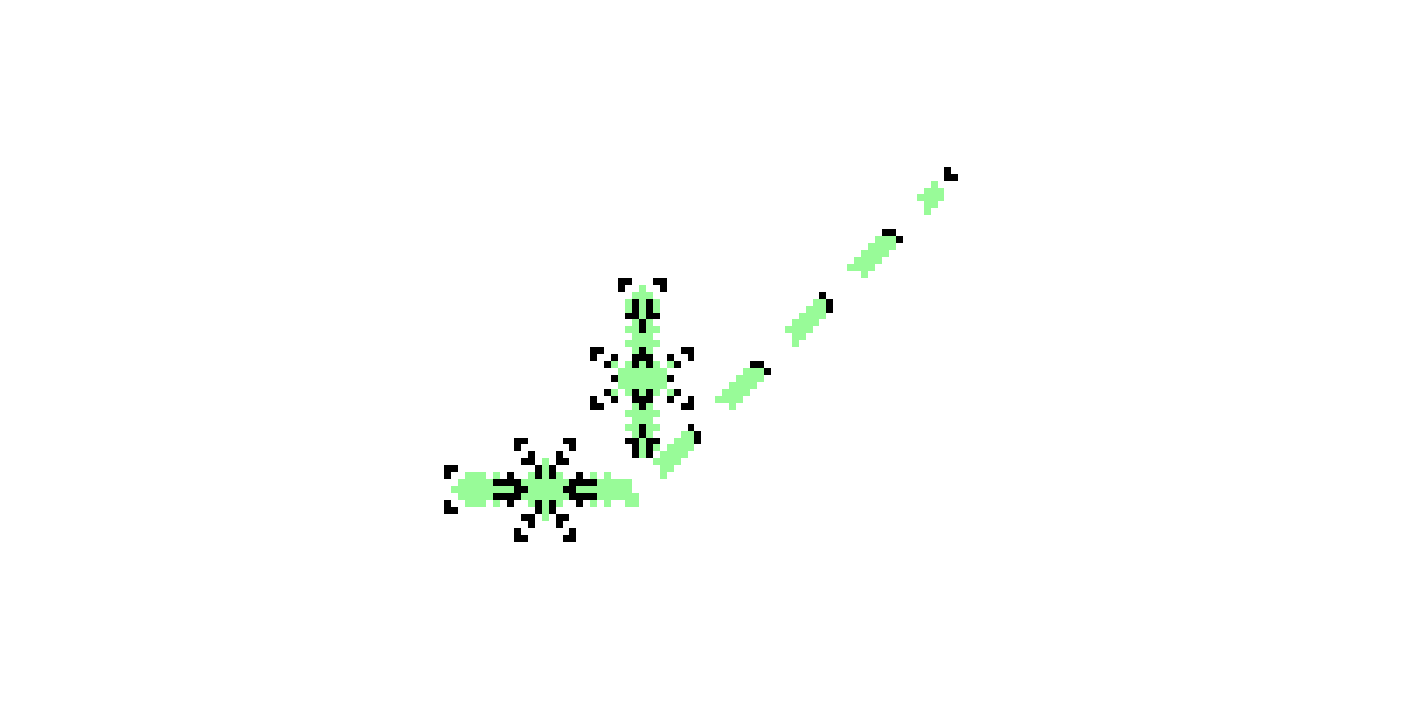}}\\  
\end{minipage}
\hfill
\begin{minipage}[c]{.42\linewidth}
\caption[Rake Ga]
{\textsf{
A new lately found compact Ga glider-gun\cite{BlinkerSpawn}, comparable in size to
the basic Ga glider-gun in figure~~\ref{GGc-GGa-basic}(b), but with double the period
(38 time-steps), is constructed by
colliding two GGc glider-streams at 90$^{\circ}$. Dynamic trails=20.
\label{Ggas}}}  
\end{minipage}
\end{minipage}
\end{center}
\vspace{-2ex}
\end{figure}

Gosper found the game-of-Life glider-gun\cite{Gardner1970,Berlekamp1982}.
The anisotropic X-Rule gliders-guns were constructed from reflecting/bouncing oscillators
in its isotropic precursor by G\'omez\cite{Gomez2015}, the search for
a glider-gun in the Precursor-Rule itself having been abandoned at that time.
However, since the publication of \cite{Gomez2015} and its announcement on the
ConwayLife forum\cite{ConwayLife-forum}, a member, Arie Paap\cite{Wildmyron}, 
discovered the first two glider-guns in the Precursor-Rule --- GG2a shooting the G2a glider
(two Ga's combined) shown in figure~\ref{GG2a}, 
followed by the ``basic'' GGc in figure~\ref{GGc-GGa-basic}(a). 
A number of other Gc and Ga glider-guns
were later announced in the forum, described in section~\ref{Other glider-guns},
together with a diversity of other complex structures.
However, the ``basic'' glider-gun that we apply
to demonstrate logical gates is GGa --- latterly constructed by G\'omez\cite{jmgomez} by 
colliding two GGc glider-streams head on (figure~\ref{GGc-GGa-basic}(b)).
For a while this was the smallest Ga glider-gun,
but a comparably compact gun with double the period has lately been found\cite{BlinkerSpawn}
by colliding two GGc glider-streams at 90$^{\circ}$ (figure~\ref{Ggas}).

In the next section the basic glider-gun GGa will be harnessed to demonstrate
the logical gates, NOT, AND, and OR, to show that the Precursor-Rule is
logically universal.    

\section{Logical Universality}
\label{Logical Universality} 

Traditionally the proof for universality in cellular automata is based on the
Turing Machine or an equivalent mechanism, but in another approach by
Conway\cite{Berlekamp1982}, a cellular automata is
universal in the full sense if it is capable of the following,

\begin{enumerate}
\item Data storage or memory.
\item Data transmission requiring wires and an internal clock.
\item Data processing requiring a universal set of logic gates NOT, AND, and OR, to satisfy
  negation, conjunction and disjunction.
\end{enumerate} 

This paper is confined to proving condition 3 only, for universality in the logical sense.
To demonstrate universality in the full sense as for the
Game-of-Life, it would be necessary to also prove conditions 1 and 2,
or to prove universality in terms of the Turing Machine, as was done
by Randall\cite{Randall2002} for the Game-of-Life.
\clearpage

\subsection{Logical Gates}
\label{Logical Gates}

Logical universality in the Precursor-Rule, as in the Game-of-Life,
is based on Post's Functional Completeness Theorem (FCT)\cite{Francis90}. 
This theorem guarantees that it is
possible to construct a conjunctive (or disjunctive) normal form formula
using only the logical gates NOT, AND and OR.

Using a specific right-angle collision, two Ga gliders can self-destruct
leaving no residue. Applying this between GGa glider-gun streams,
and a Ga glider/gap sequence with the correct
spacing and phases representing a ``string'' of information, 
its possible to build logical gates. Gates NOT, AND and OR
are illustrated in figures~\ref{gatenot} to \ref{gateor}. 
Note that the AND and OR gates include intermediate 
NOT and NOR gates\cite{EdCoxon2016}, explained in the captions.

Gaps in a string are indicated by grey circles, dynamic trails=10
are included, and eaters are positioned to eventually
stop gliders.

\subsubsection{Logical Gate NOT} 
\label{Logical Gate NOT} 
\begin{figure}[h]
 \begin{center}
\fbox{\includegraphics[height=.7\linewidth, bb=21 97 161 369, clip=]{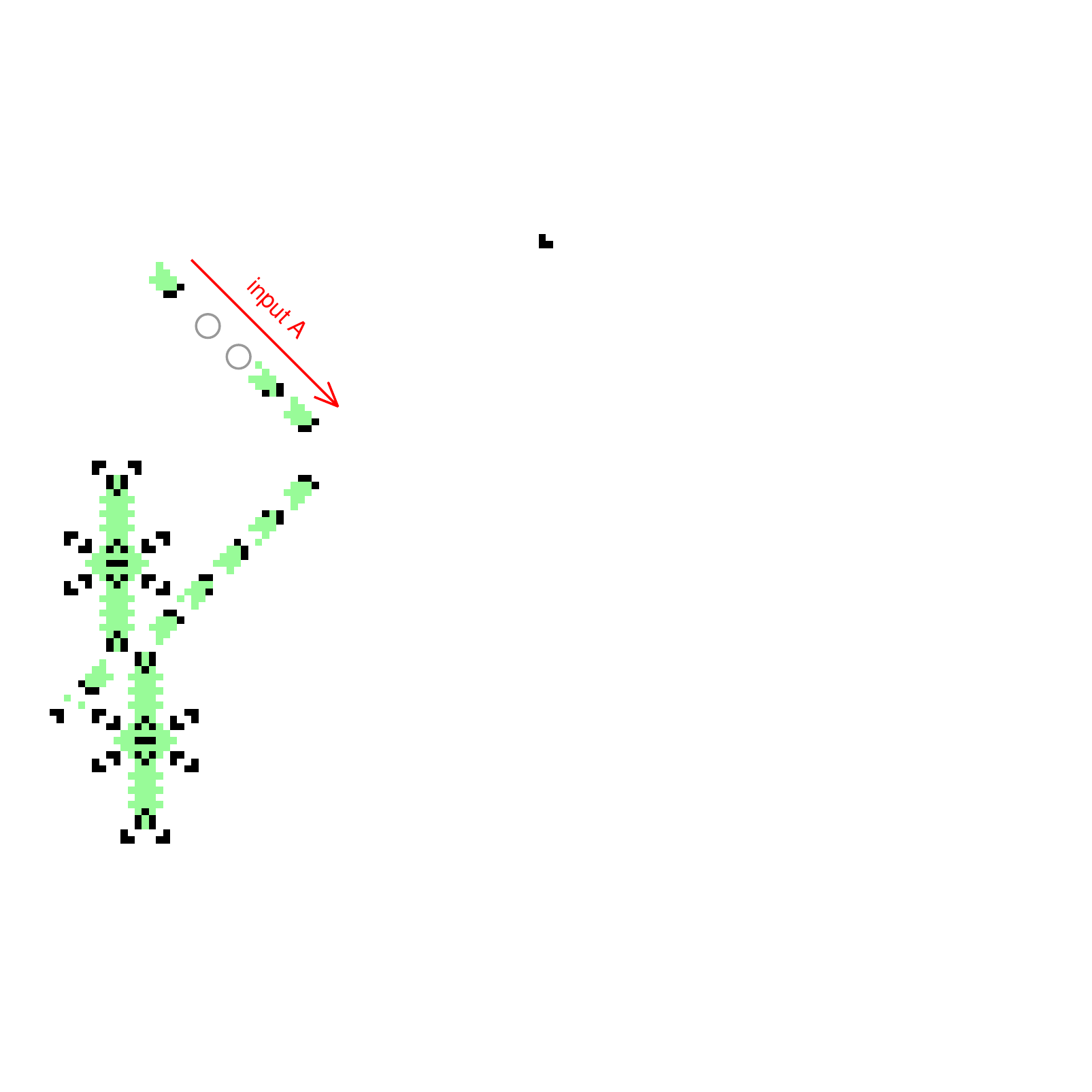}}
\hfill
\fbox{\includegraphics[height=.7\linewidth, bb=21 97 243 369, clip=]{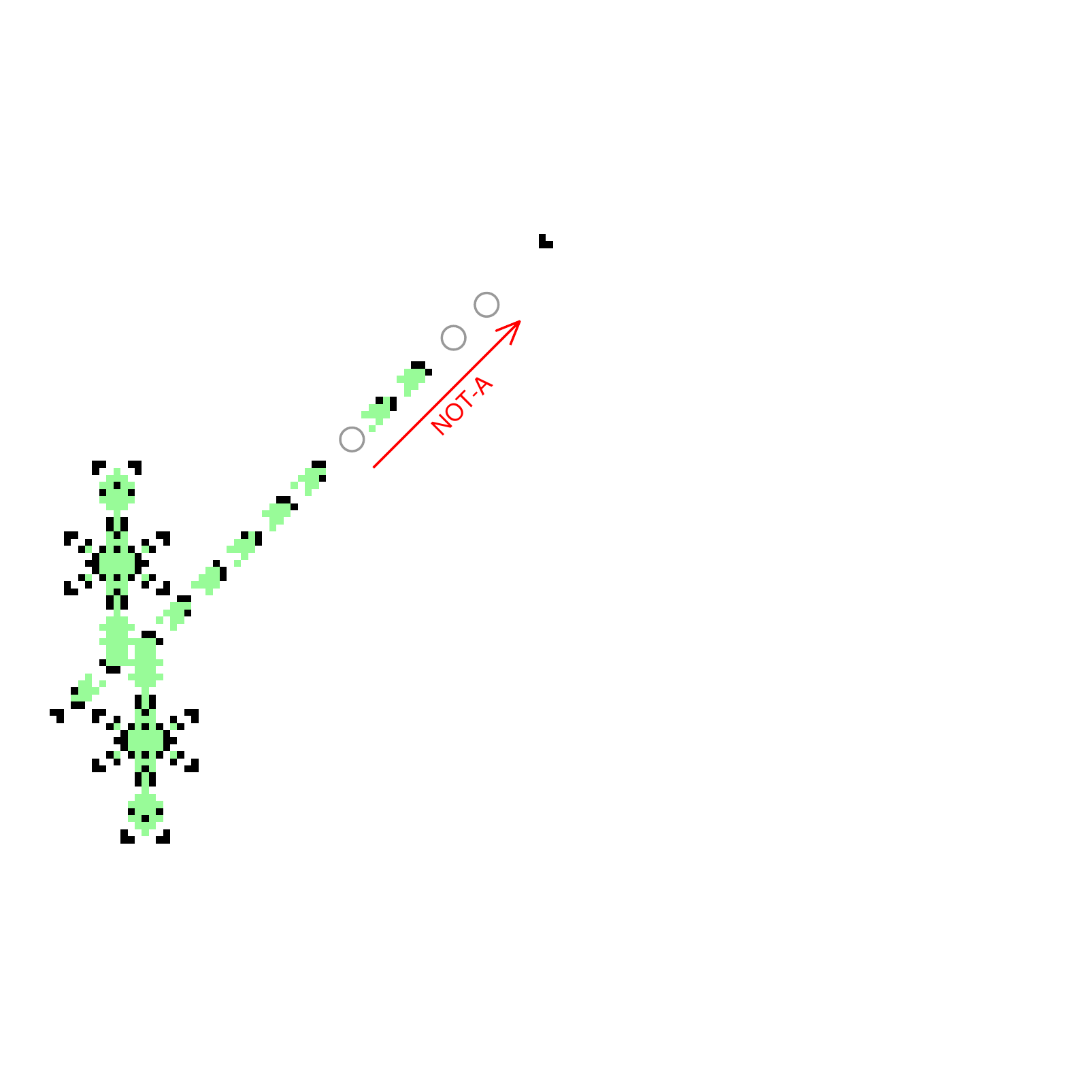}}
 \end{center} 
 \vspace{-3ex}
 \caption[Logical gate NOT] {\textsf{
 An example of the NOT gate: input string A (11001) moving SE interacts with a GGa glider-stream
 moving NE, resulting in NOT-A (00110) moving NE after 102 time-steps.
Any Ga input string can be substituted for A.
\label{gatenot}
}}
\end{figure}

\enlargethispage{5ex}
The logical gate NOT (1$\rightarrow$0 and 0$\rightarrow$1), also called 
an ``inverter'', requires one GGa glider-gun interacting with
a string of Ga gliders/gaps, illustrated as before/after
snapshots in figure~\ref{gatenot}.\\

\clearpage

\subsubsection{Logical gate AND (also NOR)}
\label{Logical gate AND (also NOR)}
\begin{figure}[h]
\begin{center}
\fbox{\includegraphics[height=.8\linewidth, bb=16 30 170 424, clip=]{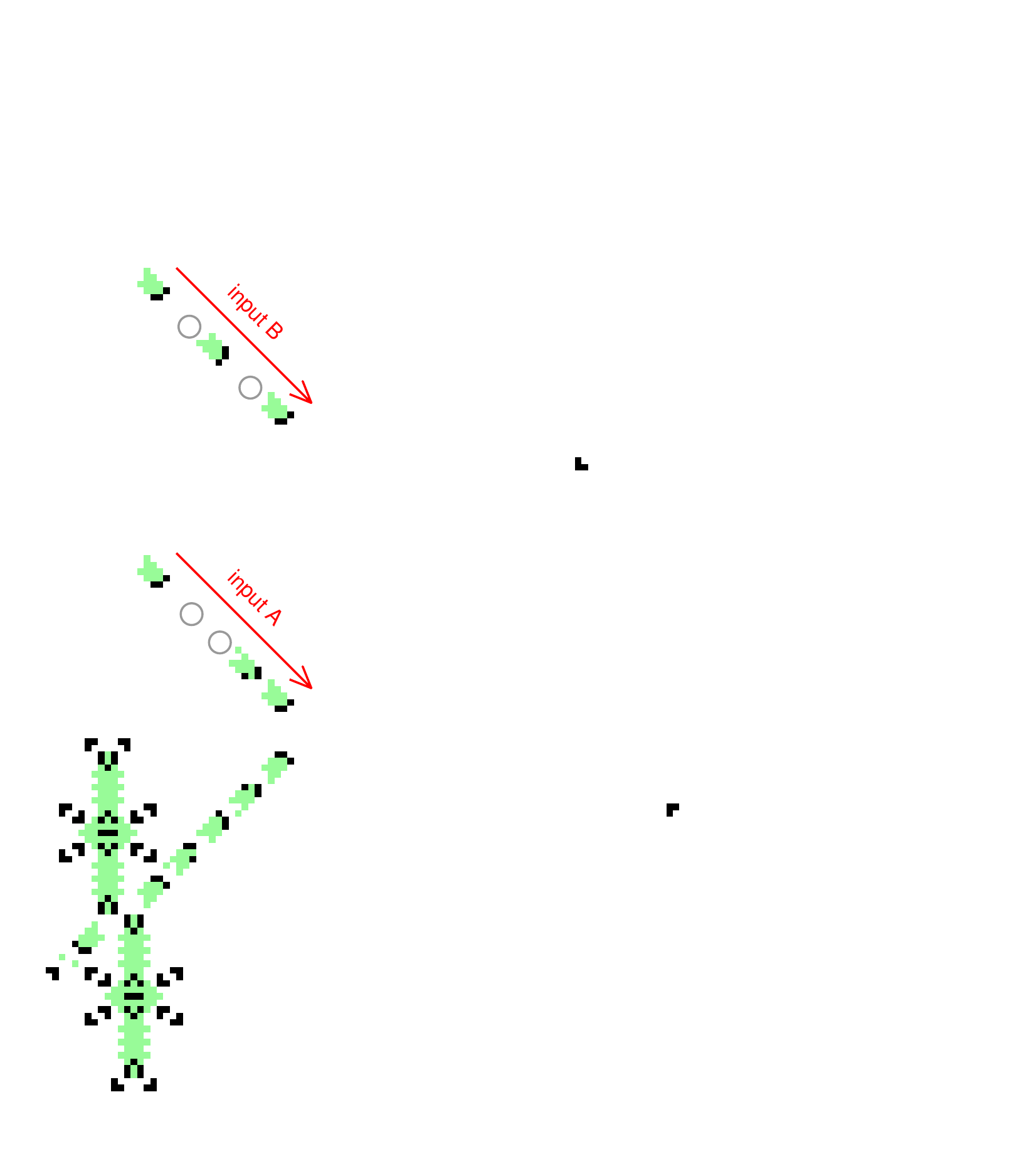}} 
\hfill
\fbox{\includegraphics[height=.8\linewidth, bb=16 30 323 424, clip=]{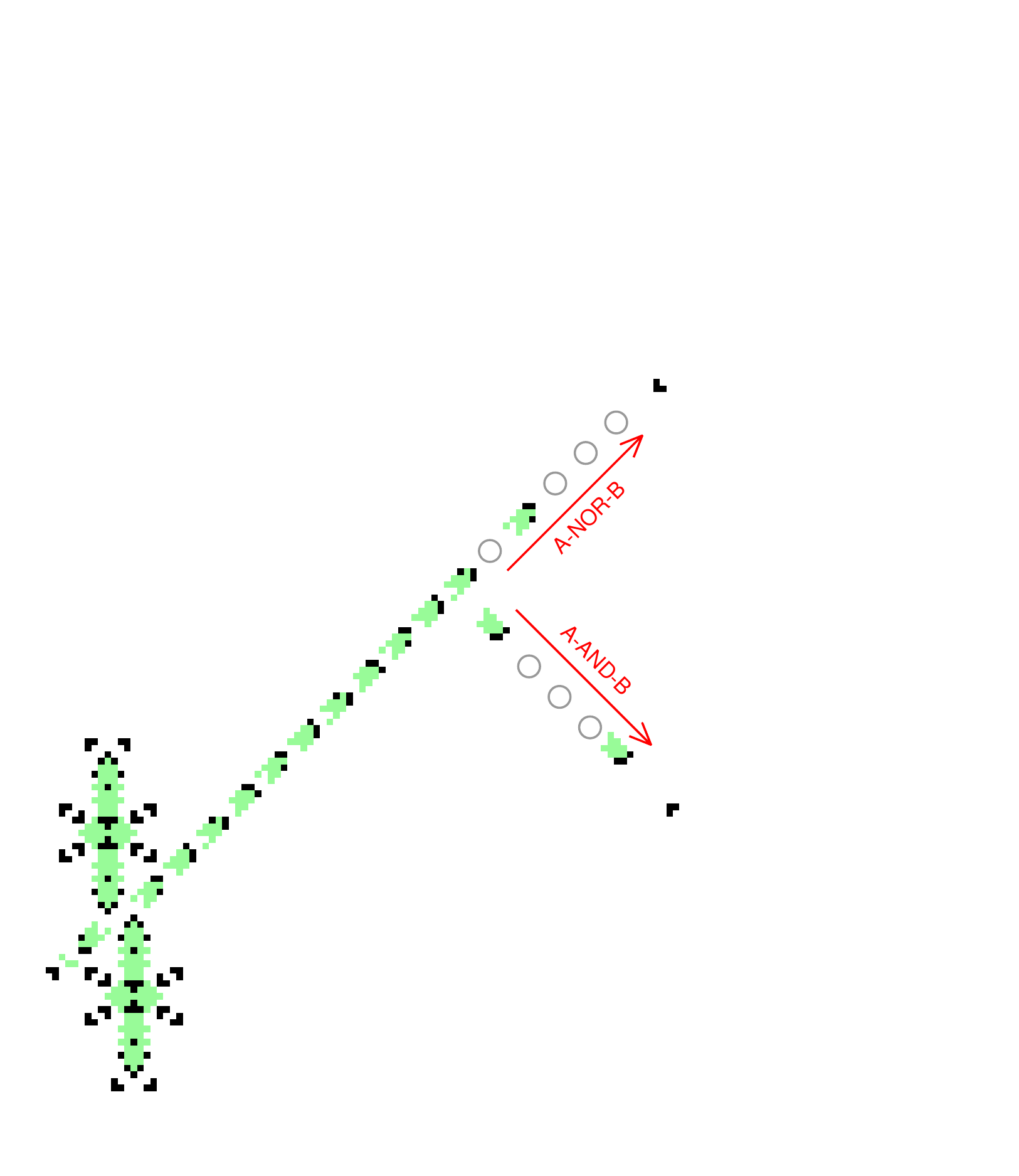}}
\end{center} 
\vspace{-3ex}
\caption[Logical gate AND]
 {\textsf{An example of the AND gate: input strings A (11001) and B (10101) 
both moving SE interact with a GGa glider-stream moving NE, resulting in 
\mbox{A-AND-B} (10001) moving SE after 208 time-steps. Any two Ga input strings
can be substituted for A and B.
The dynamics making this AND gate first
makes an intermediate \mbox{NOT-A} string 00110 (as in figure~\ref{gatenot}) 
which then interacts with string B to simultaniously
produce both the A-AND-B string moving SE described above, 
and also the \mbox{A-NOR-B} string 00010 moving NE.    
\label{gateand}
}}
\end{figure}

\enlargethispage{2ex}
The logical gate AND (1$\ast$1$\rightarrow$1, else$\rightarrow$0),
also a NOR gate (0+0$\rightarrow$1, else$\rightarrow$0),
requires one GGa glider-gun interacting with
two input strings of Ga gliders/gaps, illustrated as before/after
snapshots in figure~\ref{gateand}. Note that gate AND contains gate NOT.\\

\subsubsection{Logical Gate OR}
\begin{figure}[h!]
 \begin{center}
\fbox{\includegraphics[height=1.2\linewidth, bb=5 1 96 424, clip=]{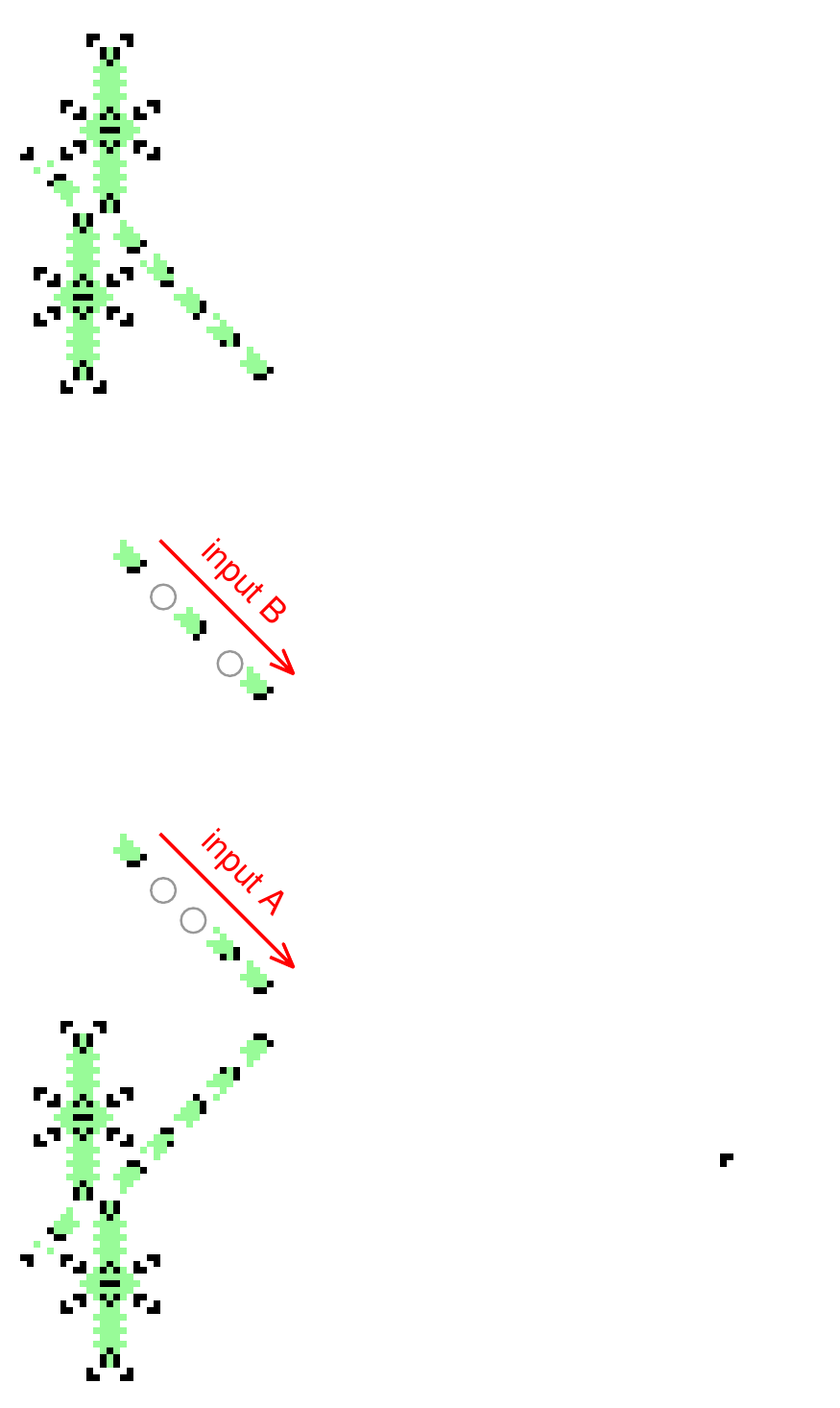}}
\hfill
\fbox{\includegraphics[height=1.2\linewidth, bb=5 5 247 424, clip=]{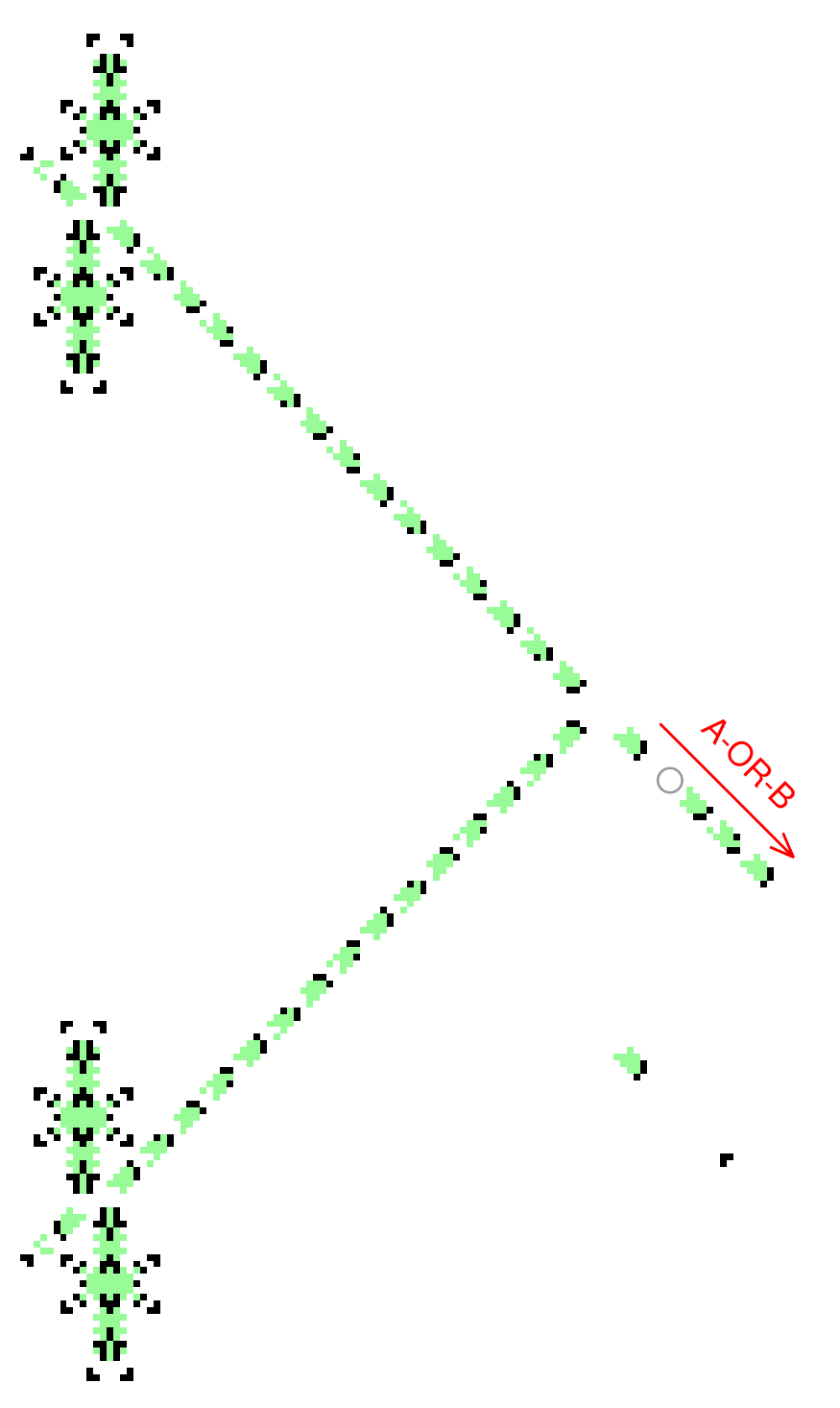}}
 \end{center}
 \vspace{-3ex}
 \caption[Logical gate OR] {\textsf{An example of the OR gate: input
     string A (11001) and B (10101) both moving SE interact with a
     GGa glider-stream moving NE, and subsequently with GGa
     glider-stream moving SE, resulting in A-OR-B (11101) moving SE
     after 302 time-steps. Any two Ga input strings can be
     substituted for A and B.
     The dynamics making this OR gate first
     makes an intermediate NOT-A string 00110 (as in figure~\ref{gatenot}), 
     which then interacts with string B to simultaniously
     produce both the AND and the NOR string as in figure~\ref{gateand}.
     The intermediate A-NOR-B string 00010 is inverted by the upper glider-gun stream
     to make NOT(A-NOR-B) which is the same as A-OR-B.
\label{gateor}
}} 
\end{figure}

The logical gate OR (0+0$\rightarrow$0, else$\rightarrow$1)
requires two GGa glider-guns interacting with
two input strings of Ga gliders/gaps, illustrated as before/after
snapshots in figure~\ref{gateor}. 
Note that gate OR contains both NOT and AND/NOR gates.

\clearpage

\section{Precursor-Rule Universe}
\label{Precursor-Rule Universe}

As well as the gliders, eaters, and collisions in
section~\ref{Gliders, Eaters, and Collisions}, the basic glider-guns
in section~\ref{Basic glider-guns}, and the logical gates built from
some of these components in section~\ref{Logical Universality}, the
Precursor-Rule is capable of an astonishing diversity of dynamical
behaviour. This section gives examples, starting with oscillators,
then various complex dynamical structures --- glider-guns, space ships,
puffer-trains, rakes, and breeders, named from the
Game-of-Life lexicon, and discovered by members of the ConwayLife
forum\cite{ConwayLife-forum}.
These kinds of structures may eventually provide the components for
universality in the full sense discussed in section~\ref{Logical Universality}.

\subsection{Variable length/period oscillators}
\label{Variable length/period oscillators}
\begin{figure}[htb]
\begin{center}
\fbox{\includegraphics[width=.55\linewidth,bb= 5 5 363 42, clip=]{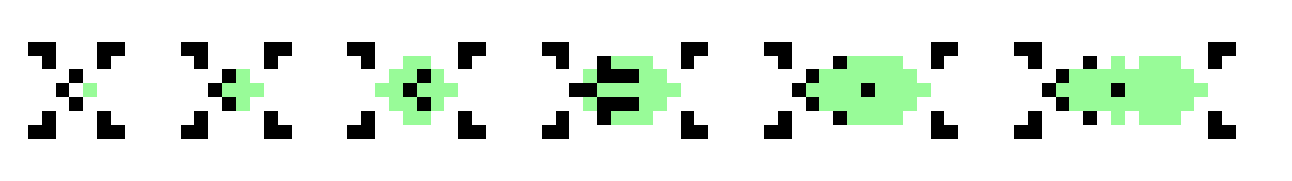}} 
\end{center}
\vspace{-3ex}
\caption[simple reflecting oscillators (SRO)]
{\textsf{Simple reflecting oscillators (SRO) --- a Gc glider bouncing between stable reflectors.
The period $p$ depends on the gap between reflectors $g$.
For the SROs above,  $g/p$ = 3/2, 4/6, 6/14, 8/22, $\dots$.
From $g/p$=4/6, as $g$ increases by 2, $p$ increases by 8.
\label{SROs}
}}  
\end{figure}

\begin{figure}[htb]
\begin{center}
\fbox{\includegraphics[width=.7\linewidth,bb= 4 4 361 34, clip=]{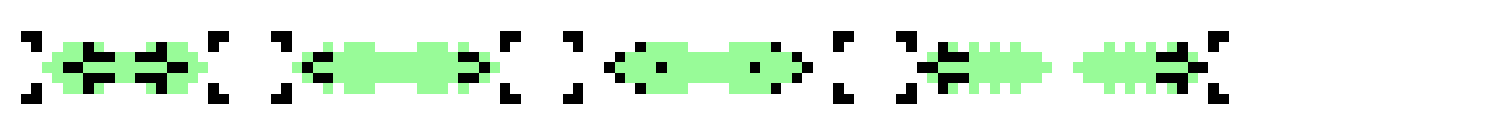}} 
\end{center}
\vspace{-3ex}
\caption[reflecting/bouncing oscillators (RBO)]
{\textsf{Reflecting/bouncing oscillators (RBO) --- two Gc gliders
bouncing off each other while reflecting between stable reflectors.
The period $p$ depends on the gap between reflectors $g$.
For the RBOs above,  $g/p$ = 16/22, 20/30, 24/38, 28/46.
As $g$ increases by 4, $p$ increases by 8.
\label{RBOs}
}}  
\end{figure}

\begin{figure}[htb]
\begin{center}
\fbox{\includegraphics[width=1\linewidth,bb= 31 19 368 51, clip=]{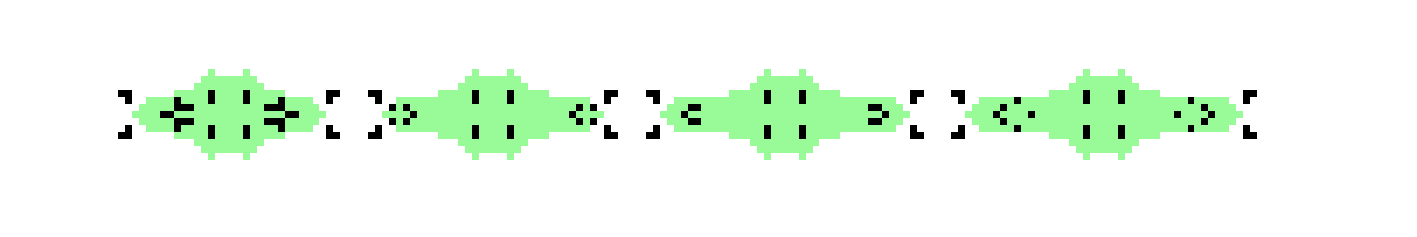}} 
\end{center} 
\vspace{-3ex}
\caption[complex RBOs]
{\textsf{Complex Reflecting/bouncing oscillators (cRBO) 
have additional central reflectors \cite{Wildmyron}, 
resulting in complex internal dynamics within the green zones.
For the cRBOs above,  $g/p$ = 28/41, 32/49, 36/57, 40/65.
As $g$ increases by 4, $p$ increases by 8.
\label{cRBOs}
}}  
\end{figure}

Here we show oscillators between stable double reflectors, where 
the gap can be increased by modular intervals, which increases the period
(figures~\ref{SROs}, \ref{RBOs} and \ref{cRBOs}).
The double reflectors can be made of
\raisebox{-1.7ex}{\includegraphics[height=3.7ex,bb=2 -1 20 20, clip=]{Ea1}},
\raisebox{-1.7ex}{\includegraphics[height=3.7ex,bb=2 -1 20 20, clip=]{Ea5}} or
\raisebox{-1.7ex}{\includegraphics[height=3.7ex,bb=2 -1 15 20, clip=]{Ea6}}
interchangeably. The dynamics are indicated by green trails.

\subsection{Other oscillators}
\label{Other oscillators} 
\begin{figure}[htb]
\begin{center}
\begin{minipage}[c]{1\linewidth}
\fbox{\includegraphics[width=1\linewidth,bb= 5 38 340 81, clip=]{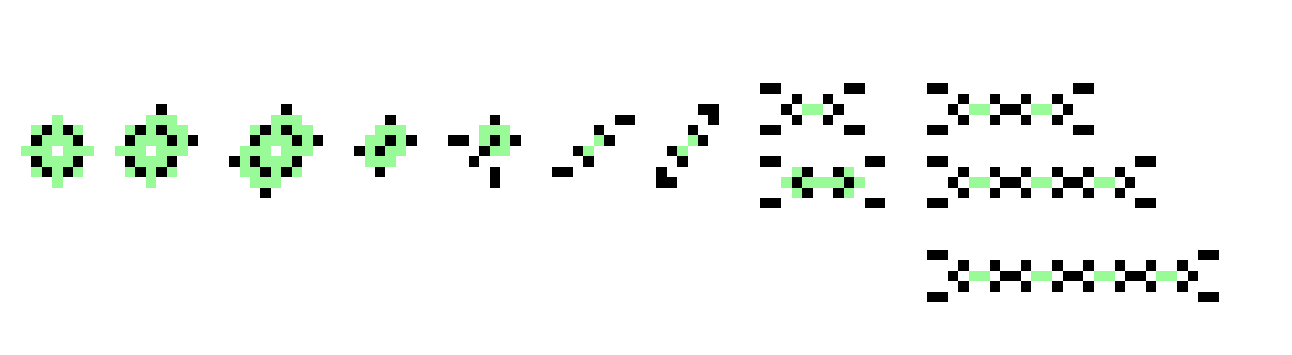}}\\ 
\textsf{P2\color{white}---------------------------------------------------------------------\color{black}$g$\color{white}-----------------\color{black}$h$}
\end{minipage}\\[2ex]
\begin{minipage}[c]{.9\linewidth}
\begin{minipage}[c]{.096\linewidth}
\fbox{\includegraphics[width=1\linewidth,bb= 10 6 48 53, clip=]{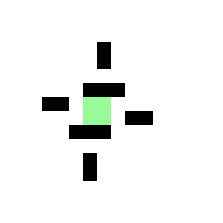}}\\ 
\textsf{P3}
\end{minipage}
\hfill
\begin{minipage}[c]{.35\linewidth}
\fbox{\includegraphics[width=1\linewidth,bb= 5 5 136 56, clip=]{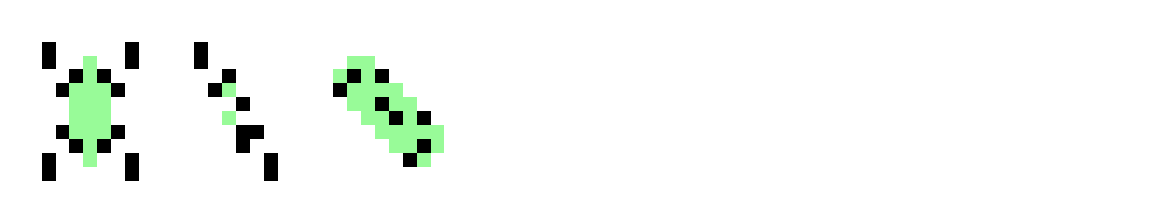}}\\ 
\textsf{P4}
\end{minipage}
\hfill
\begin{minipage}[c]{.1\linewidth}
\fbox{\includegraphics[width=1\linewidth,bb=  11 7 58 63, clip=]{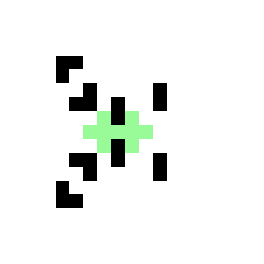}}\\ 
\textsf{P5}
\end{minipage}
\hfill
\begin{minipage}[c]{.23\linewidth}
\fbox{\includegraphics[width=1\linewidth,bb= 3 3 94 52, clip=]{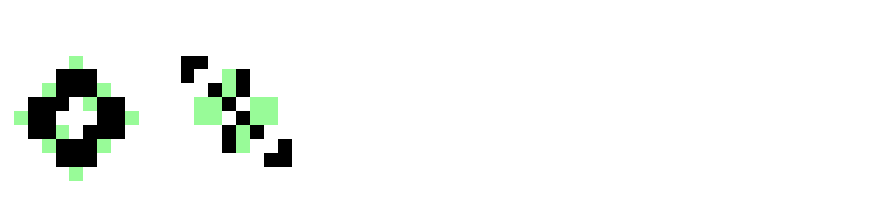}}\\ 
\textsf{P6}
\end{minipage}
\end{minipage}\\[2ex]
\begin{minipage}[c]{1\linewidth}
\begin{minipage}[c]{.26\linewidth}
\fbox{\includegraphics[width=1\linewidth,bb= 22 18 133 77, clip=]{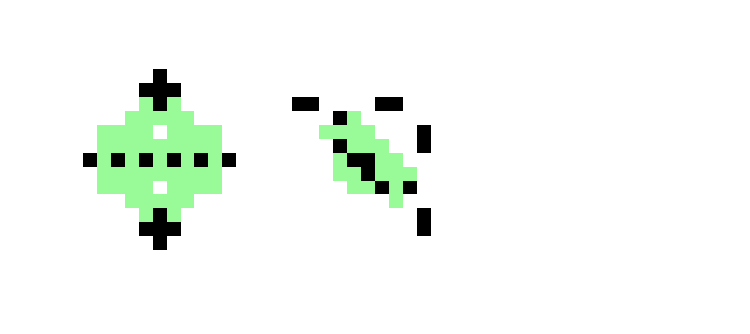}}\\ 
\textsf{P8}
\end{minipage}
\hfill
\begin{minipage}[c]{.26\linewidth}
\fbox{\includegraphics[width=1\linewidth,bb= 5 17 116 71, clip=]{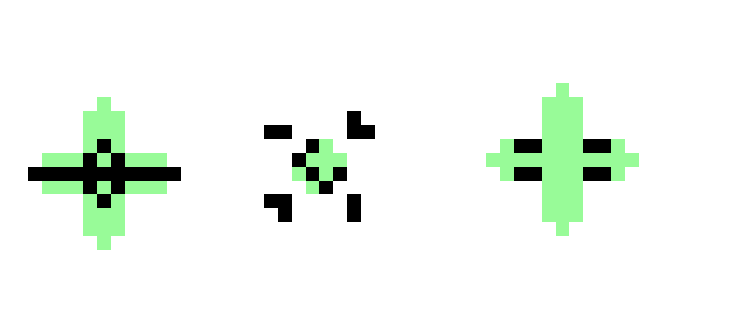}}\\ 
\textsf{P12}
\end{minipage}
\hfill
\begin{minipage}[c]{.12\linewidth}
\fbox{\includegraphics[width=1\linewidth,bb= 32 29 87 100, clip=]{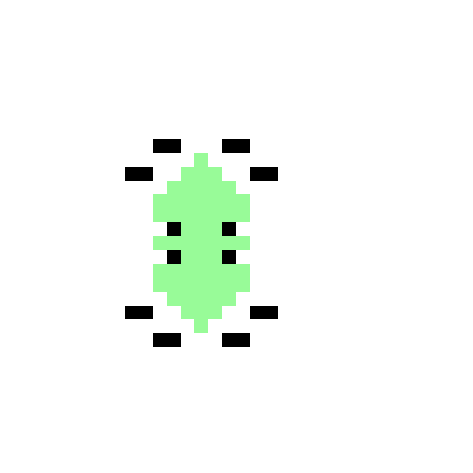}}\\ 
\textsf{P25}
\end{minipage}
\hfill
\begin{minipage}[c]{.2\linewidth}
\fbox{\includegraphics[width=1\linewidth,bb= 20 47 109 88, clip=]{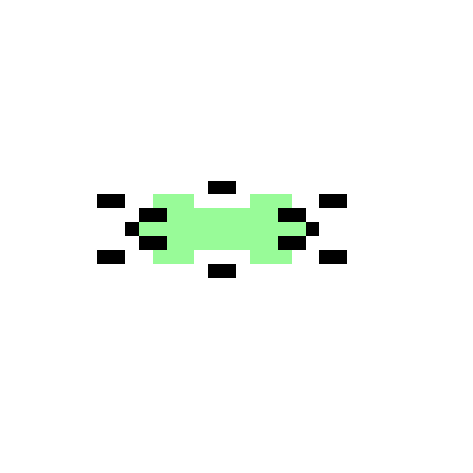}}\\ 
\textsf{P27}
\end{minipage}
\end{minipage}
\end{center}
\vspace{-3ex}
\caption[Period 2 oscillators]
{\textsf{Oscillators with periods P2 ($g$ and $h$ are extendable) and P3, P4, P5, P6, P8, P12, P25 and P27.
Dynamics occurs within the green zones.
\label{oscP2-27}
}}
\end{figure}

As well as the variable length/period oscillators in 
section~\ref{Variable length/period oscillators}, 
there are many other oscillators
with various periods, most found at the ConwayLife forum.
Some are illustrated in figure~\ref{oscP2-27}.

\subsection{Other glider-guns}
\label{Other glider-guns}  
\vspace{-8ex}
\begin{figure}[htb]
\begin{center}
\begin{minipage}[t]{.9\linewidth}
\parbox[t]{.52\linewidth}{ \vspace{-23ex}    
\caption[glider-gun GG2a]
{\textsf{Glider-gun GG2a (discovered by \cite{Wildmyron}) 
shoots 4-phase G2a (double Ga) gliders diagonally in 4 directions
with a period of 24. The gliders
are stopped by strategically placed eaters. Dynamic trails=15 \label{GG2a}}}}
\hfill
\begin{minipage}[t]{.4\linewidth}
\fbox{\includegraphics[width=1\linewidth,bb= 17 23 288 294, clip=]{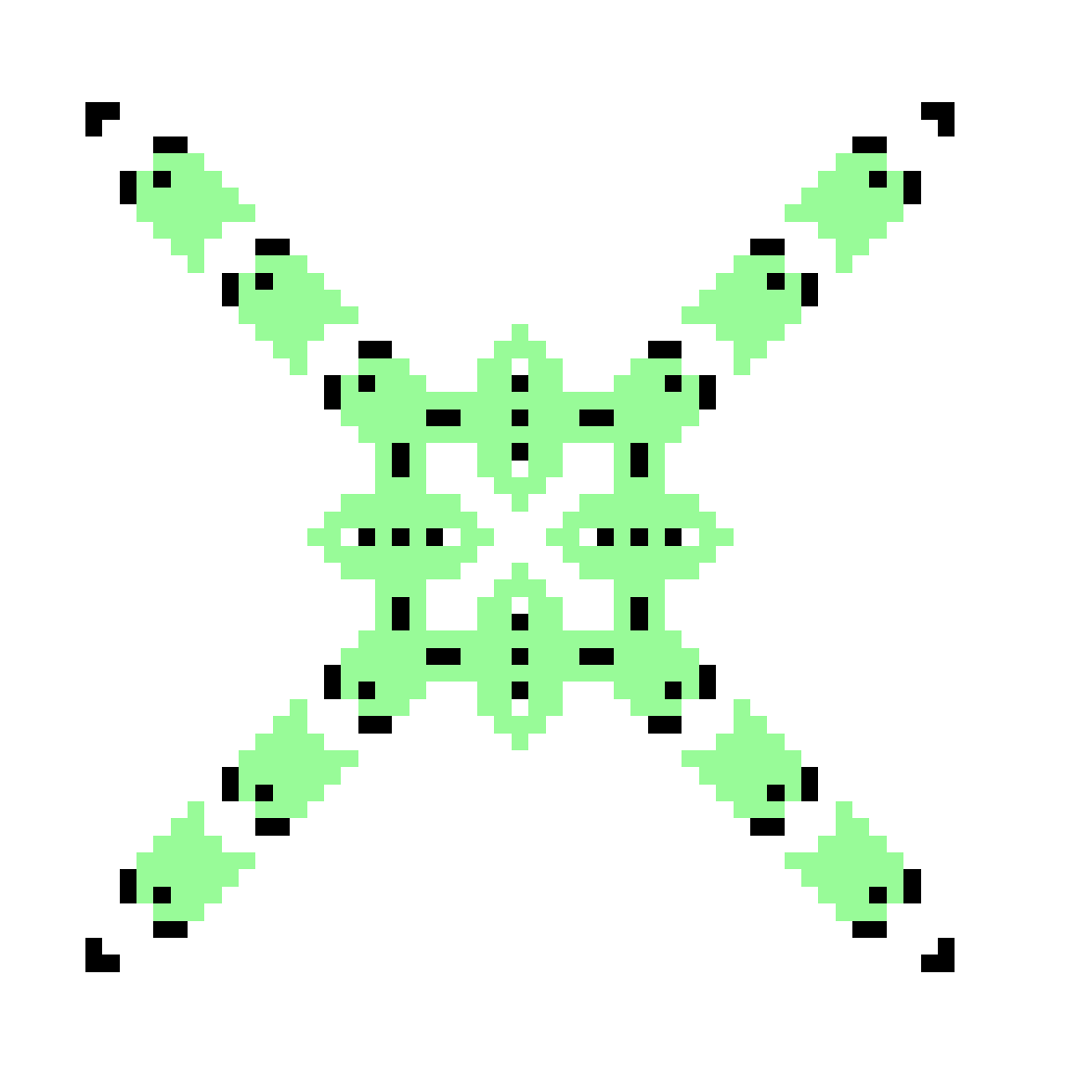}} 
\end{minipage}
\hfill 
\end{minipage}
\end{center}
\vspace{-3ex} 
\end{figure}

\noindent The first two glider-guns in the Precursor-Rule (discovered by \cite{Wildmyron}), 
opened the floodgates for further discovery.
The second was the basic GGc (figure~\ref{GGc-GGa-basic}(a)) already discussed.
The first ``QuadGG2a'' shoots G2a, double Ga gliders (figure~\ref{GG2a}) and
is significant because it provides the 
building blocks for meta-glider-guns made from interacting simpler glider-guns,
including the Ga glider-gun discovered by\cite{Wright}, shown in figure~\ref{metaGG}(b).

\subsubsection{Meta-glider-guns}
\label{Meta-glider-guns} 
 
\begin{figure}[htb]
\begin{minipage}[b]{1\linewidth}
\begin{small}
\begin{minipage}[b]{.43\linewidth}
\fbox{\includegraphics[width=1\linewidth,bb= 25 33 340 271, clip=]{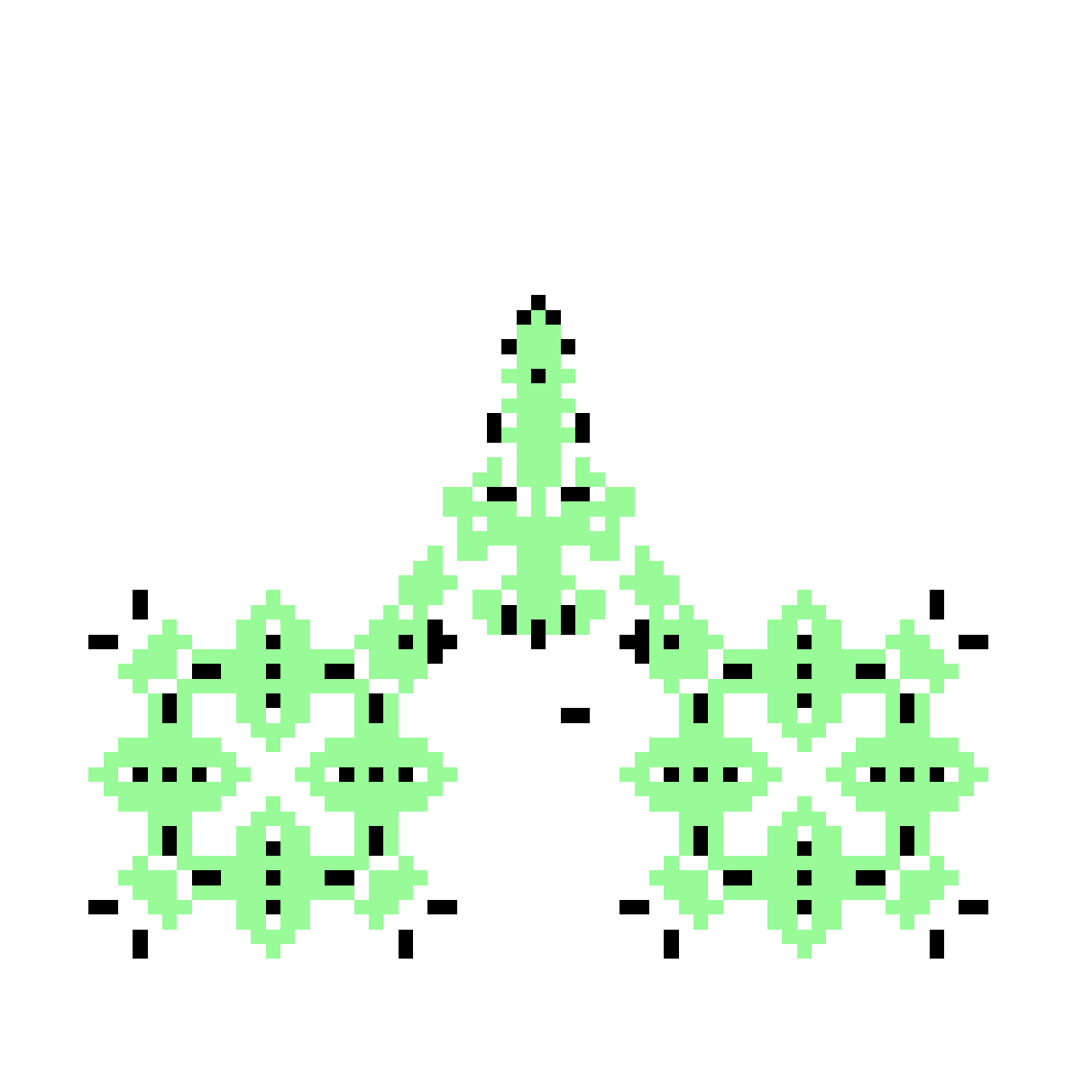}}\\ 
\textsf{(a) Gc glider-gun ``QuadGG2a'' \cite{Thunk} shoots Gc gliders North 
as shown with a period of 408 time-steps.
The dynamics are very complex --- during this interval three Gc gliders are
also shot South at time-steps 133, 307 and 355.}
\end{minipage}
\hfill
\begin{minipage}[b]{.53\linewidth}
\fbox{\includegraphics[width=1\linewidth,bb= 14 11 393 340, clip=]{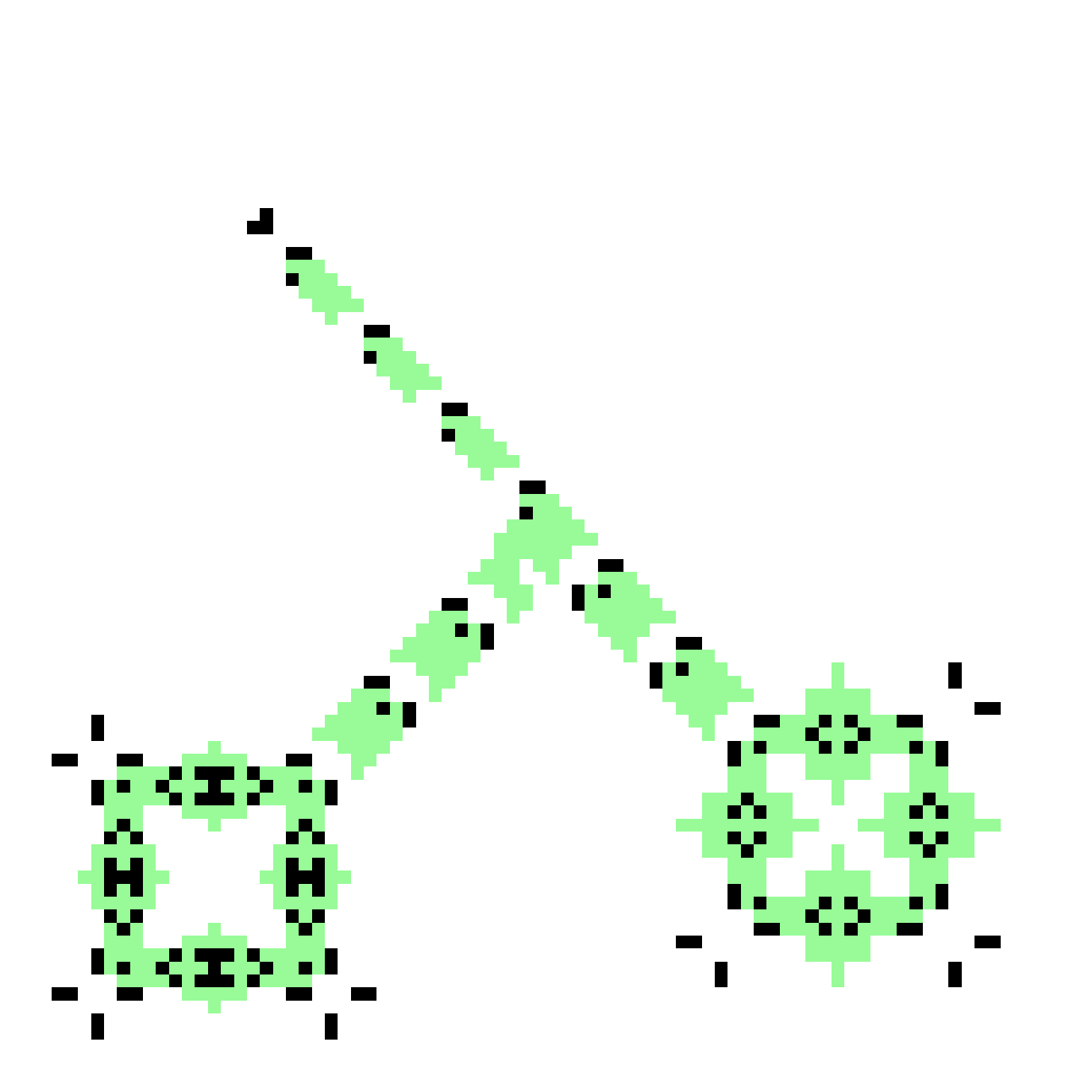}}\\ 
\textsf{(b) Ga glider-gun\cite{Wright}) shoots Ga gliders with a period of 24 time-steps, 
shown shooting NW.}
\end{minipage}\\[3.5ex]
\begin{minipage}[b]{.44\linewidth}
\fbox{\includegraphics[width=1\linewidth,bb= 40 65 359 373, clip=]{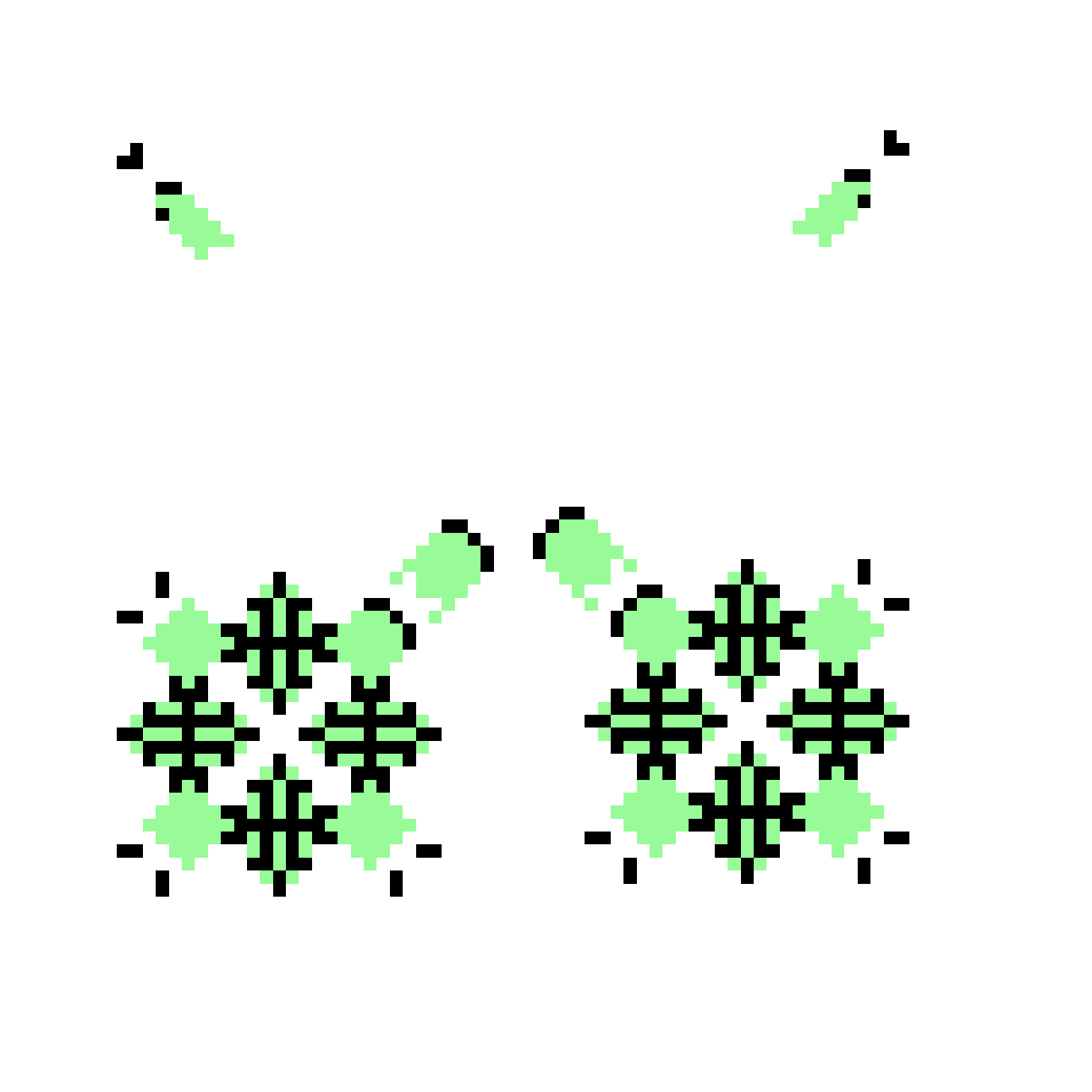}}\\ 
\textsf{(c) Ga glider-gun\cite{jmgomez} shoots
two Ga gliders with a period of 120, shown shooting NW and NE.}
\end{minipage}
\hfill
\begin{minipage}[b]{.49\linewidth}
\fbox{\includegraphics[width=1\linewidth,bb= 15 5 401 382, clip=]{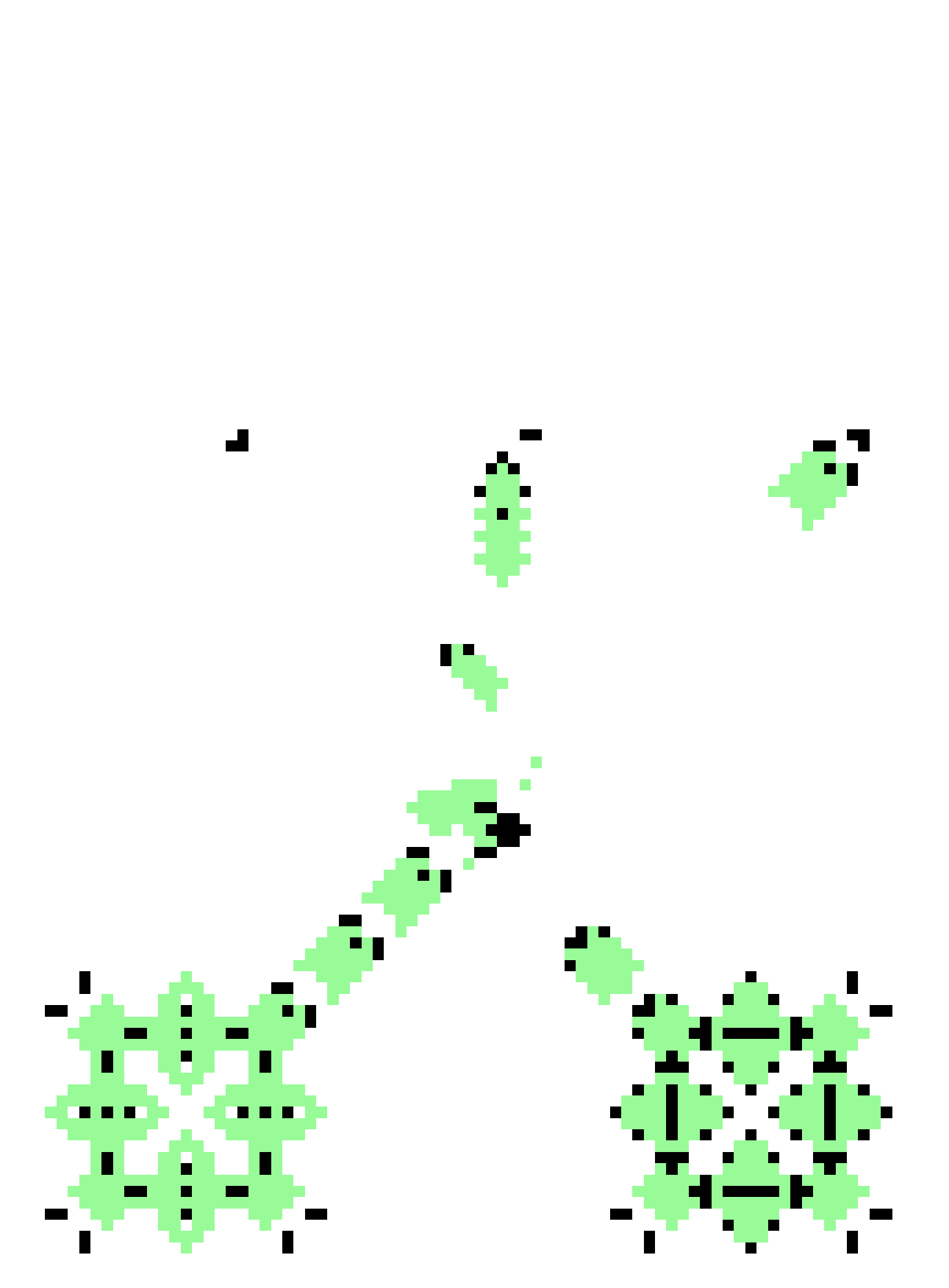}}\\ 
\textsf{(d) Multiple glider-gun\cite{Thunk}  shoots
3 different glider types, Ga, Gc and G2a, with a period of 144, shown shooting NW, North, and NE.}
\end{minipage}
\end{small}
\end{minipage}
\vspace{-2ex}
\caption[meta-glider-guns]
{\textsf{Meta-glider-guns built from two interacting GG2a glider-guns 
(figure~\ref{GG2a}),
the outcomes of different 90$^{\circ}$ collisions between two G2a gliders.
Dynamic trails=15.
\label{metaGG}
}}
\end{figure}

Figure~\ref{metaGG} shows 4 examples of meta-glider-guns --  they 
need to be seen in action to appreciate their extraordinary dynamics.

\clearpage

\begin{figure}[htb]
\begin{center}
\fbox{\includegraphics[width=.9\linewidth,bb= 21 11 441 356, clip=]{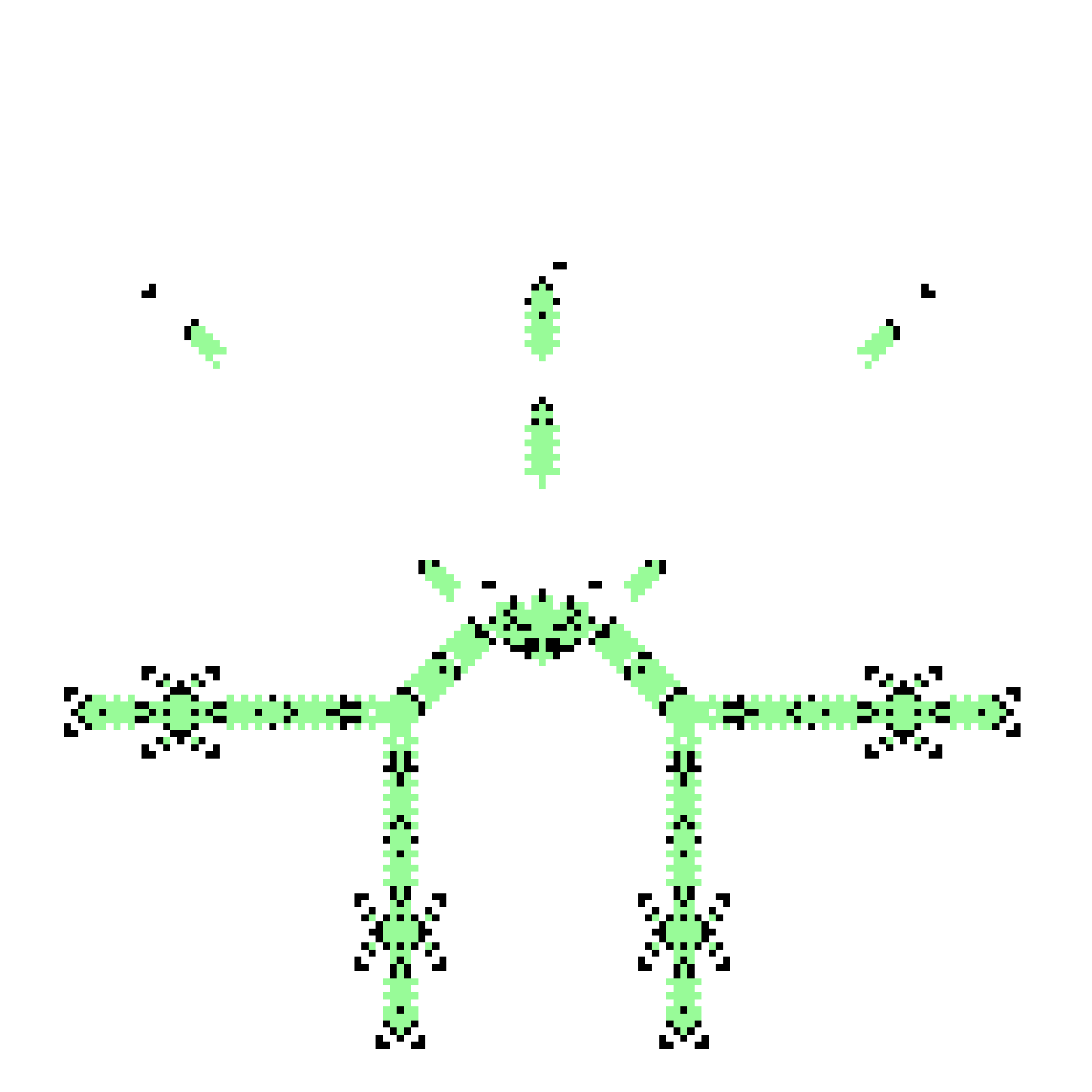}} 
\end{center}
\vspace{-3ex}
\caption[multiple meta-glider-gun]
{\textsf{Multiple meta-glider-gun\cite{jmgomez} is built from 4 interacting
basic GGc glider-guns (figure~\ref{GGc-GGa-basic}(a)) and produce 2 intermediate G2a glider-streams
colliding at 90$^{\circ}$. The system's period is 133 time-steps, but within this interval 2 Ga gliders
are released (shown moving NW and NE), and finally two Gc gliders (shown moving North).
\label{jm-multi}
}}  
\vspace{-2ex}
\end{figure}

\subsubsection{Multi-glider-guns}
\label{Multi-glider-guns} 

Figure~\ref{metaGG}(d) demonstrates a multi-glider-gun
in that it shoots more than one glider type. In
figure~\ref{jm-multi} we show another multi-glider-gun, also a meta-glider-gun
because it is constructed from 4 interacting GGc glider-gun sub-units, discovered by
\cite{jmgomez}.  Sending different glider types simultaneously from
the same gun is arguably novel in relation to the Game-of-Life. Of
course, any of the glider streams can be blocked by strategically
positioned eaters.

\subsubsection{Variable period Gc glider-guns}
\label{Variable period Gc glider-guns}

\begin{figure}[htb] 
\begin{small}
\begin{minipage}[b]{1\linewidth} 
\fbox{\includegraphics[width=1\linewidth,bb= 6 4 423 110, clip=]{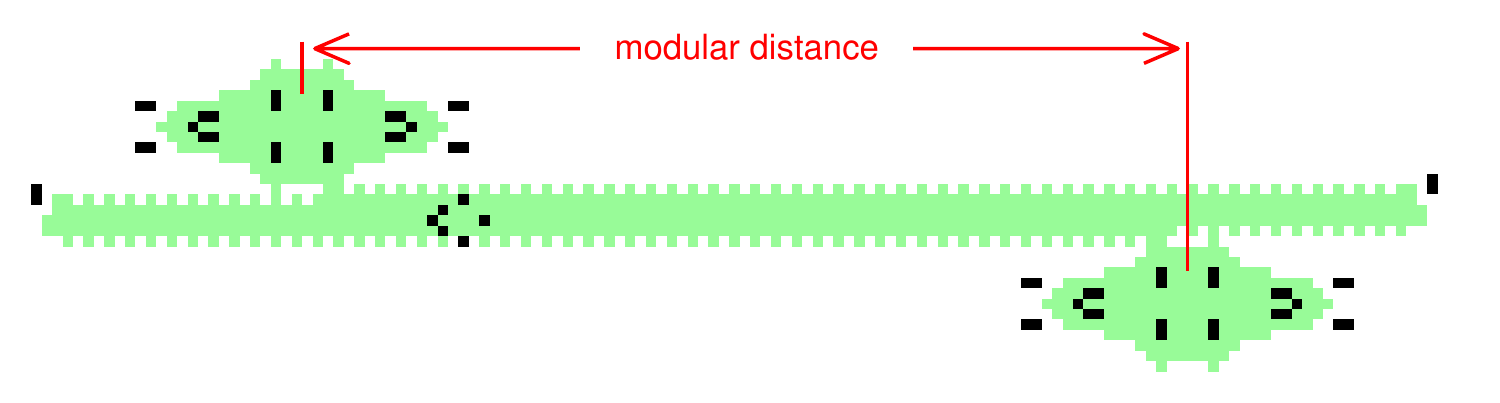}}\\ 
\textsf{(a) GGcV glider-gun at time-step 0, one Gc glider moving West.}\\[2ex]
\fbox{\includegraphics[width=1\linewidth,bb= 6 4 423 103, clip=]{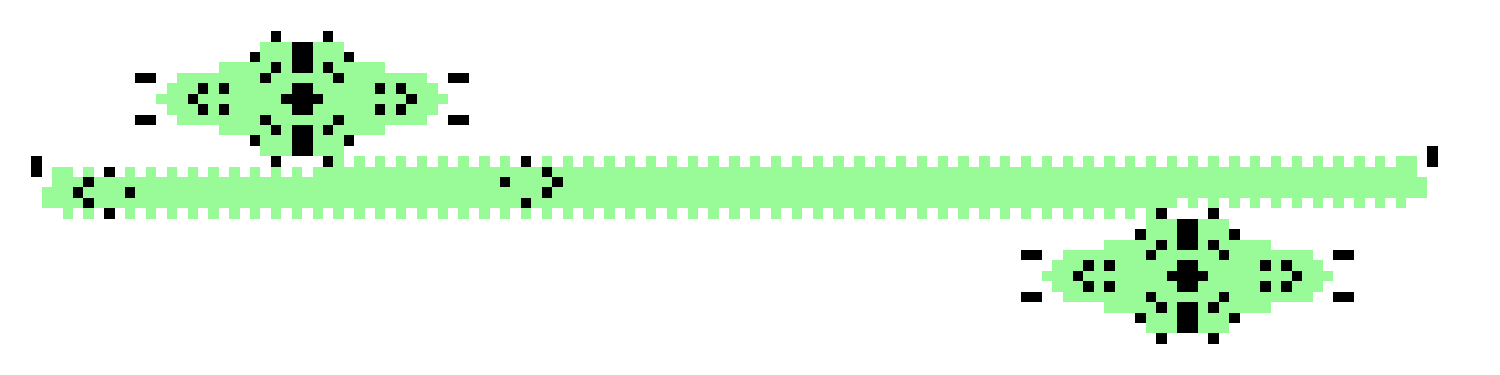}}\\
\textsf{(b) time-step 68, the Westward glider keeps going, and a new Gc glider moves East.}\\
\end{minipage}
\end{small}
\begin{minipage}[c]{1\linewidth}
\begin{minipage}[c]{.39\linewidth}
\includegraphics[width=1\linewidth]{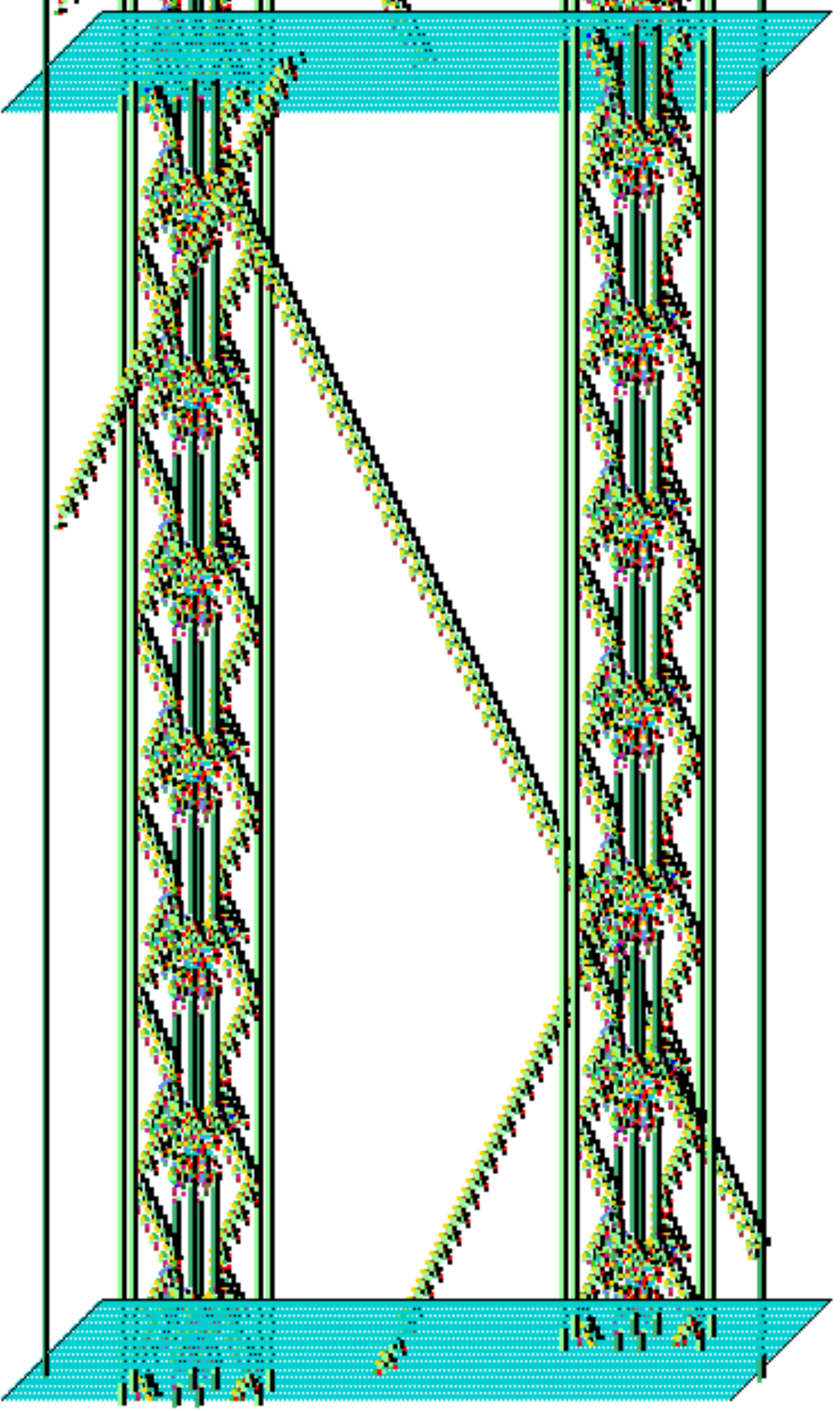} 
\end{minipage}
\hspace{-4ex}
\begin{minipage}[c]{.05\linewidth}
\includegraphics[width=1\linewidth]{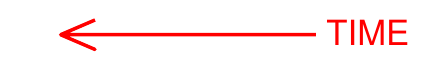}
\end{minipage}
\hfill
\begin{minipage}[c]{.57\linewidth}
\caption[GGcV Variable period Gc glider-guns]
{\textsf{\underline{\it Above}: snapshots (a) and (b) of the smallest GGcV 
variable period Gc glider-gun are shown
with a continuous green dynamic trace to indicate dynamic regions and glider movement.
If (a) is at time-step 0 with a glider moving West, a new glider is created as the Westward
glider brushes past the left cRBO.  (b) shows the resulting two gliders at time-step 68.
The new glider moving East will interact with the right cRBO to continue the cycle, so
Gc gliders will be shot West and East --- here they are stopped by eaters.\\[2ex]
\underline{\it Left}: the GGcV glider-gun as a 2d space-time pattern, with about 270 time-steps
stacked below each other in an isometric projection\cite{Wuensche2016,Wuensche-DDLab}
showing the paths of gliders and also an impression of the internal cRBO dynamics.
\label{GGcV}
}}
\end{minipage}
\end{minipage}
\end{figure}

\enlargethispage{3ex}
A system of variable period glider-guns, GGcV, for glider Gc was 
created by \cite{Wildmyron}. Built from his complex reflecting/bouncing 
oscillators (cRBOs), the system is demonstrated with the smallest 
cRBO (gap/period of 28/41 in figure~\ref{cRBOs}).
Two cRBOs are positioned as in figure~\ref{GGcV}(a), with the distance
between their centers $d$=85 cells. A Gc glider pointing West is introduced,
and its interaction brushing past the pulsating
cRBO creates a new Gc glider moving East as in figure~\ref{GGcV}(b), 
which interacts with the second cRBO repeating the cycle. 

The result is glider-gun GGcV shooting Gc gliders East and West. The whole structure
has a period of 328 time-steps.
The distance between cRBO centers $d$ can be adjusted by modular amounts to increase the period.
To date the following have been demonstrated, $d/p$ = 85/328, 167/656, 249/984, 331/1312.
The series continues $d$+82/$p$+328.
Similar structures can also be built with the larger cRBOs in figure~\ref{cRBOs}.

\clearpage

\subsection{Spaceships and puffer-trains}
\begin{figure}[htb]
\begin{center}
\begin{minipage}[c]{1\linewidth}
(a) \begin{minipage}[c]{.3\linewidth}
\fbox{\includegraphics[width=1\linewidth,bb= 28 82 152 132, clip=]{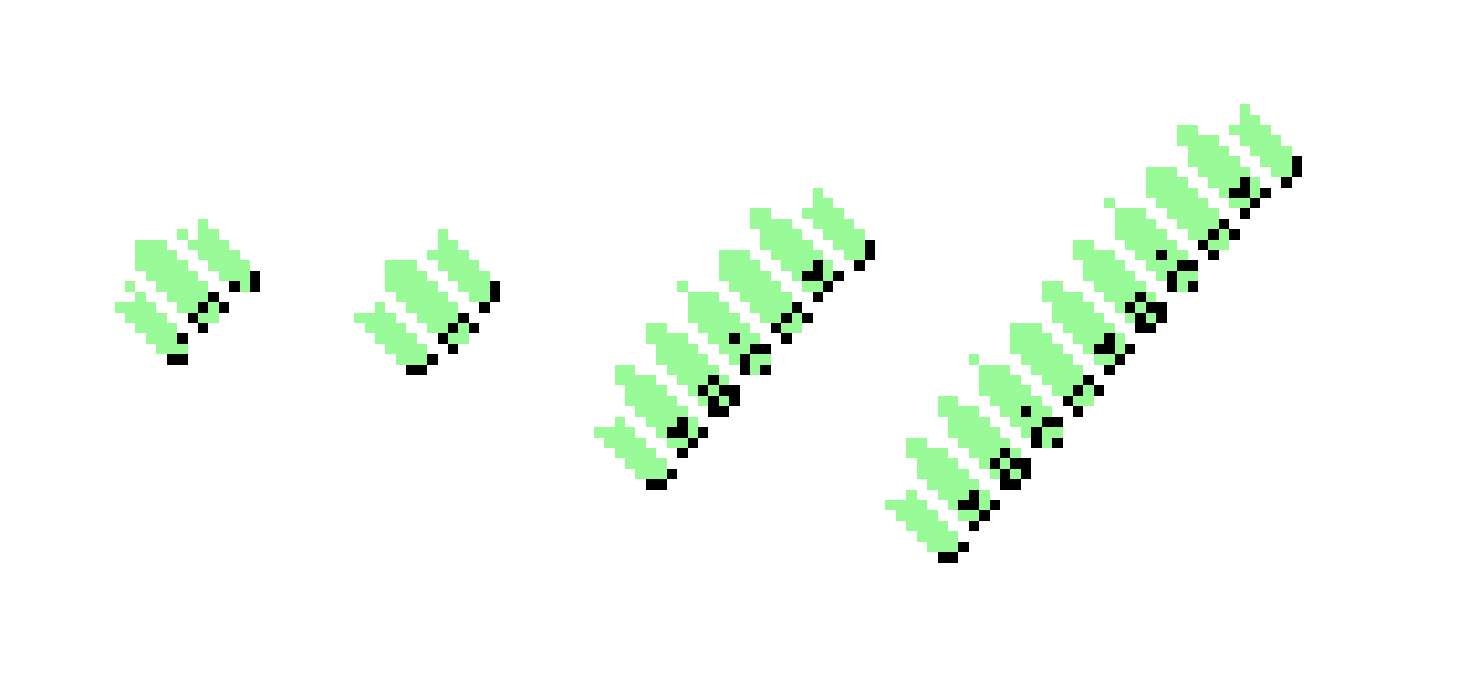}} 
\end{minipage}
\hfill
(b) \begin{minipage}[c]{.53\linewidth}
\fbox{\includegraphics[width=1\linewidth,bb= 169 29 382 167, clip=]{comGa}}\\ 
\end{minipage}
\end{minipage}
\end{center}
\vspace{-5ex}
\caption[Ga related spaceships]
{\textsf{(a) small spaceships built from a subunit between two Ga gliders\cite{Wildmyron},
and (b) extend to any length by inserting more subunits\cite{Scorbie}.
Shown here moving SE in 4-phases ($p$=4, $s$=$c/4$). Dynamic trails=20.
\label{comGa}
}}  
\vspace{-3ex}
\end{figure}

\begin{figure}[htb]
\begin{center}
(a)\begin{minipage}[c]{.44\linewidth}
\fbox{\includegraphics[width=1\linewidth,bb= 73 28 284 124, clip=]{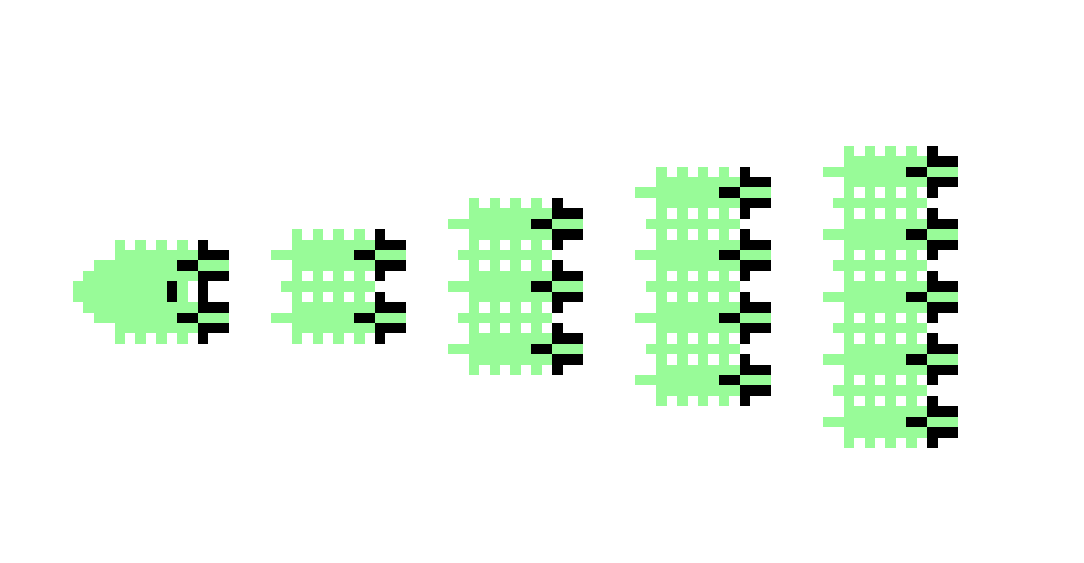}}  
\end{minipage}\\[1ex]
\begin{minipage}[c]{1\linewidth}
\fbox{\includegraphics[width=1\linewidth,bb= 99 61 420 128, clip=]{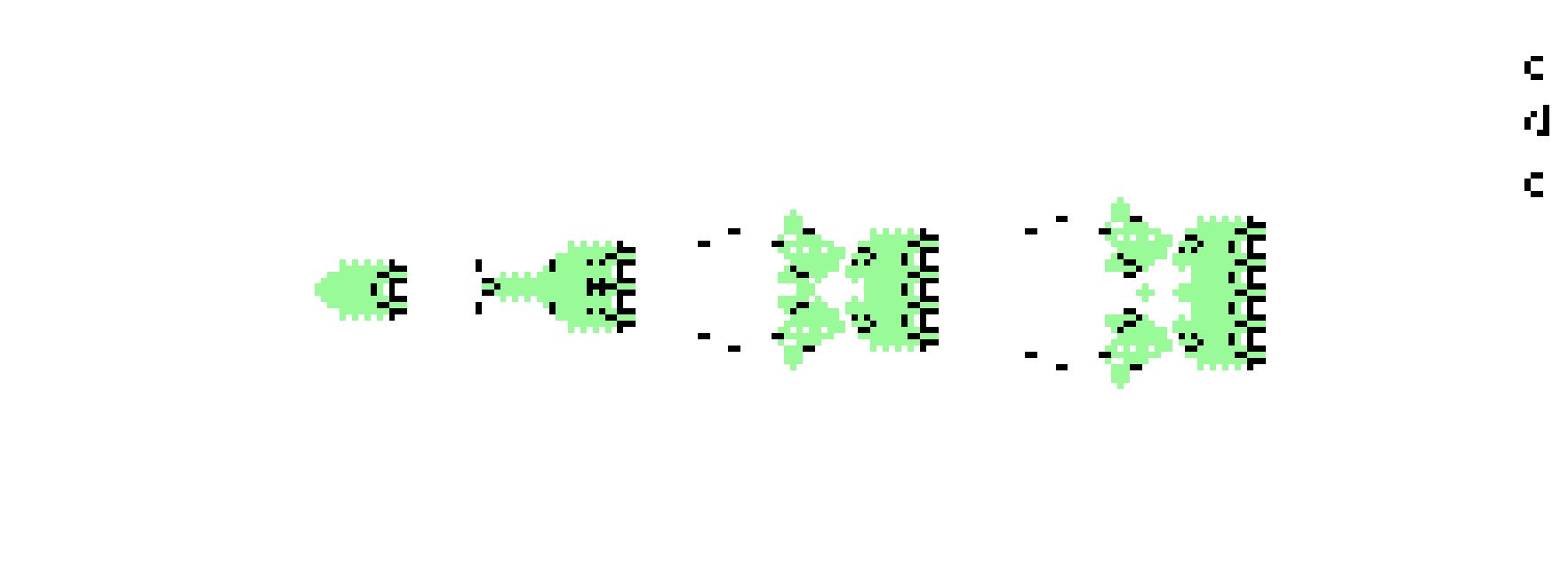}}\\ 
\color{white}--\color{black}(b) $p$=8
\color{white}----------\color{black}(c) $p$=24
\color{white}------------------\color{black}(d) $p$=24
\color{white}----------------------\color{black}(e) $p$=24
\end{minipage}
\end{center}
\vspace{-3ex}
\caption[Gc related spaceships and puffer-trains]
{\textsf{
Spaceships and puffer-trains can be built from Gc glider subunits.
Two or more Gc gliders separated by
one cell (at their widest phase) form stable spaceships (a)\cite{Saka}.
However, if the Gc gliders touch (b - e) the tail becomes unstable.
(b) with two Gc's forms a spaceship\cite{Saka}, (c) with 3 Gc's forms a spaceship after
a Gc glider is first ejected\cite{Wildmyron}, and (d, e) with 4 and 5 Gc's form
puffer-trains\cite{Wildmyron}.
All are shown moving East ($s$=$c/2$) with dynamic trails=16. 
The leading edges of Gc gliders transit through the usual 4-phases.
The periods of the structures are noted.
\label{GcShips}
}}  
\end{figure}

\begin{figure}[htb]
\begin{center}
\begin{minipage}[c]{.8\linewidth}
\fbox{\includegraphics[width=1\linewidth,bb= 15 22 337 88, clip=]{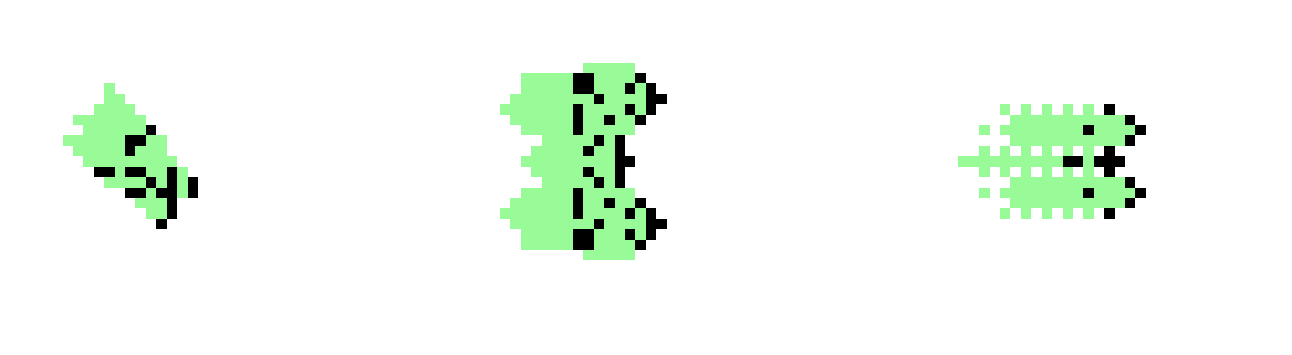}}\\  
\color{white}--------\color{black}(a)
\color{white}----------------------------\color{black}(b)
\color{white}--------------------------------\color{black}(c)
\end{minipage}
\end{center}
\vspace{-4ex}
\caption[other spaceships]
{\textsf{
Three spaceships. (a)\cite{Wildmyron} moving SE ($p$=6, $s$=$c/6$), 
(b)\cite{Wildmyron} moving East ($p$=4, $s$=$c/2$), (c)\cite{Danieldb} moving East ($p$=4, $s$=$c/2$)
with dynamic trails=20.
\label{Ships-3}
}}  
\vspace{-2ex}
\end{figure}

\begin{figure}[htb]
\begin{center}
\begin{minipage}[c]{1\linewidth}
\fbox{\includegraphics[width=1\linewidth,bb= 19 29 447 83, clip=]{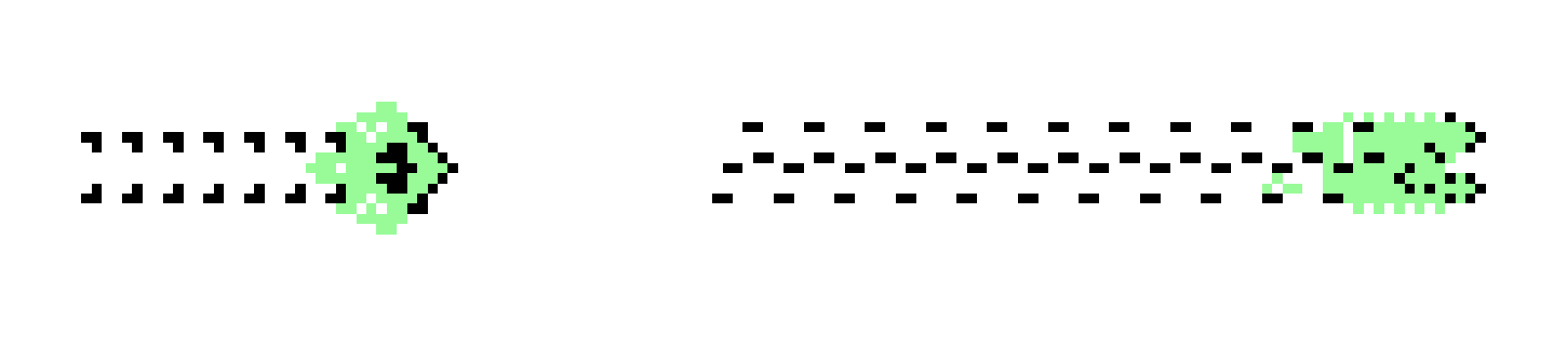}}\\  
\color{white}-----------------------\color{black}(a)
\color{white}------------------------------------------------------------------\color{black}(b)
\end{minipage}
\end{center}
\vspace{-4ex}
\caption[other puffer-trains]
{\textsf{
Puffer-trains discovered by \cite{Wildmyron}.
(a) is slow moving, advancing just 4 spaces in its period $p$=23, $s$=$c/5.75$, 
and (b) $p$=12, $s$=$c/2$.
Shown moving East with dynamic trails=20.
\label{PufW45}
}}  
\end{figure}

A spaceship is a mobile periodic pattern
larger than a simple glider --- a puffer-train is similar but leaves debris in
its wake. The combined Ga gliders in figure~\ref{Ga combined gliders SE}
could be classified as spaceships.
Figures~\ref{comGa} to \ref{PufW45} show examples of spaceships and puffer-trains,
some suggested at the ConwayLife forum,
and relating to combinations of either Ga or Gc gliders, as well as other patterns.
The period $p$ and speed $s$ are indicated.

\enlargethispage{10ex}
\clearpage

\subsection{Rakes}
\begin{figure}[htb] 
\begin{center}
\begin{minipage}[c]{1\linewidth}
\fbox{\includegraphics[width=1\linewidth,bb= 41 80 376 202, clip=]{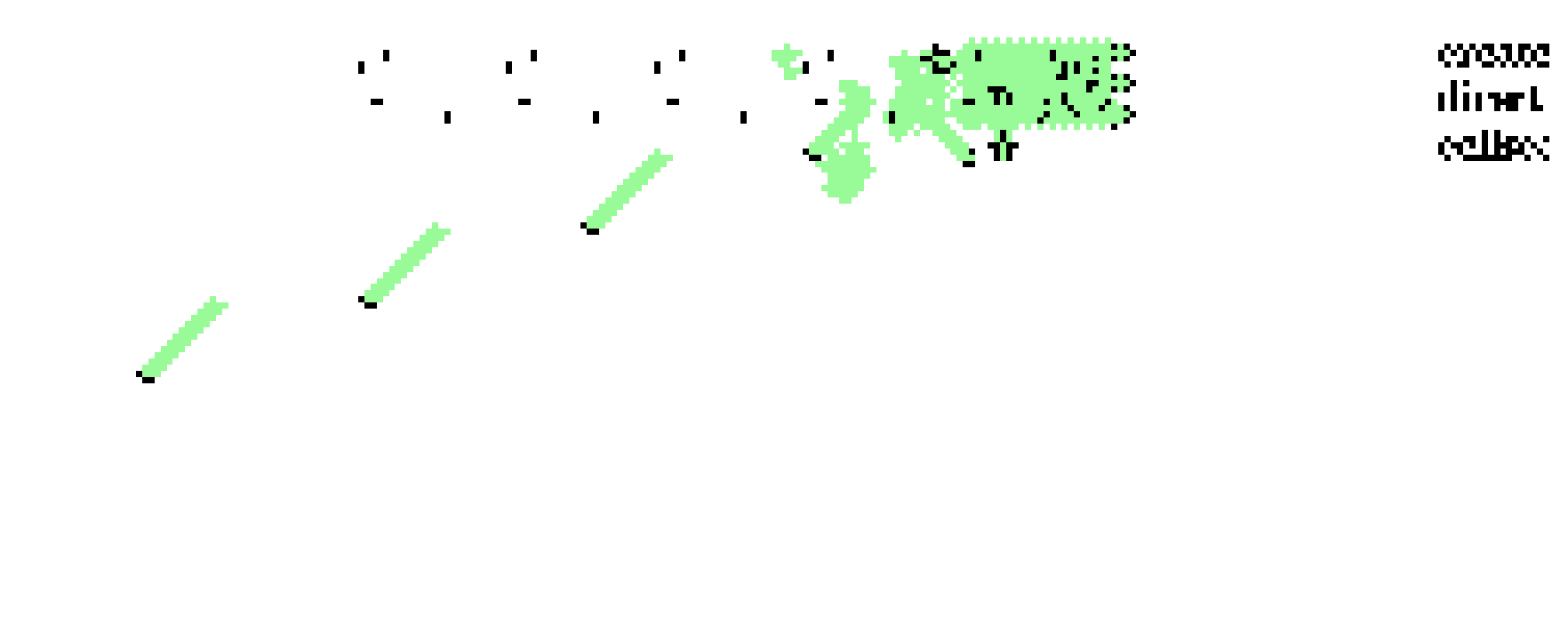}}\\  
\end{minipage}
\end{center}
\vspace{-4.5ex}
\caption[Meta-Rakes of glider Ga]
{\textsf{
A combination between a puffer-train and a rake\cite{Awesome}, $p$=48, $s$=$c$/2, sends
Ga gliders SW and leaves debris. Dynamic trails=50.
\label{RPawe3}
}}
\end{figure}

\begin{figure}[htb]
\begin{center}
\begin{minipage}[c]{.8\linewidth} 
\color{white}--------------------------------------------\color{black}(a)\\
\fbox{\includegraphics[width=1\linewidth,bb= 54 81 432 229, clip=]{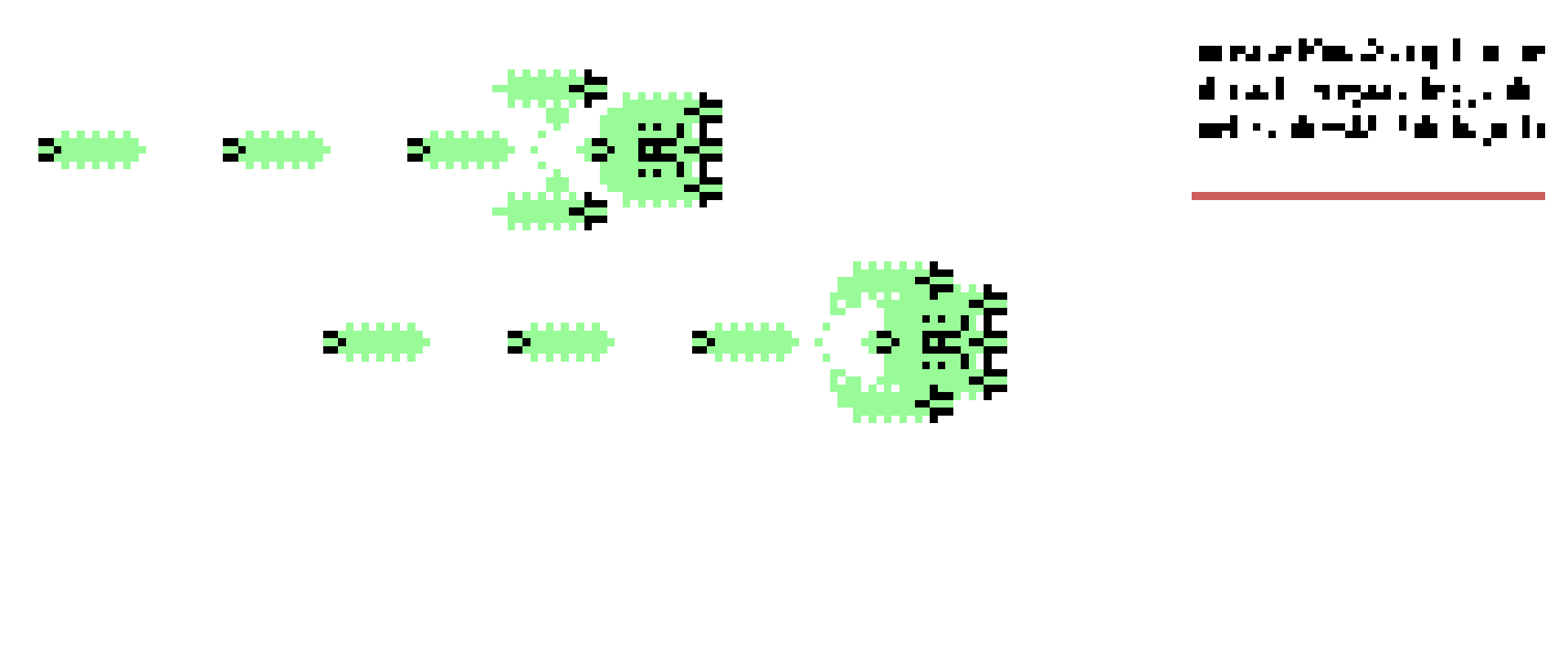}}\\  
\color{white}--------------------------------------------------------------------\color{black}(b)
\end{minipage}
\end{center}
\vspace{-3.5ex}
\caption[Rake Gc]
{\textsf{
Two similar rakes based on a core of three touching Gc subunits move East ($p$=24, $s$=$c/2$),
shed Ga gliders moving West. Although rake (a)\cite{Thunk} is slightly different from 
rake (b)\cite{Scorbie} --- a similar phase is show.
Dynamic trails=20.
\label{RakeGcB}
}}  
\end{figure}

\begin{figure}[htb]
\begin{center}
\fbox{\includegraphics[width=1\linewidth,bb= 83 107 453 393, clip=]{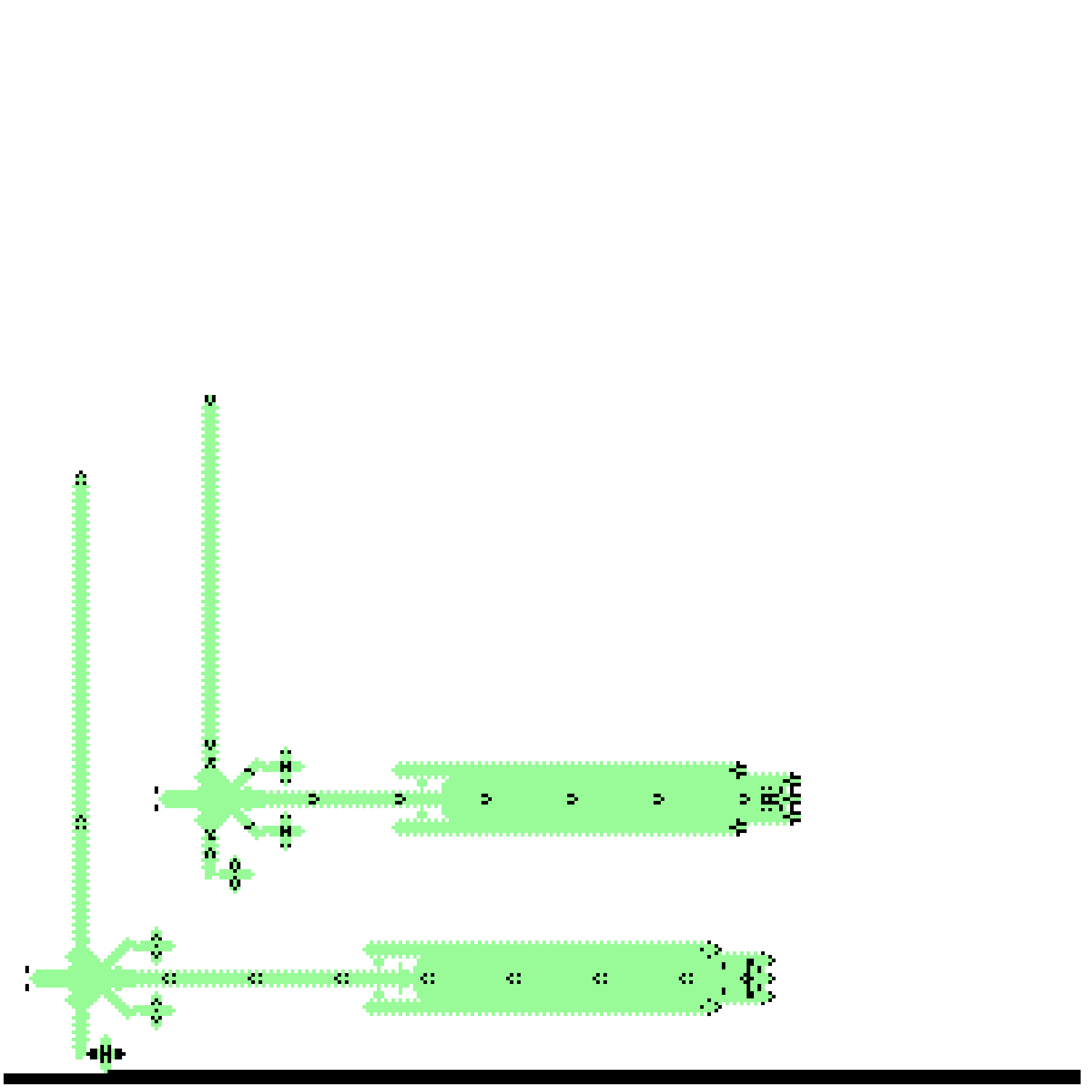}} 
\end{center}
\vspace{-3ex}
\caption[Rake Gc + static]
{\textsf{A rake moving East as in as in figure~\ref{RakeGcB}(a)
sheds Gc gliders moving West,
which interact with stationary oscillating structures, eventually sending  
Gc gliders North every 192 time-steps\cite{Thunk}.
A similar structure was discovered by \cite{Scorbie} based on figure~\ref{RakeGcB}(b).
Dynamic trails=190.
\label{RakeGcTS}
}}
\end{figure}

\begin{figure}[htb]
\begin{center}
\begin{minipage}[c]{1\linewidth}
\fbox{\includegraphics[width=1\linewidth,bb= 46 45 435 165, clip=]{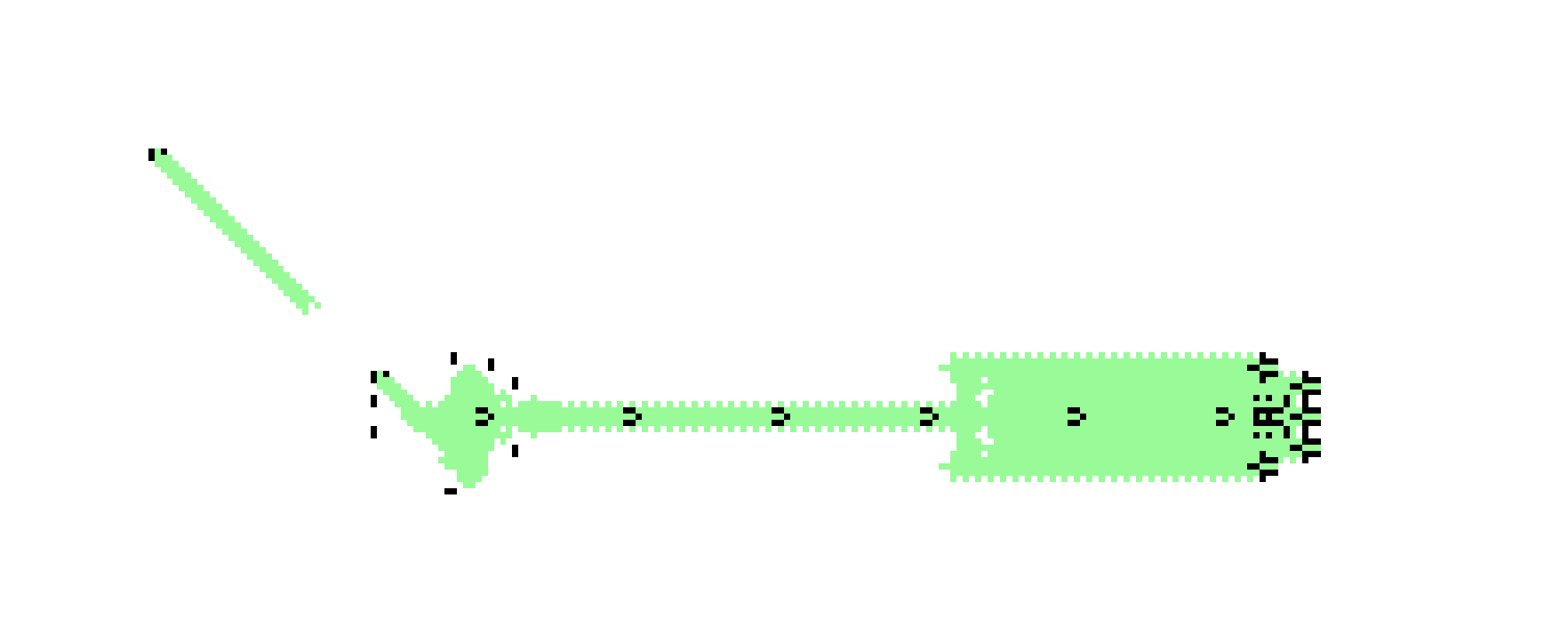}}\\  
\end{minipage}
\end{center}
\vspace{-4.5ex}
\caption[Rake Ga interacts with static structure]
{\textsf{A rake moving East as in figure~\ref{RakeGcB}(b)
sheds Gc gliders moving West,
which interact with a static structure, eventually sending
Ga gliders NW every 144 time-steps\cite{Scorbie}. Dynamic trails=100.
\label{RakeSc16ps}
}}
\end{figure}

\begin{figure}[htb]
\begin{center}
\begin{minipage}[c]{1\linewidth} 
\fbox{\includegraphics[width=1\linewidth,bb= 123 65 379 144, clip=]{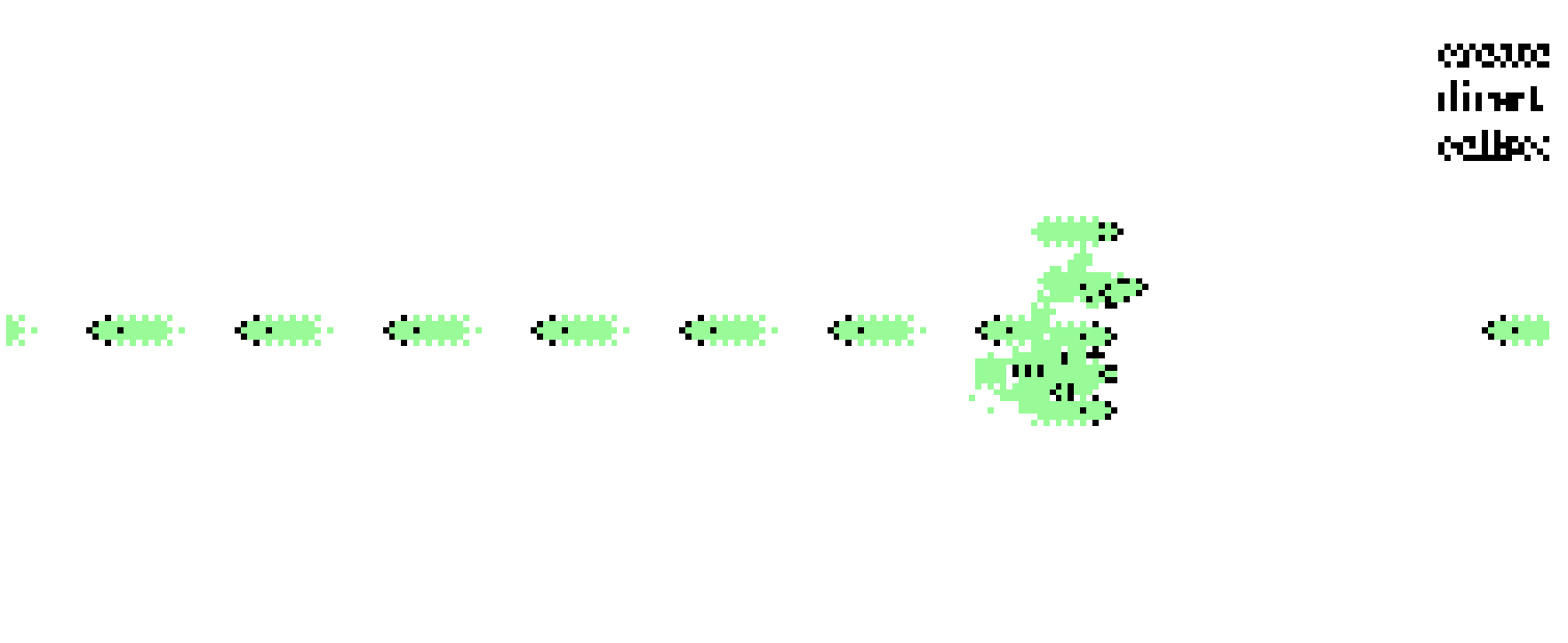}}\\  
\end{minipage}
\end{center}
\vspace{-4.5ex}
\caption[Large Rake Gc]
{\textsf{A rake with a complicated body moving East
($p$=12 $s$=$c/2$) sheds Gc gliders moving West\cite{Wildmyron}.
Dynamic trails=20.
\label{wildm8}
}}
\end{figure}

\begin{figure}[htb]
\begin{center}
\begin{minipage}[c]{1\linewidth}
\begin{minipage}[c]{.47\linewidth} 
   \begin{minipage}[c]{.2\linewidth} 
      \includegraphics[width=1\linewidth,bb= 5 4 59 158]{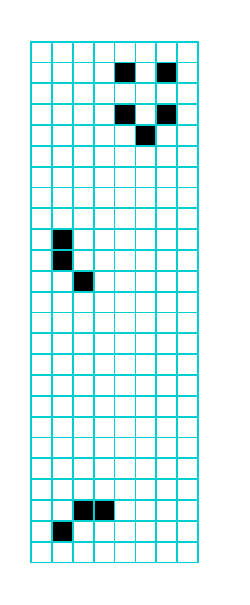}\\[-5.5ex]
           \begin{center} (a) \end{center}
   \end{minipage} 
   \begin{minipage}[c]{.36\linewidth}
      \includegraphics[width=1\linewidth,bb= 5 4 107 99]{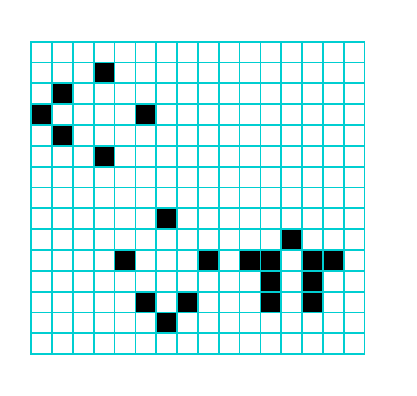}\\[-5.5ex]
            \begin{center} (b) \end{center}
   \end{minipage}
   \begin{minipage}[c]{.27\linewidth} 
    \fbox{\includegraphics[width=1\linewidth,bb= 109 90 161 199, clip=]{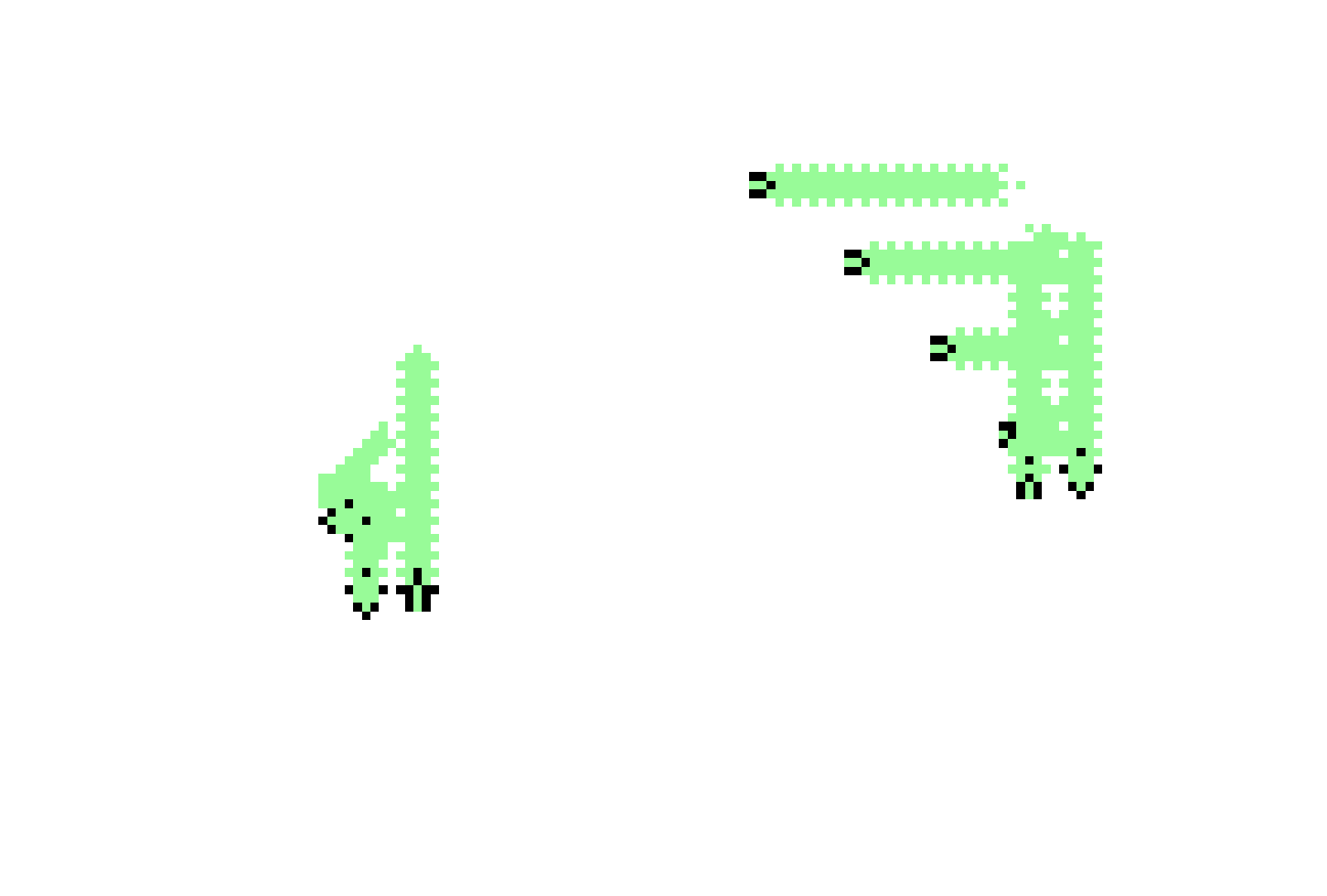}}\\[-4ex]
            \begin{center} (c)  \end{center}
   \end{minipage} 
\end{minipage}
\hfill
\begin{minipage}[c]{.45\linewidth}
    \fbox{\includegraphics[width=1\linewidth,bb= 214 90 392 261, clip=]{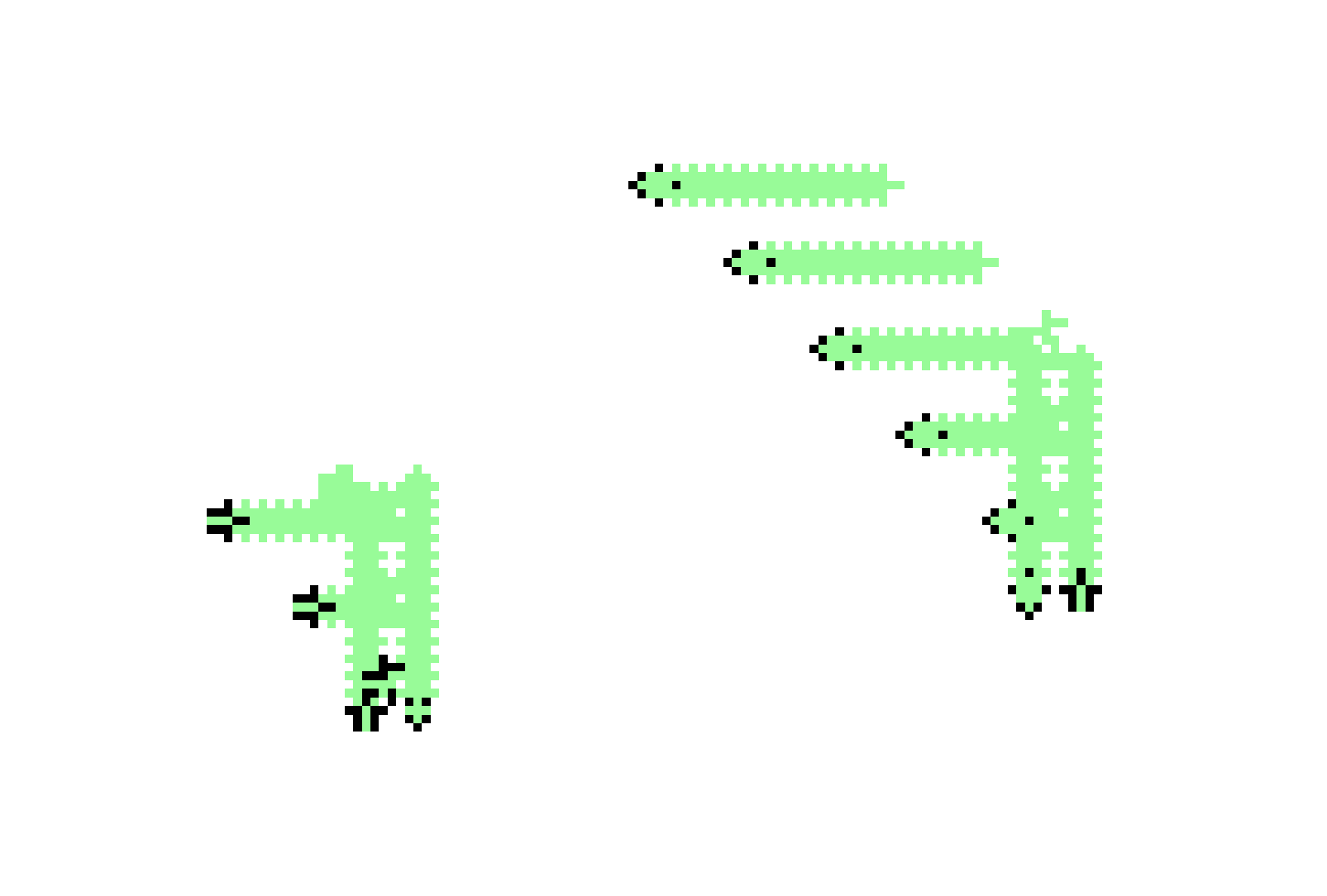}}\\[-4ex]
           \begin{center} (d)  \end{center}
\end{minipage}
\end{minipage}
\end{center}
\vspace{-3.5ex}
\caption[Rake Ga]
{\textsf{
Rake ``RGc90'' ($p$=20, $s$=$c/2$) sheds Gc gliders at 90$^{\circ}$.
RGc90 can be started from (a) a seed found by \cite{Thunk}, which evolves to (b)
a seed found by \cite{Wildmyron}. (c) show the transit from (a) to (b) in 53 time-steps.
(d) shows the subsequent evolution for 80 time-steps, 
where the rake moves South shedding Gc gliders moving West.
Dynamic trails=53.
RGc90 is reproduced by the breeders in section~\ref{Breeders}.
\label{T+S80}
}}  
\end{figure}

A rake is a mobile periodic pattern that
sheds a succession of gliders in its wake, 
including Ga or Gc gliders, or a combination of both, a sort of mobile glider-gun.
In this sense a rake is an adapted spaceship, or a puffer-train if debris is
also left in the wake.
A rake can also be a sub-component in building a ``meta-rake''.
Several rakes of various type and complexity have been discovered
at the ConwayLife forum\cite{ConwayLife-forum}.
Figures~\ref{RakeGaA} to \ref{T+S80} provide details.\\

\begin{figure}[htb]
\begin{center}
\fbox{\includegraphics[width=.57\linewidth,bb= 74 158 335 338, clip=]{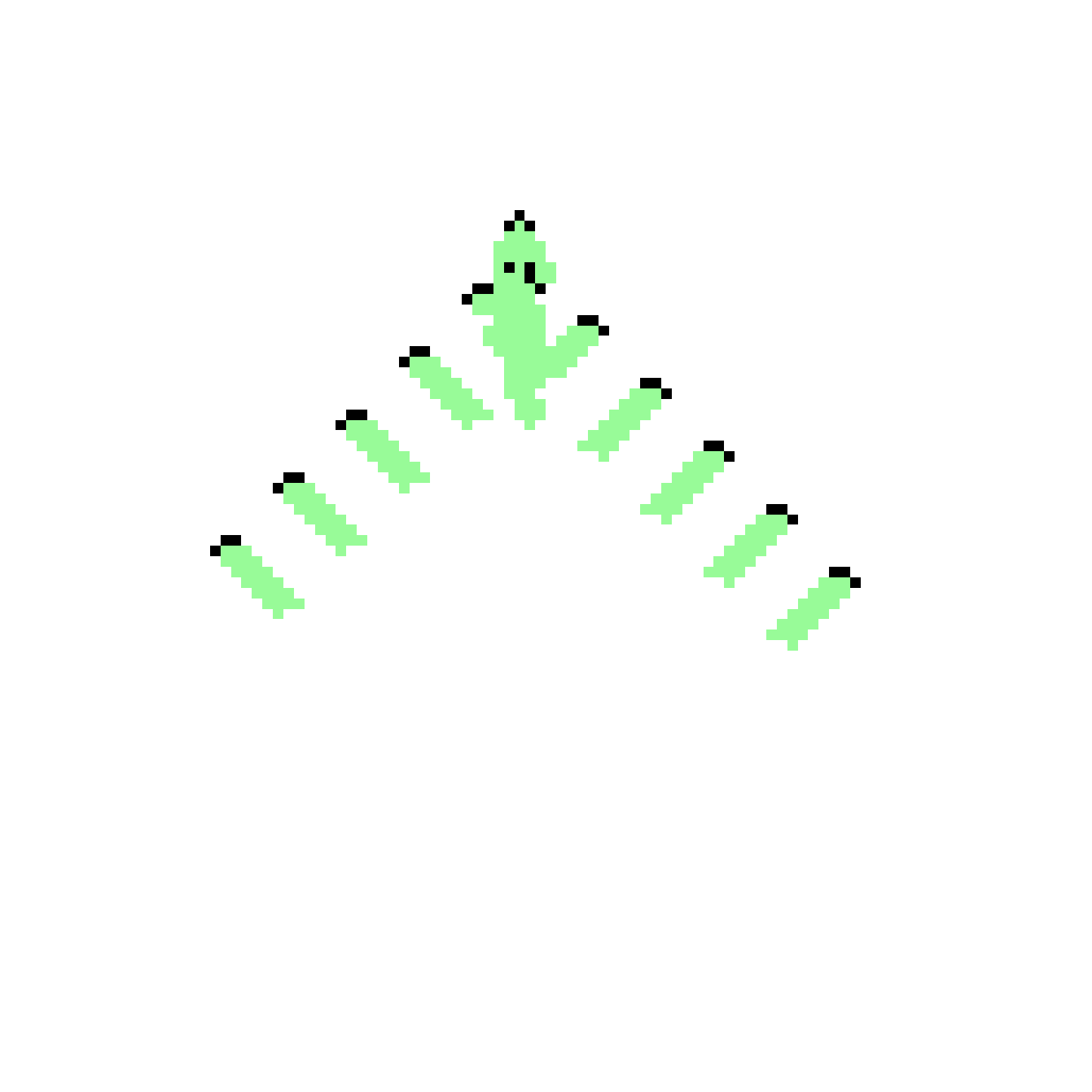}}\\
\caption[Rake Ga]
{\textsf{
A rake\cite{Wildmyron} moving North ($p$=24, $s$=$c/2$) 
sheds two Ga glider streams alternately
NW and NE. Dynamic trails=24.
\label{RakeGaW4b}
}}
\end{center}
\end{figure}

\begin{figure}[htb]
\begin{center}
\begin{minipage}[c]{1\linewidth} 
\fbox{\includegraphics[width=1\linewidth,bb= 52 20 350 104, clip=]{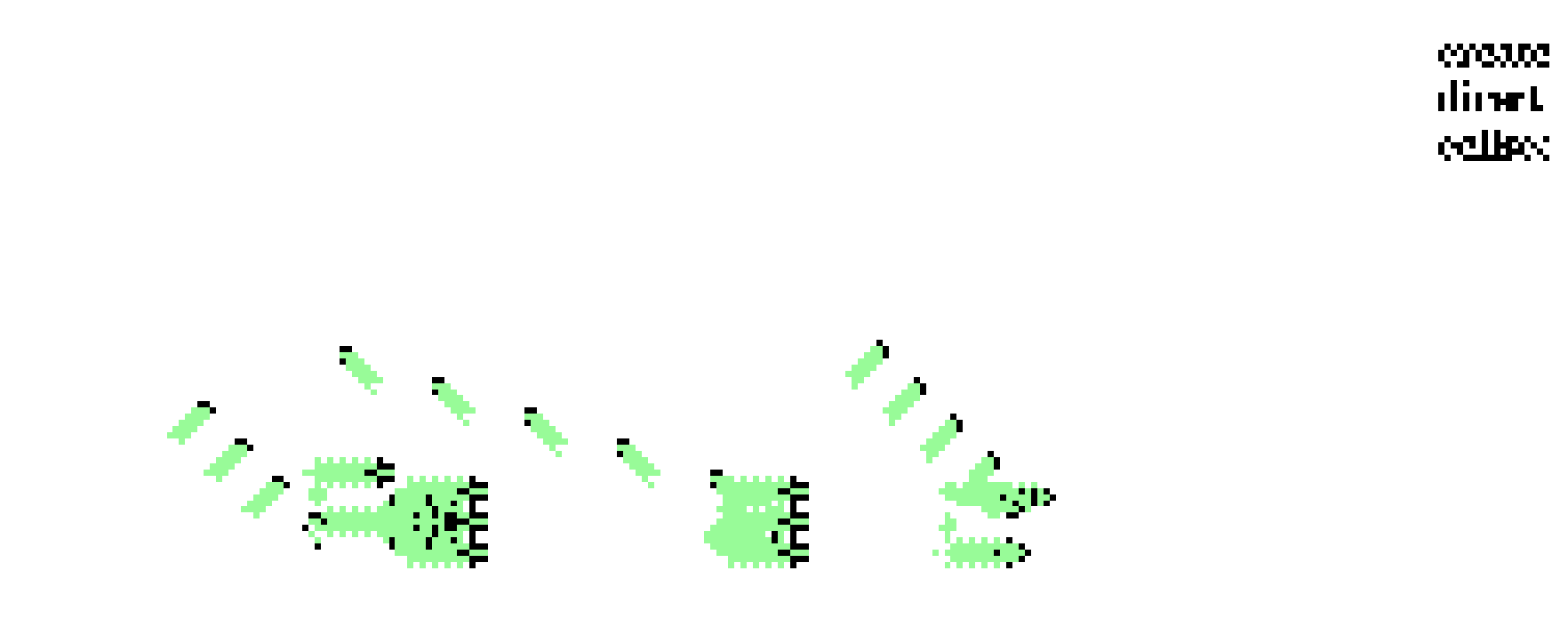}}\\  
\color{white}-----------------------\color{black}(a) $p$=24
\color{white}-------------------------\color{black}(b) $p$=20
\color{white}---------------\color{black}(c) $p$=24
\end{minipage}
\end{center}
\vspace{-3.5ex}
\caption[Rake Ga]
{\textsf{
Rakes moving East shedding Ga gliders. (a) NE \cite{Danieldb} and (b)  NW \cite{Wildmyron}
based on three touching Gc subunits, shed intermediate Gc gliders
which interact with debris, but also clean it up.
(c) \cite{Wildmyron} is the same as the rake in figure~\ref{RakeGaW4b} 
but an accompanying
Gc glider cuts off one Ga stream leaving Ga gliders moving NE.
Periods as noted. Dynamic trails=20.
\label{RakeGaA}
}}
\end{figure}

\clearpage

\clearpage

\subsection{Breeders}
\label{Breeders}

\begin{figure}[htb]
\fbox{\includegraphics[width=1\linewidth,bb= 138 146 382 288, clip=]{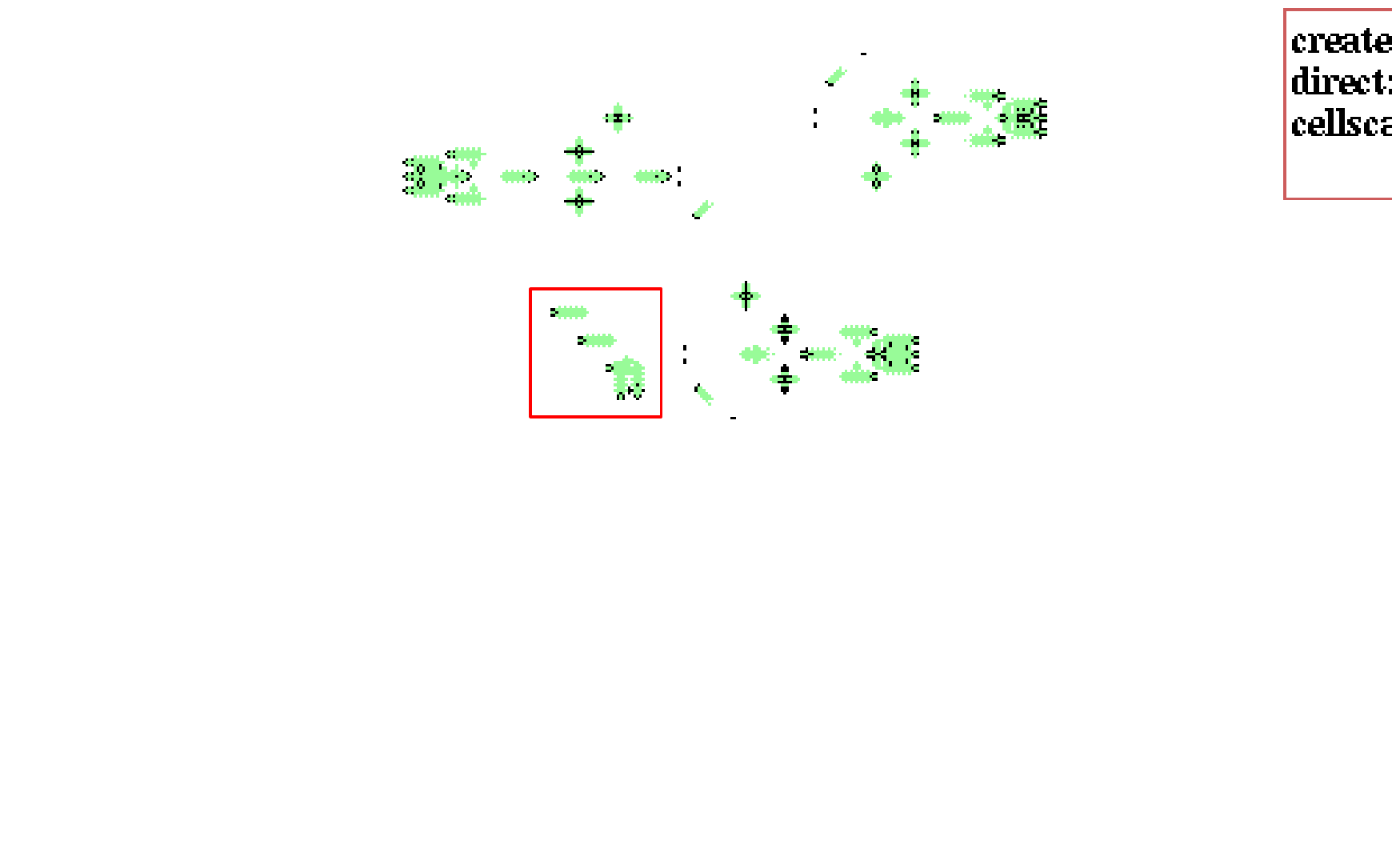}}\\[1ex]
\fbox{\includegraphics[width=1\linewidth,bb= 83 187 353 288, clip=]{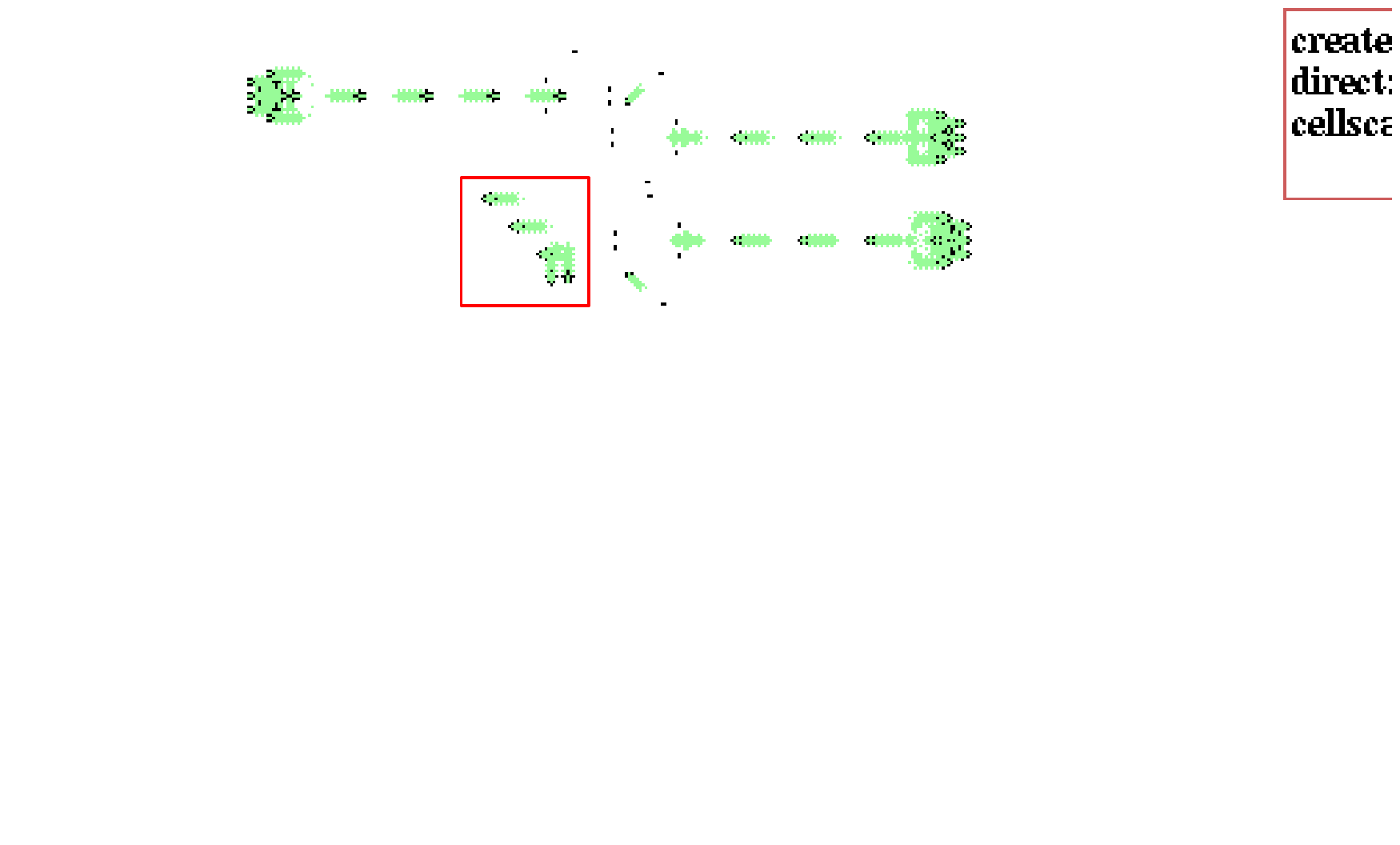}}
\vspace{-4ex}
\caption[Two alternative breeders from rakes]
{\textsf{Two alternative breeders that reproduces the rake ``RGc90'' (red outline) 
from figure~\ref{T+S80}(d), transiting via the seed in figure~\ref{T+S80}(a),
and reproduced every 192 time-steps. The breeders were constructed by \cite{Thunk} from interactions between
three rakes from figure~\ref{RakeGcB}(a) --- the rakes on the right are modified to eject Ga gliders.
Dynamic trails=20. 
\label{BreeT0}
}}
\end{figure}

\begin{figure}[htb] 
\fbox{\includegraphics[width=1\linewidth,bb= 98 136 248 282, clip=]{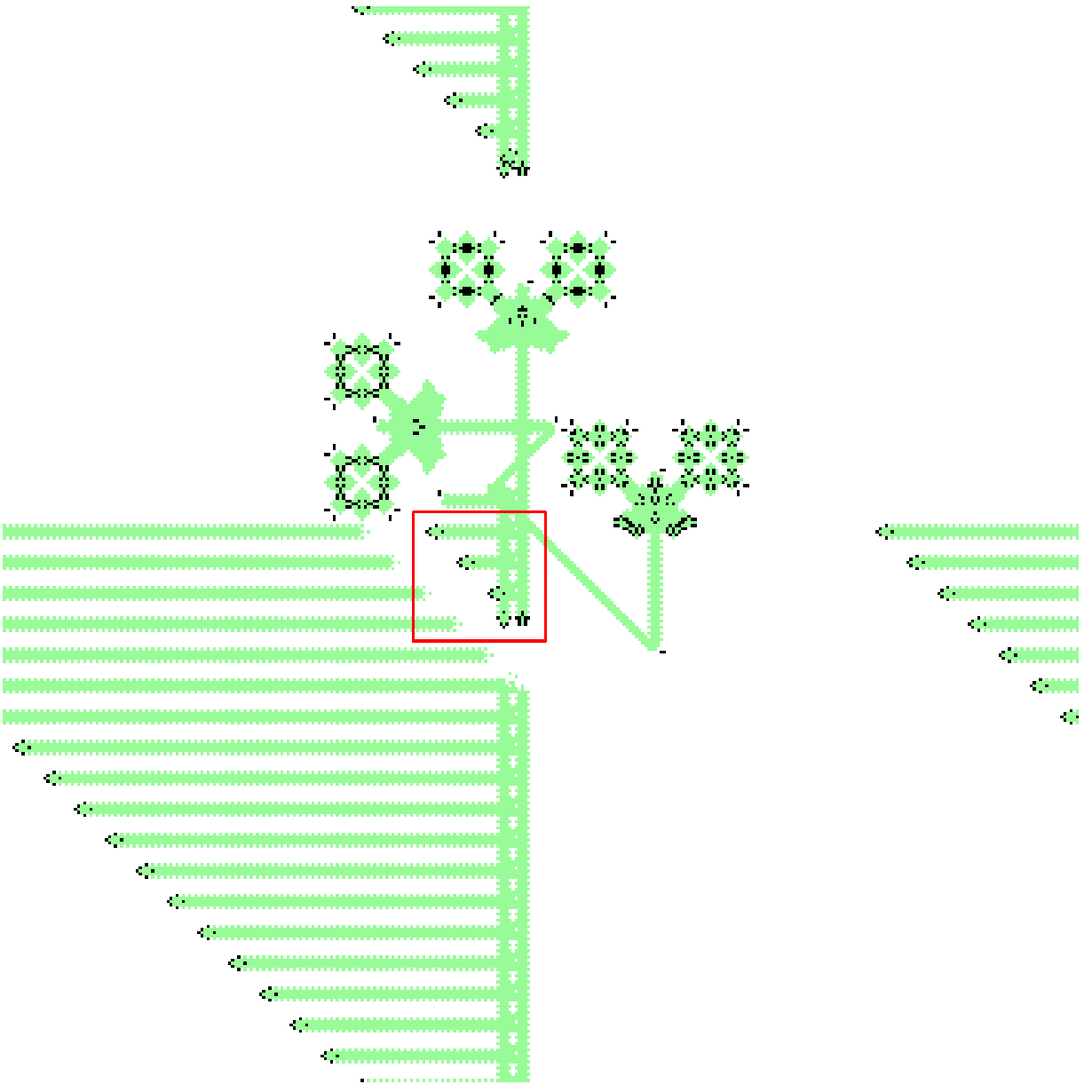}}
\vspace{-3ex}
\caption[A breeder from glider-guns] {\textsf{A breeder made from
    meta-glider-guns by \cite{BlinkerSpawn} for the rake ``RGc90'' (red
    outline) from figure~\ref{T+S80}(d), transiting via the seed in
    figure~\ref{T+S80}(a), and reproduced every 409
    time-steps. This static breeder is constructed from three
    interacting ``QuadGG2a'' glider-guns
    (figure~\ref{metaGG}(a). Dynamic trails=360.
    Its feasible to represent such long trails because the
    breeder itself is static (it does not expand as in
    figure~\ref{BreeT0}).  The horizontal trails in the lower-left
    are the ends of the dynamic trails from the RGc90 rake that was
    ejected previously.
\label{BreeTb0}
}}
\end{figure}

\begin{figure}[htb] 
\fbox{\includegraphics[width=1\linewidth,bb= 96 192 493 406, clip=]{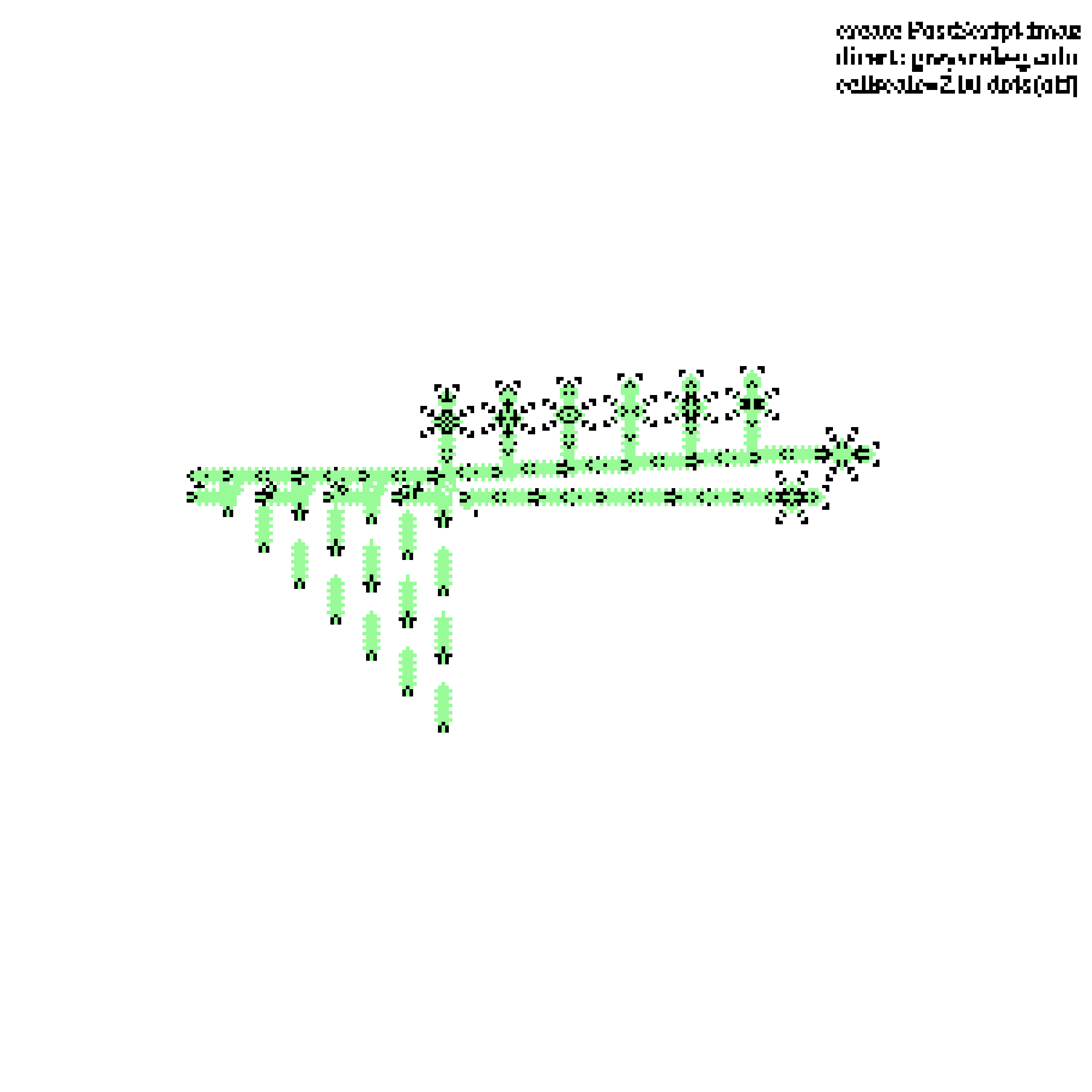}}
\vspace{-3ex}
\caption[A breeder from 8 glider-guns] {\textsf{A breeder from 8 GGc
    glider-guns built by \cite{BlinkerSpawn}.  Along the top, 6 GGc's
    are aligned horizontally, one slightly above the next, from left
    to right, sending 6 Gc glider-streams South.  The two remaining
    GGc's are positioned above each other on the right and eject Gc
    gliders West.  The upper Westward glider-stream is deflected
    downward (by the 6 top GGc's) and the two Westward glider-streams
    finally merge into a complex combined glider-stream, creating a sort of
    rake that continues to sends Gc glider-streams South (at
    90$^{\circ}$) all along its length, starting at the tip.  An eater
    below the merging zone is an essential component.  Dynamic trails=20.
\label{blink33}
}}
\end{figure}

\enlargethispage{6ex}
\noindent Whereas a glider-gun or a rake ejects a stream of gliders,
a breeder is a pattern that ejects a stream of
glider-guns, rakes, or puffer-trains, in various combinations.
Breeders are said to exhibit unbounded quadratic growth
by creating multiple copies of a second object, each of which creates multiple
copies of a third object\cite{ConwayLife-breeder}.
The significance of breeders lies in their high level of complexity
demonstrating open ended pattern evolution, 
as well as providing further components for computation and memory. 
Breeders have been constructed in the Precursor-Rule
at the ConwayLife forum\cite{ConwayLife-forum}.
In figures~\ref{BreeT0} to \ref{blink33}, we show four examples of breeders
ejecting rakes.

\clearpage

\section{Concluding remarks}
\label{Concluding remarks}

The X-Rule's isotropic precursor was the original cellular automaton
that we studied in our search for universal computation, but at the
time we found it necessary to modify the rule to find glider-guns ---
we were then able to demonstrate logical universality in the
anisotropic X-rule\cite{Gomez2015}.  After announcing the X-rule and
its precursor on ConwayLife\cite{ConwayLife-forum}, members of the
forum applied their considerable know-how in Game-of-Life pattern
search to discover glider-guns in the isotropic Precursor-Rule. 
As well as glider-guns, many other important complex dynamical
mechanisms have been constructed and we have presented a selection,
which incidentally shows the power of diversified search in a dedicated community.

Armed with glider-guns, we were able to construct the logical gates
and demonstrate logical universality in the Precursor-Rule, which
like the Game-of-Life exhibits an extraordinary diversity of dynamics,
but according to a rule not based on birth/survival logic. The results
documented in this paper are an initial exploration --- further
interesting dynamics can be discovered possibly including
memory functions required for universality in the Turing sense. The
dynamics are open-ended and impossible to pin down within a
sufficiently large space-time.

Although the Game-of-Life has accumulated a vast compendium of
behaviour, it can be argued that the Precursor-Rule has a more diverse
range of basic gliders and glider-guns, providing a richer diversity
of the fundamental particles from which more complex structures can be
built.  
\clearpage

The Precursor-Rule belongs to the ordered zone in rule-space within
the input-entropy scatter-plot (figure~\ref{scatter-plot}), the zone
with low values of entropy variability and mean entropy.  It seems
that the ingredients that enable logical universality --- gliders,
eaters, and crucially glider-guns --- are more likely to occur in this
zone, rather than in the ``complex'' zone with high entropy
variability, were activity tends to overwhelm stability. Although 2D
binary cellular automata rules supporting glider-guns are exceedingly
rare, it appears nonetheless that many such rules are to be found in
this zone, which begs the question, what are the underlying principles
for the existence of glider-guns?

\subsection{SansDomino rule-space}
\label{SansDomino rule-space}

\begin{figure}[htb]
\begin{center}
\begin{minipage}[c]{1\linewidth}
\begin{minipage}[c]{.6\linewidth} 
\fbox{\includegraphics[width=1\linewidth,bb= 92 115  299 257, clip=]{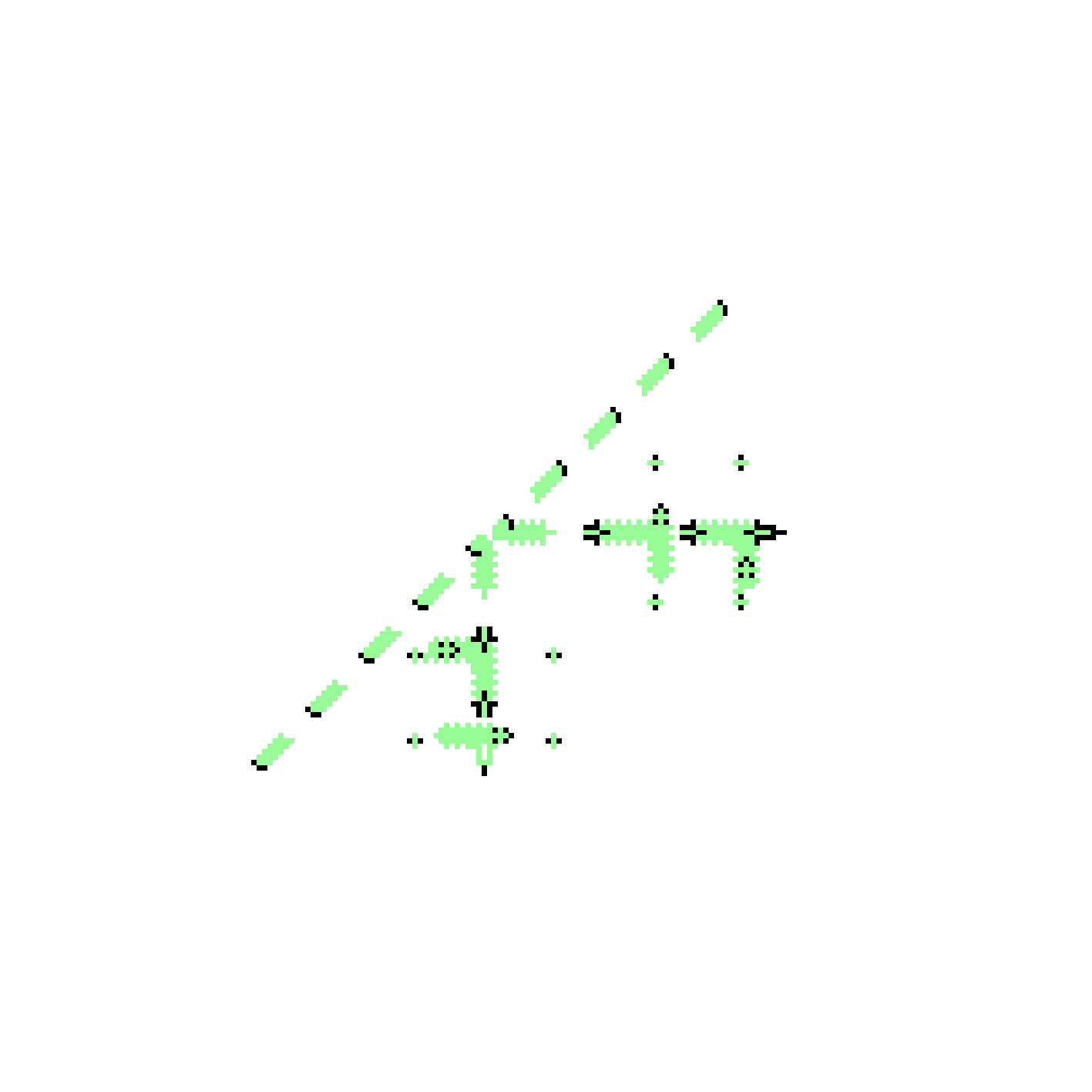}}\\ 
\end{minipage}
\hfill
\begin{minipage}[c]{.32\linewidth}
\caption[Rake Ga]
{\textsf{A glider-gun from ``SansDomino'' rule-space\cite{Tropylium} based on birth/survival (B2/S14)
featuring identical Ga and similar Gc gliders to the Precursor-Rule.
Dynamic trails=20.
\label{GGS14-3}}}  
\end{minipage}
\end{minipage}
\end{center}
\vspace{-2ex}
\end{figure}

We have recently become aware of ``SansDomino'' rule-space\cite{Tropylium}
where rules can support gliders similar to Ga and Gc gliders in the Precursor-Rule,
with analogous glider-guns shooting these gliders. These rules are based on modified
birth/survival (B2/S13 and B2/S14) with the exception that
an adjoining pair of 1s in the outer neighborhood
outputs zero. An example of such a modified B2/S14 glider-gun is shown in figure~\ref{GGS14-3}.
It will be important to investigate the relationship between the Precursor-Rule and
`SansDomino'' rules.

\section{Acknowledgements}
\label{Acknowledgements}

Experiments were done with Discrete Dynamics Lab
\cite{Wuensche2016,Wuensche-DDLab}, Mathematica and Golly.  The Precursor-Rule was
found during a collaboration at June workshops in 2013 and 2014 at
the DDLab Complex Systems Institute in Ariege, France, and also at the
Universidad Aut\'onoma de Zacatecas, M\'exico, and in London, UK.
Later patterns were discovered during interactions with the ConwayLife
forum\cite{ConwayLife-forum} where many people made important
contributions.  J.M. G\'omez Soto also acknowledges his residency at
the DDLab Complex Systems Institute, and financial support from the
Research Council of Zacatecas (COZCyT).

\end{document}